\documentclass[aps,superscriptaddress,nofootinbib,eqsecnum,tightenlines,floatfix]{revtex4}
\usepackage{epsfig,rotating}
\usepackage{amsmath,amssymb}
\usepackage{dsfont}
\numberwithin{equation}{section}
\usepackage{graphics}
\usepackage{rotate}
\usepackage{psfrag}
\begin{document}
\newcommand{\D}{\displaystyle} 
\newcommand{\T}{\textstyle} 
\newcommand{\SC}{\scriptstyle} 
\newcommand{\SSC}{\scriptscriptstyle} 
\newcommand{\be}{\begin{equation}}
\newcommand{\ee}{\end{equation}}
\newcommand{\avg}[1]{\langle #1 \rangle}
\newcommand{\vx}{{\boldsymbol{x}}}
\newcommand{\rv}{{\boldsymbol{r}}}
\newcommand{\vq}{\ensuremath{\vec{q}}}
\newcommand{\pv}{\ensuremath{\vec{p}}}
\def\AJ{{\it Astron. J.} }
\def\ARAA{{\it Annual Rev. of Astron. \& Astrophys.} }
\def\ApJ{{\it Astrophys. J.} }
\def\ApJL{{\it Astrophys. J. Letters} }
\def\ApJS{{\it Astrophys. J. Suppl.} }
\def\ApP{{\it Astropart. Phys.} }
\def\AA{{\it Astron. \& Astroph.} }
\def\AAR{{\it Astron. \& Astroph. Rev.} }
\def\AAL{{\it Astron. \& Astroph. Letters} }
\def\AASu{{\it Astron. \& Astroph. Suppl.} }
\def\AN{{\it Astron. Nachr.} }
\def\IJMP{{\it Int. J. of Mod. Phys.} }
\def\JGR{{\it Journ. of Geophys. Res.}}
\def\JHEP{{\it Journ. of High En. Phys.} }
\def\JPhG{{\it Journ. of Physics} {\bf G} }
\def\MNRAS{{\it Month. Not. Roy. Astr. Soc.} }
\def\Nature{{\it Nature} }
\def\NewAR{{\it New Astron. Rev.} }
\def\NJPh{{\it New Journ. of Phys.} }
\def\PASP{{\it Publ. Astron. Soc. Pac.} }
\def\PhFl{{\it Phys. of Fluids} }
\def\PLB{{\it Phys. Lett.}{\bf B} }
\def\PhysRep{{\it Phys. Rep.} }
\def\PR{{\it Phys. Rev.} }
\def\PRD{{\it Phys. Rev.} {\bf D} }
\def\PRL{{\it Phys. Rev. Letters} }
\def\RMP{{\it Rev. Mod. Phys.} }
\def\Science{{\it Science} }
\def\ZfA{{\it Zeitschr. f{\"u}r Astrophys.} }
\def\ZfN{{\it Zeitschr. f{\"u}r Naturforsch.} }
\def\etal{{\it et al.}}
\hyphenation{mono-chro-matic sour-ces Wein-berg
chang-es Strah-lung dis-tri-bu-tion com-po-si-tion elec-tro-mag-ne-tic
ex-tra-galactic ap-prox-i-ma-tion nu-cle-o-syn-the-sis re-spec-tive-ly
su-per-nova su-per-novae su-per-nova-shocks con-vec-tive down-wards
es-ti-ma-ted frag-ments grav-i-ta-tion-al-ly el-e-ments me-di-um
ob-ser-va-tions tur-bul-ence sec-ond-ary in-ter-action
in-ter-stellar spall-ation ar-gu-ment de-pen-dence sig-nif-i-cant-ly
in-flu-enc-ed par-ti-cle sim-plic-i-ty nu-cle-ar smash-es iso-topes
in-ject-ed in-di-vid-u-al nor-mal-iza-tion lon-ger con-stant
sta-tion-ary sta-tion-ar-i-ty spec-trum pro-por-tion-al cos-mic
re-turn ob-ser-va-tion-al es-ti-mate switch-over grav-i-ta-tion-al
super-galactic com-po-nent com-po-nents prob-a-bly cos-mo-log-ical-ly
Kron-berg Berk-huij-sen}

\def\refalt@jnl#1{{\rm #1}}

\def\aj{\refalt@jnl{AJ}}                   
\def\araa{\refalt@jnl{ARA\&A}}             
\def\apj{\refalt@jnl{ApJ}}                 
\def\apjl{\refalt@jnl{ApJ}}                
\def\apjs{\refalt@jnl{ApJS}}               
\def\ao{\refalt@jnl{Appl.~Opt.}}           
\def\apss{\refalt@jnl{Ap\&SS}}             
\def\aap{\refalt@jnl{A\&A}}                
\def\aapr{\refalt@jnl{A\&A~Rev.}}          
\def\aaps{\refalt@jnl{A\&AS}}              
\def\azh{\refalt@jnl{AZh}}                 
\def\baas{\refalt@jnl{BAAS}}               
\def\jrasc{\refalt@jnl{JRASC}}             
\def\memras{\refalt@jnl{MmRAS}}            
\def\mnras{\refalt@jnl{MNRAS}}             
\def\pra{\refalt@jnl{Phys.~Rev.~A}}        
\def\prb{\refalt@jnl{Phys.~Rev.~B}}        
\def\prc{\refalt@jnl{Phys.~Rev.~C}}        
\def\prd{\refalt@jnl{Phys.~Rev.~D}}        
\def\pre{\refalt@jnl{Phys.~Rev.~E}}        
\def\prl{\refalt@jnl{Phys.~Rev.~Lett.}}    
\def\pasp{\refalt@jnl{PASP}}               
\def\pasj{\refalt@jnl{PASJ}}               
\def\qjras{\refalt@jnl{QJRAS}}             
\def\skytel{\refalt@jnl{S\&T}}             
\def\solphys{\refalt@jnl{Sol.~Phys.}}      
\def\sovast{\refalt@jnl{Soviet~Ast.}}      
\def\ssr{\refalt@jnl{Space~Sci.~Rev.}}     
\def\zap{\refalt@jnl{ZAp}}                 
\def\nat{\refalt@jnl{Nature}}              
\def\iaucirc{\refalt@jnl{IAU~Circ.}}       
\def\aplett{\refalt@jnl{Astrophys.~Lett.}} 
\def\apspr{\refalt@jnl{Astrophys.~Space~Phys.~Res.}}
\def\bain{\refalt@jnl{Bull.~Astron.~Inst.~Netherlands}}
\def\fcp{\refalt@jnl{Fund.~Cosmic~Phys.}}  
\def\gca{\refalt@jnl{Geochim.~Cosmochim.~Acta}}   
\def\grl{\refalt@jnl{Geophys.~Res.~Lett.}} 
\def\jcp{\refalt@jnl{J.~Chem.~Phys.}}      
\def\jgr{\refalt@jnl{J.~Geophys.~Res.}}    
\def\jqsrt{\refalt@jnl{J.~Quant.~Spec.~Radiat.~Transf.}}
\def\memsai{\refalt@jnl{Mem.~Soc.~Astron.~Italiana}}
\def\nphysa{\refalt@jnl{Nucl.~Phys.~A}}   
\def\physrep{\refalt@jnl{Phys.~Rep.}}   
\def\physscr{\refalt@jnl{Phys.~Scr}}   
\def\planss{\refalt@jnl{Planet.~Space~Sci.}}   
\def\procspie{\refalt@jnl{Proc.~SPIE}}   

\let\astap=\aap
\let\apjlett=\apjl
\let\apjsupp=\apjs
\let\applopt=\ao

\title{\Large Towards the Chalonge 16th Paris Cosmology Colloquium 2012:

\medskip

HIGHLIGHTS and CONCLUSIONS

of the Chalonge 15th Paris Cosmology Colloquium 2011 

\medskip

\Large FROM COLD DARK MATTER TO WARM DARK MATTER IN THE STANDARD MODEL OF THE UNIVERSE. THEORY AND OBSERVATIONS:

\medskip
  
Ecole Internationale d'Astrophysique Daniel Chalonge

Observatoire de Paris

in the historic Perrault building, 20-22 July 2011.}

\author{\Large \bf   H. J. de Vega$^{(a,b)}$,  M.C. Falvella$^{(c)}$, N. G. Sanchez$^{(b)}$}

\date{\today}

\affiliation{$^{(a)}$ LPTHE, Universit\'e
Pierre et Marie Curie (Paris VI) et Denis Diderot (Paris VII),
Laboratoire Associ\'e au CNRS UMR 7589, Tour 24, 5\`eme. \'etage, 
Boite 126, 4, Place Jussieu, 75252 Paris, Cedex 05, France. \\
$^{(b)}$ Observatoire de Paris,
LERMA. Laboratoire Associ\'e au CNRS UMR 8112.
 \\61, Avenue de l'Observatoire, 75014 Paris, France.\\
$^{(c)}$ Italian Space Agency and MIUR, Viale Liegi n.26,
00198 Rome, Italy.}


\begin{abstract}

The Chalonge 15th Paris Cosmology Colloquium 2011 was held on 20-22 July in the historic 
Paris Observatory's Perrault building, in the
Chalonge School spirit combining real cosmological/astrophysical data and hard theory
predictive approach connected to them in the Warm Dark Matter Standard Model of the Universe:
News and reviews from {\it Herschel}, QUIET, Atacama Cosmology Telescope (ACT), 
South Pole Telescole (SPT), {\it Planck}, PIXIE, the JWST, UFFO, KATRIN and MARE experiments; 
astrophysics, particle physics and nuclear physics warm dark matter (DM) searches and galactic observations, 
related theory and simulations,
with the aim of synthesis, progress and clarification.  Philippe Andre, Peter Biermann, Pasquale Blasi, 
Daniel Boyanovsky, Carlo Burigana, Hector de Vega, Joanna Dunkley, Gerry Gilmore, 
Alexander Kashlinsky, Alan Kogut, Anthony Lasenby,  John Mather, Norma  Sanchez, Alexei Smirnov,  
Sylvaine Turck-Chieze present here their highlights of the Colloquium. Ayuki Kamada and Sinziana 
Paduroiu present here their poster highlights. $\Lambda$WDM (Warm Dark Matter) is progressing 
impressively 
over $\Lambda$CDM whose galactic scale crisis and decline are staggering. 
The International School Daniel Chalonge issued an statement of strong support to the 
{\it James Webb Space Telescope} (JWST). The  Daniel Chalonge Medal 2011 was awarded to 
John C. Mather, Science PI of the JWST.  Summary and conclusions by H. J. de Vega, 
M. C. Falvella and N. G. Sanchez stress among other points: 

\medskip
 
{\bf (I)} Data confirm primordial CMB gaussianity. Inflation effective theory predicts
negligible primordial non-gaussianity, negligible scalar index running  
and a tensor to scalar ratio $ r \sim 0.05-0.04 $ at reach/border line of next CMB observations;
the present data with this theory clearly prefer new inflation and the `cosmic banana' region in the 
$r$ vs $n_s$
diagram; early fast-roll inflation is generic and provides the lowest multipoles depression and 
TE oscillations.
Impressive new CMB high-$l$ data were released from ACT and SPT, the CMB power spectrum displays now 
$9$ clear peaks, various cosmic parameter degeneracies are now resolved, ACT observed primordial 
helium and detected the CMB lensing, CMB alone can provide now evidence for dark energy. WMAP7 
Sunyaev-Zeldovich (SZ) amplitudes  are smaller of 0.5-0.7 than expected with respect to X-ray models, 
relaxed and non-relaxed clusters need distinction; the {\it Planck} first released results did not 
found such SZ depression. The Primordial Inflation Explorer (PIXIE) is planned to test inflation and 
to robustly detect CMB tensor (B-mode) polarisation (primordial gravitational waves) to limit 
$ r < 10^{-3} $ at more than 5 sigma. PIXIE also allows to test keV warm dark-matter and reionization. 
On the other hand, cluster peculiar flows of about $600-1000 ~ km/sec$ and coherent length 
$ \sim 750 ~ Mpc $ seem to be measured and would require if confirmed a fast-roll inflationary 
explanation. \\
 
{\bf (II)} Significant progresses are made in dSph galaxy observations: Very low star formation rates 
and very low baryonic feedback are found. Baryon feedback does not affect DM structure, CDM does not 
produce realistic galaxies with realistic feedback and star formation recipes. Very precise kinematics 
of stars in dSph is probing directly DM density profiles: cusped profiles are excluded at the 
$ 95 \%  CL $. Cored (non cusped) DM halos and warm (keV scale mass) DM are strongly favored from 
theory and astrophysical observations, they naturally produce the observed small scale structures; 
keV sterile neutrinos of mass  between $1$ and $10$ keV are the most serious candidates. Progresses 
in computing the distribution function for WDM sterile neutrinos as well as $\Lambda$WDM simulations 
with them were reported. Wimps (heavier than 1 GeV) are strongly disfavoured combining theory with 
galaxy observations. In addition,  
the observed cosmic ray positron and electron excesses are explained by natural astrophysical 
processes, while annihilating/decaying cold dark matter models are strongly disfavored. On the 
other hand, non-annihilating CDM wimps in the sun are excluded. The sun seismic model can be also 
used to constrain WDM (sterile neutrinos).  \\

{\bf (III)} {\it Herschel} remarkably observed a profusion of interstellar pc-scale filaments sharing 
a common width $ \sim 0.1 pc$ and a critical mass per unit length $ \sim 15\, M_\odot$/pc, above which 
they are gravitationally unstable. This filamentary structure is intimately connected to the cloud core 
formation process. Remarkably, the independently   observed interstellar surface density (colum density)
 and galaxy surface density have almost the same universal value, irrespective of size and composition, 
morphology and system types.

\medskip

Putting all together, evidence that $\Lambda$CDM does not work at small scales is staggering. 
$ \Lambda$WDM simulations with 1 keV particles reproduce remarkable well the observations, the 
distribution of Milky Way satellites in the range above $\sim 40$ kpc, sizes of local minivoids and 
velocity functions. Overall, $\Lambda$WDM  and keV scale DM particles deserve dedicated
astronomical and laboratory experimental searches, theoretical work and simulations. KATRIN experiment 
in the future
could perhaps adapt its set-up to look to keV scale sterile neutrinos. It will be a a fantastic 
discovery to 
detect dark matter in a beta decay.  Photos of the Colloquium are included.

\end{abstract}

\maketitle

\newpage

\tableofcontents

\newpage

\section{Purpose of the Colloquium, Context and Introduction}

The main aim of the series `Paris Cosmology Colloquia', in the framework of the International 
School of Astrophysics  {\bf `Daniel Chalonge'}, is to put together real data : cosmological, 
astrophysical, particle physics, nuclear physics data, and hard theory predictive approach 
connected to them in the framework of the   Standard Model of the Universe. 

\medskip

The Chalonge Paris Cosmology Colloquia 
bring together physicists, astrophysicists and astronomers from the world over. Each year these 
Colloquia are more attended and appreciated both by PhD students, post-docs, senior participants 
and lecturers. 
The format of the Colloquia is intended to act as a true working meeting and laboratory of ideas, 
and allow easy and fruitful mutual contacts and communication.

\medskip

The discussion sessions are as important as the lectures themselves.

\bigskip

The {\bf 15th Paris Cosmology Colloquium 2011}  was devoted to `FROM COLD DARK 
MATTER TO WARM DARK MATTER IN THE STANDARD MODEL OF THE UNIVERSE: THEORY AND OBSERVATIONS'. 

\bigskip

The  Colloquium took  place during three full days  (Wednesday July 20, Thursday 21 and 
Friday July 22) at the parisian campus of  Paris Observatory (HQ), in the historic 
Perrault building.

\bigskip

Never as in this period, the Golden 
Age of Cosmology, the major subjects of the Daniel Chalonge School were so timely and in 
full development: Recently, Warm (keV scale) Dark Matter (WDM) emerged impressively over CDM (Cold Dark Matter) as the leading Dark Matter candidate. Astronomical evidence that Cold Dark Matter (LambdaCDM) and its proposed tailored cures do not work at small scales is staggering. LambdaWDM solves naturally the problems of LambdaCDM and agrees remarkably well with the observations at small as well as large and cosmological scales. LambdaWDM numerical simulations naturally  {\it agree} with observations at all scales, in contrast to LambdaCDM simulations which only agree at large scales.

\bigskip

In the context of the new Dark Matter situation represented by Warm (keV scale) Dark Matter which implies novelties in the astrophysical, cosmological and keV particle and nuclear physics context, this 15th Paris Colloquium 2011 was devoted to the LambdaWDM Standard Model of the Universe.

\bigskip

Strong {\bf Support to the James Webb Space Telescope} (JWST) was provided during this Colloquium and a Statement of Support to 
the JWST from the Chalonge School was written and made public during the Open Session and largely advertised worldwide: after the Colloquium. See section IV devoted to the JWST Support in these Highlights.

\bigskip

An {\bf Open Session} took place before the end of the Colloquium with two Lectures:  The James Webb Space Telescope (JWST) by John Mather, Science PI of the JWST,  and the Ultra-Fast Flash Observatory (UFFO) Pathfinder by George Smoot, both Nobel Laureats of physics and of the Daniel Chalonge Medal.

\medskip

The {\bf Daniel Chalonge Medal 2011} was awarded to John Mather during the Open Session of the Colloquium for his multiple achievements, COBE, and his brilliant activity in the JWST. See section V devoted to the Daniel Chalonge Medal 2011 in these Highlights.

\bigskip

\begin{center}

{\bf The Main topics of this Colloquium included} :  

\end{center}

\medskip

Observational and theoretical progress on the nature of dark matter : keV scale Warm Dark Matter.\\ 
\medskip
Dark energy: cosmological constant: the quantum energy of the cosmological vacuum. \\
\medskip
Large and small scale structure formation in agreement with observations at large scales and small (galactic) scales.\\
\medskip
The ever increasing problems of LambdaCDM. \\
\medskip
The emergence of Warm (keV scale) Dark Matter from theory and observations.\\ 
\medskip
Neutrinos in astrophysics and cosmology.\\
\medskip
The new serious dark matter candidate: Sterile neutrinos at the keV scale. \\ 
\medskip
Neutrino mass bounds from cosmological data and from high precision beta decay experiments.\\
\medskip
The analysis of the CMB+LSS+SN data with the effective (Ginsburg-Landau) effective theory of inflation: New Inflation (double well inflaton potential) strongly favored by the CMB + LSS + SN data.\\ 
\medskip
The presence of the lower bound for the primordial gravitons (non vanishing tensor to scalar ratio r) with the present CMB + LSS + SN data. CMB polarization and forecasts for Planck. \\
\medskip
CMB measurements. The Atacama Cosmology Telescope. The {\it Planck} mission, its science perspectives and results.\\ 
\medskip
Recent {\it Herschel} results from the ISM to the IMF. \\
\medskip
Recent Herschel-SPIRE Legacy Survey (HSLS) results for cosmology and dark Matter.\\
\medskip
The James Webb Space Telescope: mission and science

\bigskip

\begin{center}

{\bf Context, CDM Crisis and CDM Decline:} 

\end{center}

\bigskip

On large cosmological scales, CDM agrees in general with observations but CDM does not agree with observations on galaxy scales and small scales. Over most of twenty years, increasing number of cyclic arguments and ad-hoc mechanisms or recipes were-and continue to be- introduced in the CDM galaxy scale simulations, in trying to deal with the CDM small scale crisis: Cusped profiles and overabundance of substructures are predicted by CDM. Too many satellites are predicted by CDM simulations while cored profiles and no such overabundant substructures are seen by astronomical observations. Galaxy formation within CDM is increasingly confusing and in despite of the proposed cures, does not agree with galaxy observations.

\bigskip

On the CDM particle physics side, the situation is no less critical: So far, all the dedicated experimental searches after most of twenty years to find the theoretically proposed CDM particle candidate (WIMP) have failed. The CDM indirect searches (invoking CDM annihilation) to explain cosmic ray positron excesses, are in crisis as well, as wimp annihilation models are plagued with growing tailoring or fine tuning, and in any case, such cosmic rays excesses are well explained and reproduced by natural astrophysical process and sources. The so-called and repeatedly invoked 'wimp miracle' is nothing but been able to solve one equation with three unknowns (mass, decoupling temperature, and annihilation cross section) within wimp models theoretically motivated by SUSY model building twenty years ago (at that time those models were fashionable and believed for many proposals).

\bigskip

After more than twenty years -and as often in big-sized science-, CDM research has by now its own internal inertia: growing simulations involve large super-computers and large and long-time planned experiments, huge number of people, (and huge budgets); one should not be surprised in principle, if a fast strategic change would not yet operate in the CDM and wimp research, although its interest would progressively decline.

\bigskip

In contrast to the CDM situation, the WDM research situation is progressing fast, the subject is new and 
WDM essentially {\it works}, naturally reproducing the observations over all the scales, small as well as 
large and cosmological scales ($\Lambda$WDM). WDM became a hot topic and the subject of many doctoral and post-doctoral
researches.

\bigskip

\begin{center}

{\bf Format and Exhibitions}

\end{center}

\bigskip

All Lectures are  plenary and followed by a discussion. 
Enough time is provided to the discussions.  

\begin{center}

Informations of the Colloquium are available on

\medskip
 
 {\bf http://www.chalonge.obspm.fr/colloque2011.html}

\end {center}

\bigskip

Informations on the previous Paris Cosmology Colloquia and  
on the Chalonge school events are available at  

\begin{center}

 {\bf http://chalonge.obspm.fr}

(lecturers, lists of participants, lecture files and photos during the Colloquia).

\end {center}

\bigskip

This Paris Chalonge Colloquia series started in 1994 at the Observatoire de Paris. 
The series cover selected topics of high current interest in the interplay between 
cosmology, astrophysics and fundamental physics. 
The purpose of this series is an updated understanding, from a fundamental, conceptual and unifying view, of the progress 
and current problems in the early universe, cosmic microwave background radiation, large and small scale 
structure and neutrinos in astrophysics and the interplay between them. Emphasis is given to the 
mutual impact of fundamental physics, astrophysics and cosmology, both at theoretical and experimental 
-or observational- levels. 

\bigskip

Deep understanding, clarification, synthesis, a careful interdisciplinarity within a 
fundamental physics and unifying approach, are goals of this series of Colloquia.

\bigskip

Sessions leave enough time for private discussions and to enjoy 
the beautiful parisian campus of Observatoire de Paris (built on orders from Colbert and to 
plans by Claude Perrault from 1667 to 1672).

\bigskip

Sessions take  place in the Cassini Hall, on the meridean of Paris, in 'Salle du Conseil'
(Council Room) under the portraits of Laplace, Le Verrier, Lalande, Arago, Delambre and Louis XIV, and in the 
'Grande Galerie' (the Great Gallery), in the historic Perrault building ('B\^atiment Perrault') 
of Observatoire de Paris HQ.

\bigskip

An {\bf Exhibition} at the `Grand Galerie' (Great Gallery) and "Salle Cassini" (Cassini Hall) 
retraced the 20 years of activity of the Chalonge School and {\bf "The Golden Days" in 
`Astrofundamental Physics': The Construction of the Standard Model Of the Universe}, 
as well as {\bf `The World High Altitude Observatories Network: World Cultural and Scientific Heritage'}.

\medskip
 
The books and proceedings of the School since its creation, as well as historic 
Daniel Chalonge material, instruments and the Daniel Chalonge Medal were on exhibition at the 
Grande Galerie.

\bigskip

The exhibitions were prepared by Maria Cristina Falvella, Alba Zanini, Fran\c{c}ois Sevre, 
Nicole Letourneur and Norma G. Sanchez.

\medskip

Photos and Images by : Kathleen Blumenfeld, Letterio Pomara, Jean Mouette,
Fran\c{c}ois Sevre, Fran\c{c}ois Colas, Nadia Blumenfeld, Gerard Servajean,
Sylvain Cnudde, graphic design by Emmanuel Vergnaud. 

\bigskip

After the Colloquium, a visit of the Perrault building took place guided by Professor Suzanne Debarbat 

\bigskip

More information on the Colloquia of this series can be found in the Proceedings
(H.J. de Vega and N. Sanchez, Editors) published by World Scientific Co. since 1994 and by 
Observatoire de Paris, and the Chalonge School Courses published by World Scientific Co 
and by Kluwer Publ Co. since 1991.

\bigskip

We address here the recent turning point in the research of Dark Matter represented by 
Warm Dark Matter (WDM) putting together astrophysical, cosmological, particle and nuclear physics WDM,
astronomical observations, theory and WDM numerical simulations which naturally reproduce the observations, 
as well as the experimental search for the WDM particle candidates (sterile neutrinos).

\medskip

This 15th Paris Chalonge Colloquium 2011 enlarges, strentghs and unifies with new topics, lecturers and deep discussions the issues discussed 
and pre-viewed in the Warm Dark Matter Chalonge Meudon Workshop 2011 in June 2011.

The Highlights and Conclusions of the 14th Paris Chalonge Colloquium 2010 and of the Chalonge Meudon Workshop 2011 are a useful and complementary introduction to these Highlights and can be read with benefit, they are available at:

\begin{center}

http://arxiv.org/abs/1009.3494

\medskip

http://arxiv.org/abs/1109.3187

\end{center}

\bigskip

\begin{center}

{\bf Summary and Conclusions:}

\end{center}

\bigskip

This 15th Paris Colloquium 2011 addressed the last progresses made in the STANDARD MODEL OF THE 
UNIVERSE-WARM DARK MATTER with both theory and observations. In the tradition of the Chalonge School, 
an effort of clarification and synthesis is  made by combining in a conceptual framework, theory, 
analytical, observational and numerical simulation results which reproduce observations, both in 
astrophysics and particle and nuclear physics, keV sterile neutrinos being
today the more serious candidates for Dark Matter.

\bigskip

\begin{center} 

{\bf The subject have been approached in a fourfold coherent way:}

\end{center}

\bigskip

(I) Conceptual context, the standard cosmological model includes inflation

\bigskip

(II) CMB observations and astronomical observations,  cosmological, large, intermediate and  small galactic scales, and interstellar medium observations, linked to structure formation at the different scales.

\bigskip

(III) WDM theory, models and WDM numerical simulations which reproduce observations at large and small (galactic) scales.

\bigskip

(IV) WDM particle candidates, keV sterile neutrinos: particle models and astrophysical constraints on them

\bigskip

Philippe Andre, Peter Biermann, Pasquale Blasi, Daniel Boyanovsky, Carlo Burigana, Hector de Vega, 
Joanna Dunkley, Gerry Gilmore, Ayuki Kamada, Alexander Kashlinsky, Alan Kogut, Anthony Lasenby,  
John Mather, Sinziana Paduroiu, Norma Sanchez, Alexei Smirnov, Sylvaine Turck-Chieze 
present here their highlights of the Colloquium. 

\medskip

Summary and conclusions are presented at the end by H. J. de Vega, M. C. Falvella and N. G. Sanchez in 
three subsections: 

\medskip

\begin{center}

A. General view and clarifying remarks. 

\medskip

B. Conclusions. 

\medskip

C. The present context and future in the Dark Matter research and Galaxy formation

\bigskip

The Summary of the Conclussions stress among other points :

\end{center}

\bigskip

{\bf (I)} Data confirm primordial CMB gaussianity. Inflation effective theory predicts
negligible primordial non-gaussianity, negligible scalar index running  
and a tensor to scalar ratio $ r \sim 0.05-0.04 $ at reach/border line of next CMB observations;
the present data with this theory clearly prefer new inflation and the `cosmic banana' region in the $r$ vs $n_s$
diagram; early fast-roll inflation is generic and provides the lowest multipoles depression and TE oscillations.
Impressive new CMB high-$l$ data were released from ACT and SPT, the CMB power spectrum displays now $9$ clear peaks, various cosmic parameter degeneracies are now resolved, ACT observed primordial helium and detected the CMB lensing, CMB alone can provide now evidence for dark energy. WMAP7 Sunyaev-Zeldovich (SZ) amplitudes  are smaller of 0.5-0.7 than expected with respect to X-ray models, relaxed and non-relaxed clusters need distinction; the {\it Planck} first released results did not found such SZ depression. The Primordial Inflation Explorer (PIXIE) is planned to test inflation and to robustly detect CMB tensor (B-mode) polarisation (primordial gravitational waves) to limit $r < 10^{-3}$ at more than 5 sigma. PIXIE also allows to test keV warm dark-matter and reionization. On the other hand, cluster peculiar flows of about $600-1000 km/sec$ and coherent length $\sim 750 Mpc $ seem to be measured and would require if confirmed a fast-roll inflationary explanation.  

\bigskip
 
{\bf (II)} Significant progresses are made in dSph galaxy observations: Very low star formation rates and very low baryonic feedback are found. Baryon feedback does not affect DM structure, CDM does not produce realistic galaxies with realistic feedback and star formation recipes. Very precise kinematics of stars in dSph is probing directly DM density profiles: cusped profiles are excluded at the $95 \%  CL$. Cored (non cusped) DM halos and warm (keV scale mass) DM are strongly favored from theory and astrophysical observations, they naturally produce the observed small scale structures; keV sterile neutrinos of mass  between $1$ and $10$ keV are the most serious candidates. Progresses in computing the distribution function for WDM sterile neutrinos as well as $\Lambda$WDM simulations with them were reported. Wimps (heavier than 1 GeV) are strongly disfavoured combining theory with galaxy observations. In addition,  
the observed cosmic ray positron and electron excesses are explained by natural astrophysical processes, while annihilating/decaying cold dark matter models are strongly disfavored. On the other hand, non-annihilating CDM wimps in the sun are excluded. The sun seismic model can be also used to constrain WDM (sterile neutrinos).  

\bigskip

{\bf (III)} {\it Herschel} remarkably observed a profusion of interstellar pc-scale filaments sharing a common width $ \sim 0.1 pc$ and a critical mass per unit length $ \sim 15\, M_\odot$/pc, above which they are gravitationally unstable. This filamentary structure is intimately connected to the cloud core formation process. Remarkably, the independently   observed interstellar surface density (colum density) and galaxy surface density have almost the same universal value, irrespective of size and composition, morphology and system types. 

\bigskip

Putting all together, evidence that $\Lambda$CDM does not work at small scales is staggering. Impressive evidence points that DM particles have a mass in the keV scale and that those keV scale particles naturally produce the small scale structures observed in galaxies. Wimps (DM particles heavier than 1 GeV) are strongly disfavoured combining theory with galaxy astronomical observations. keV scale sterile neutrinos are the most serious DM candidates and deserve dedicated 
experimental searchs and simulations. Astrophysical constraints including Lyman alpha bounds put the 
sterile neutrinos mass in the range $ 1 < m < 13$ keV. $\Lambda$WDM simulations with 1 keV particles reproduce remarkable well the observations, the distribution of Milky Way satellites in the range above $\sim 40$ kpc, sizes of local minivoids and velocity functions. Overall, $\Lambda$WDM  and keV scale DM particles deserve dedicated
astronomical and laboratory experimental searches, theoretical work and simulations. MARE -and hopefully an adapted KATRIN- experiment could provide a sterile neutrino signal. The experimental search for this
serious DM candidate appears urgent and important: It will be a fantastic discovery to 
detect dark matter in a beta decay. 

\medskip

There is an encouraging and formidable WDM work to perform ahead us, these highlights point some of the directions
where to put the effort.

\bigskip

We want to express our grateful thanks to all the sponsors of the Colloquium,  
to all the lecturers for their excellent and polished presentations, to all the 
lecturers and participants for their active participation and their contribution to 
the outstanding discussions and lively atmosphere, to the assistants, secretaries and 
all collaborators of the Chalonge School, who made this event so harmonious, wonderful and successful . 

\bigskip

We are pleased to give you appointment at the next Colloquium of this series:

\begin{center}

The 16th Paris Cosmology Colloquium 2012 devoted to 

\bigskip

THE NEW STANDARD MODEL OF THE UNIVERSE: LAMBDA WARM DARK MATTER ($\Lambda$WDM) THEORY AND OBSERVATIONS

\bigskip

Observatoire de Paris, historic Perrault building,  25, 26, 27 JULY 2012.

\bigskip

http://www.chalonge.obspm.fr/colloque2012.html

\end{center}

\bigskip

\begin{center}
                                   
With Compliments and kind regards,

\bigskip  
                                            
{\bf Hector J de Vega, $ \; $ Maria Cristina Falvella,  $ \; $  Norma G Sanchez}

\end{center}

\bigskip

\bigskip

\begin{figure}[ht]
\includegraphics[height=12cm,width=17cm]{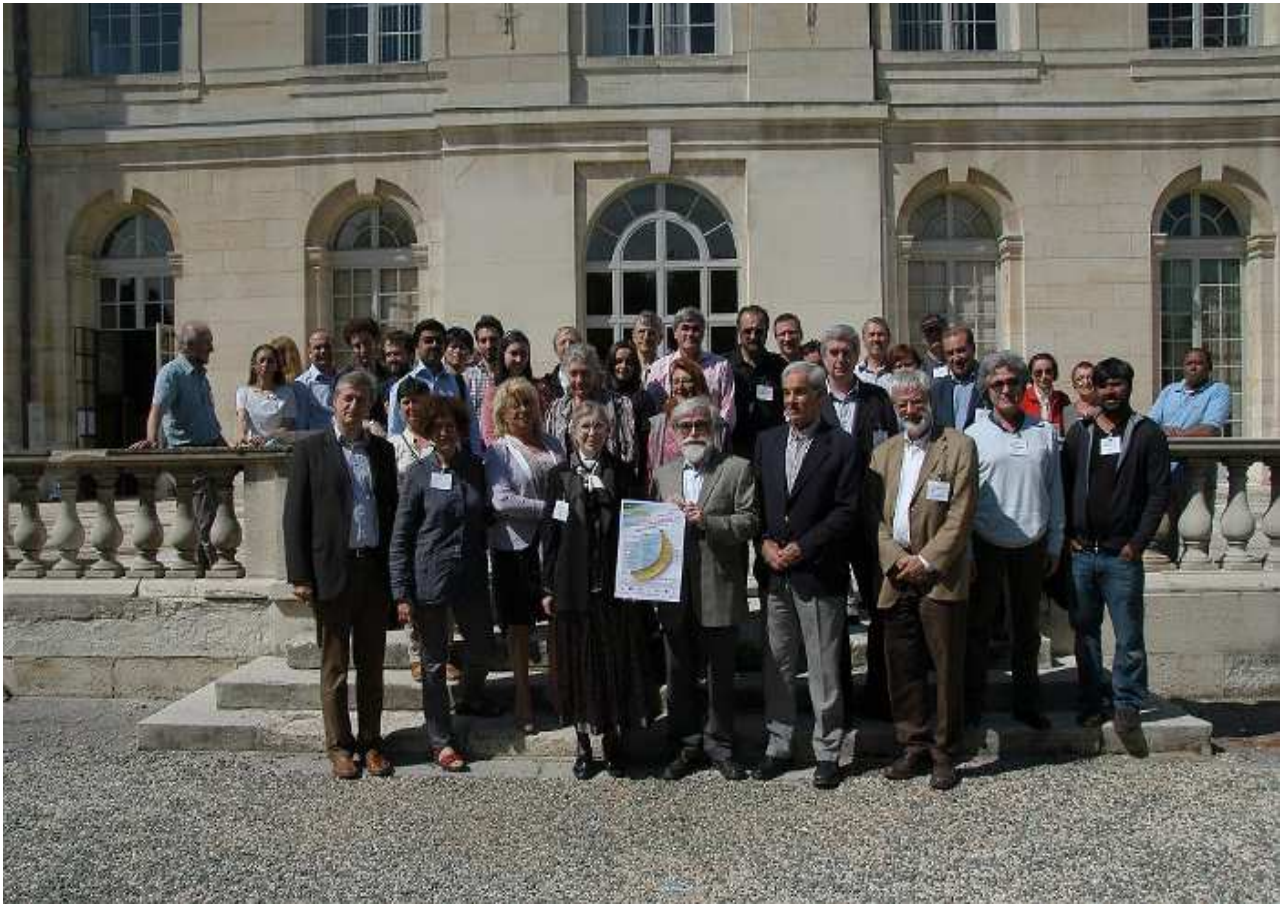}
\caption{Photo of the Group}
\end{figure}

\newpage

\begin{figure}[ht]
\includegraphics[height=20cm,width=15cm]{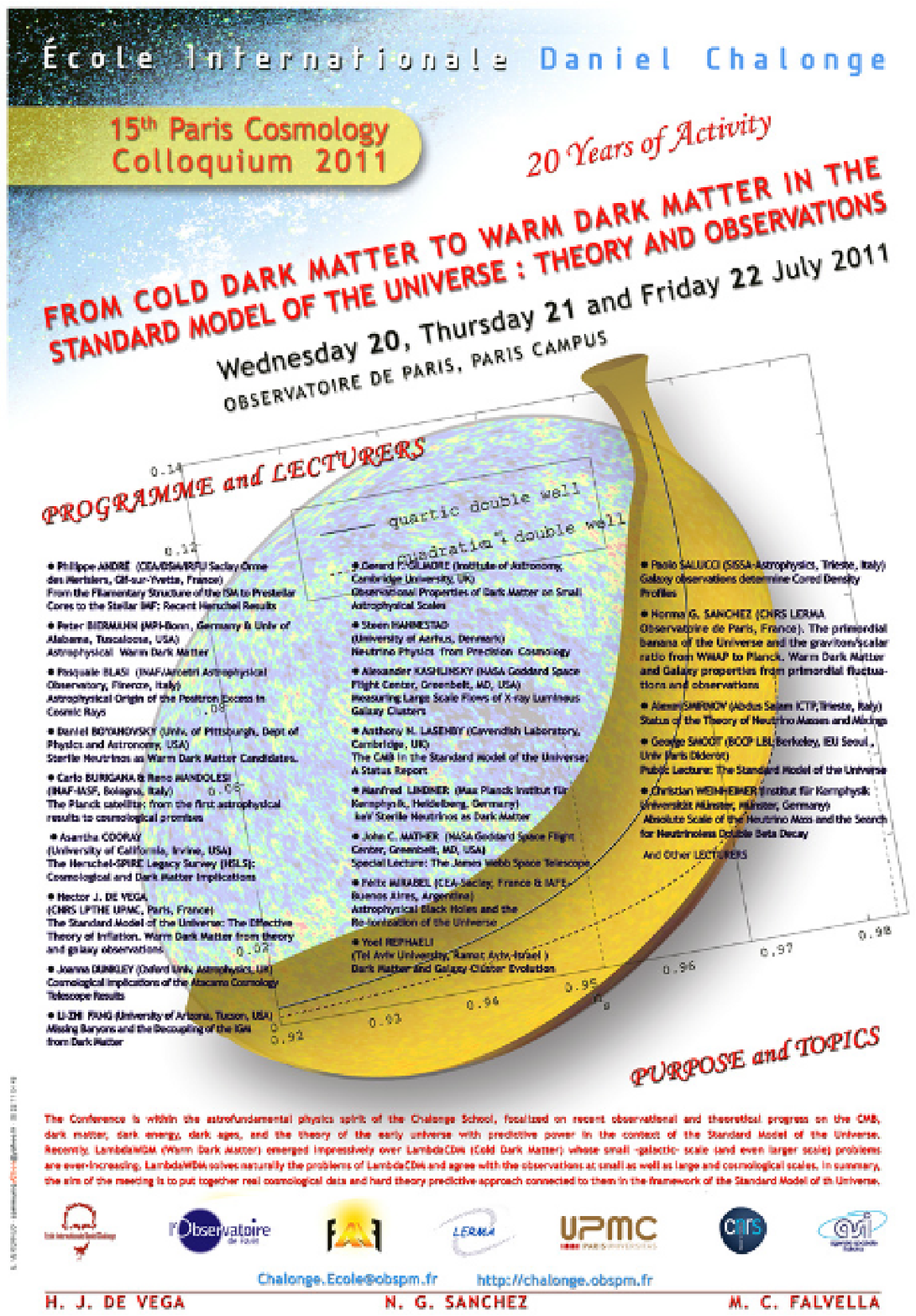}
\caption{Poster of the 14th Paris Cosmology Colloquium 2010}
\end{figure}

\newpage

\section{Programme and Lecturers}


\begin{itemize}

\item{{\bf Philippe ANDRE}(CEA/DSM/IRFU Saclay Orme des Merisiers, Gif-sur-Yvette, France)\\                             From the Filamentary Structure of the ISM to Prestellar Cores to the Stellar IMF: Recent Herschel Results}

\item{{\bf Peter BIERMANN} (MPI-Bonn, Germany \& Univ of Alabama, Tuscaloosa, USA)\\
Astrophysical Warm Dark Matter}

\item{{\bf Pasquale BLASI} (INAF/Arcetri Astrophysical Observatory, Firenze, Italy)\\
Astrophysical Origin of the Positron Excess in Cosmic Rays}

\item{{\bf Daniel BOYANOVSKY} (Univ. of Pittsburgh, Dept of Physics and Astronomy, USA)\\
Sterile Neutrinos as Warm Dark Matter Candidates.}

\item{{\bf Carlo BURIGANA \& Reno MANDOLESI} (INAF-IASF, Bologna, Italy)\\
The Planck satellite: from the first astrophysical results to cosmological promises}

\item{{\bf Asantha COORAY} (University of California,Irvine, USA)\\
The Herschel-SPIRE Legacy Survey (HSLS): Cosmological and Dark Matter Implications.}

\item{{\bf Hector J. DE VEGA} (CNRS LPTHE Univ de Paris VI, France)\\
The Standard Model of the Universe: The Effective Theory of Inflation. 
Warm Dark Matter from theory and galaxy observations.}

\item{{\bf Joanna DUNKLEY} (Oxford Univ, Astrophysics, UK)\\
Cosmological Implications of the Atacama Cosmology Telescope Results}

\item{{\bf Gerard F. GILMORE} (Institute of Astronomy, Cambridge University, UK)\\
Observational Properties of Dark Matter on Small Astrophysical Scales.}

\item{{\bf Alexander KASHLINSKY} (NASA Goddard Space Flight Center, Greenbelt, MD, USA)\\
Measuring Large Scale Flows of X-ray Luminous Galaxy Clusters}

\item{{\bf Alan KOGUT} (NASA Godard Space Flight Center, Greenbelt, MD, USA)\\
Testing the Standard Model with the Primordial Inflation Explorer}

\item{{\bf Anthony N. LASENBY} (Cavendish Laboratory, Cambridge, UK)\\
The CMB in the Standard Model of the Universe: A Status Report}

\item{{\bf Manfred LINDNER} (Max Planck Institut für Kernphysik, Heidelberg, Germany)\\
keV Sterile Neutrinos as Dark Matter}

\item{{\bf John C.MATHER} (NASA Goddard Space Flight Center, Greenbelt, MD, USA)\\
Special Lecture: The James Webb Space Telescope}

\item{{\bf F\'elix MIRABEL} (CEA-Saclay, France \& IAFE-Buenos Aires, Argentina)\\
Astrophysical Black Holes and the Re-ionization of the Universe}

\item{{\bf Norma G. SANCHEZ} (CNRS LERMA Observatoire de Paris, France)\\
The primordial banana of the Universe and the graviton/scalar ratio from WMAP to Planck. 
Warm Dark Matter and Galaxy properties from primordial fluctuations and observations.}

\item{{\bf Alexei SMIRNOV} (Abdus Salam ICTP,Trieste, Italy)\\
Status of the Theory of Neutrino Masses and Mixings}

\item{{\bf George SMOOT} (BCCP LBL Berkeley,IEU Seoul, Univ Paris Diderot USA)\\
The Standard Model of the Universe}

\item{{\bf Sylvaine TURCK-CHIEZE} (SAp/IRFU/CEA Saclay, Gif sur Yvette, France)\\
Helioseismology and Dark Matter}

\item{{\bf Christian WEINHEIMER} (Institut f\"ur Kernphysik Universit\"at M\"unster, M\"unster, Germany)\\
Absolute Scale of the Neutrino Mass and the Search for Neutrinoless Double Beta Decay.}

\end{itemize}

\section{Posters and Presentations}

\begin{itemize}

\item{{\bf Ayuki KAMADA} (IPMU, Institute for the Physics and Mathematics of the Universe,Tokyo,Japan)
keV-mass sterile neutrino dark matter and the structure of galactic haloes}

\item{{\bf Sinziana PADUROIU, Observatoire de Geneve, Switzerland} Numerical simulations with 
Warm Dark Matter: The effects of free streaming on Warm Dark Matter haloes of Galaxies}

\end{itemize}

\newpage

\section{Support to the James Webb Space Telescope}

\bigskip

\begin{center}

Ecole Internationale d'Astrophysique Daniel Chalonge

\end{center}

\bigskip

CHALONGE  SCHOOL  STATEMENT  OF SUPPORT  TO  THE  JAMES  WEBB  SPACE  TELESCOPE

\bigskip

\bigskip

On 22th July 2011, during the 15th Paris Cosmology Colloquium 2011 of the International School of 
Astrophysics Daniel Chalonge at Paris Observatory, {\bf Dr. John Mather}, Nobel Laureate 2006 for 
the outstanding results of the NASA COBE satellite, brilliantly {\bf presented the NASA Program  
James Webb Space Telescope (JWST)}, the large infrared-optimized space telescope planned 
to operate from 2018 as the best successor for the Hubble and Spitzer Space Telescopes, with an 
ambitious scientific programme which ranges from the early universe, first galaxies formation, 
first light and reionization, star formation, to the protoplanetary and planetary systems and the 
origins of life.  These subjects {\bf represent the current frontier of knowledge in modern 
astrophysics and astronomy and the most interesting new programs in the field}. 

\bigskip

{\bf One of the aims of the Chalonge School is to bring the attention into the new programs 
which will produce a clearer and deeper understanding of the Universe with both innovative 
experiments and theory}. During the 15th Paris Cosmology Colloquium 2011, the 
 Chalonge School has made the following statement in support to the James Webb Space Telescope:

\bigskip

The Ecole Internationale Daniel Chalonge considers that  JWST is an exceptional 
opportunity for the future of astrophysics and astronomy worldwide. The Ecole 
Internationale Daniel Chalonge recognizes the outstanding scientific value  of the 
JWST project, its potentiality and worldwide impact and strongly supports its 
development and successfull completion.

\bigskip

\begin{center}

Norma G. Sanchez (1)  , Héctor J. de Vega (2), Maria Cristina Falvella (3) , Alba Zanini (4)  

\end{center}

(1) Director of the International Astrophysics Daniel Chalonge School,  
CNRS, Observatoire de Paris, France. (2) CNRS, University Pierre \& Marie Curie, Paris, France. 
(3) Italian Space Agency HQ, Rome, Italy. 
(4) Istituto Nazionale di Fisica Nucleare (INFN), Turin, Italy.
 
\bigskip
 
\begin{center}

Links to the Statement and to the JWST : 

\bigskip

\bigskip

http://chalonge.obspm.fr/jwst.pdf

\bigskip

JWST Community Support from the 15th paris Cosmology Colloquium Chalonge 2011

\bigskip

http://www.aura-astronomy.org/news/2011/jwstChalonge.pdf

\bigskip

Information on The James Webb Space Telescope:

\bigskip

http://www.jwst.nasa.gov/

\bigskip

http://www.chalonge.obspm.fr/Paris11$_{}$Mather.pdf

\bigskip

Support to the JWST among the world:

\bigskip

Association of Universities for the Research in Astronomy (AURA): 

\medskip

JWST AURA Resource Center

\medskip

http://www.aura-astronomy.org/news/jwst.asp
 
\end{center}

\newpage

\section{Award of the Daniel Chalonge Medal 2011}

\bigskip

The Daniel Chalonge Medal 2011 has been awarded to {\bf Dr John C. MATHER}.   

\bigskip

The International Astrophysics School Daniel Chalonge has awarded the Daniel Chalonge Medal 2011 

to Dr. John C. Mather, Nobel prize of Physics 2006 for the outstanding results of the COBE satellite, and present Senior Project Scientist for the James Webb Space Telescope.
  
\bigskip 

Dr. John C. Mather is a Senior Astrophysicist in the Observational Cosmology Laboratory at the NASA Goddard Space  Flight Center (College Park -Maryland, USA).

\bigskip

The medal was awarded to John Mather for his huge contribution to modern cosmology, in particular for his 
outstanding effort in promoting and leading key missions for the study of the Universe,  as the COBE satellite and now the  JWST, deeply discussed in the frame of the Chalonge School and the training and formation of young physicists and astrophysicists. 

\bigskip

John Mather brilliantly presented the  JWST Program , the large infrared-optimized 
space telescope planned to operate from 2018 as the best successor for the Hubble and Spitzer Space Telescopes.  The Chalonge Medal represents  too a warm  aknowledgement and support to Dr. John Mather present and future activities in the JWST.

\bigskip

John Mather also contributed to ground observation programs leading advisory and working groups for the National Academy of Sciences, NASA, 
and the NSF (for the ALMA, the Atacama Large Millimeter Array, and for the CARA, the Center for Astrophysical 
Research in the Antarctic). 

\bigskip

As Senior Project Scientist for the JWST, John Mather successfully leads the 
science team, and represents the scientific interests within the project management. 

\bigskip

The medal was presented to John Mather on 22th July 2011 during the Open Session of the 15th Paris Cosmology Colloquium 2011 at the Observatoire de Paris HQ (historic Perrault building)  in the Cassini Hall, on the meridian of Paris, which was attended by about hundred participants from the world over, among them three laureate of the Chalonge Medal: George Smoot, Nobel laurate of physics, Anthony Lasenby and Peter Biermann. 

\bigskip

The list of the awarded Chalonge Medals is the following:

\bigskip

\begin {itemize}

\item {1991: Subramanyan Chandrasekhar, Nobel prize of physics.}
\item {1992: Bruno Pontecorvo.}
\item {2006: George Smoot, Nobel prize of physics.}
\item {2007: Carlos Frenk.}
\item {2008: Anthony Lasenby.}
\item {2008: Bernard Sadoulet.}
\item {2009: Peter Biermann.}
\item {2011: John Mather, Nobel prize of physics.}

\end{itemize}

\bigskip

The Chalonge Medal, coined exclusively for the Chalonge School by the prestigious Hotel de la Monnaie de Paris (the French Mint), is a surprise award and only eight Chalonge medals have been awarded in the 20 year school history. 

\bigskip

The Medal aknowledges science with great intellectual endeavour and a human face. True and healthy science. Outstanding gentleperson scientists. Scientists recipients of the Daniel Chalonge Medal are Ambassadors of the School.

\bigskip

See the announcement, full history, photo gallery and links  at: http://chalonge.obspm.fr

\bigskip

The Daniel Chalonge Medal 2011  http://chalonge.obspm.fr/Medal$_{}$Chalonge2011.pdf

\bigskip

Mather's Open Lecture on the JWST :

\bigskip

http://www.chalonge.obspm.fr/Paris11$_{}$Mather.pdf

\bigskip

Photo Gallery:

\bigskip

http://chalonge.obspm.fr/albumopensession2011/index.html

\bigskip

Archives Daniel Daniel Chalonge:  http://chalonge.obspm.fr/Archives$_{}$Daniel$_{}$Chalonge.html

\newpage

\section{Highlights by the Lecturers}

\begin{center}
  
More  informations on the Colloquium Lectures  are at: 

\bigskip

{\bf http://www.chalonge.obspm.fr/colloque2011.html}

\end{center}

\subsection{Philippe Andr\'e, Alexander Men'shchikov, Vera K\"onyves, Doris Arzoumanian, and Nicolas Peretto}

\vskip -0.3cm

\begin{center}

Laboratoire AIM, CEA Saclay, IRFU/Service d'Astrophysique, Gif-sur-Yvette, France

\bigskip

{\bf From the filamentary structure of the ISM to prestellar cores to the IMF: Recent $Herschel$ results  } 

\end{center}

\medskip

The immediate observational objective of the $Herschel$ Gould Belt survey ([1], [2]) 
is to image the bulk of nearby ($d \sim 500$~pc) molecular clouds 
at 6 wavelengths between 70~$\mu$m and 500~$\mu$m. 
This will provide complete samples of prestellar cores and ``Class~0'' 
protostars with well characterized luminosities, 
temperatures, and density profiles, as well as robust core mass functions
and protostar luminosity functions, in a variety of star-forming environments. 
The main scientific goal is to elucidate the physical mechanisms responsible for the 
formation of prestellar cores out 
of the diffuse interstellar medium (ISM), crucial for understanding the origin of the stellar initial 
mass function (IMF). 
Briefly, the first images from the Gould Belt survey have revealed a profusion of parsec-scale 
filaments in nearby molecular clouds 
and suggested an intimate connection between the filamentary structure 
of the ISM and the formation process of dense cloud cores  ([2], [3] -- see Fig.~1). 
Remarkably, filaments are omnipresent even in unbound, non-star-forming complexes such as the Polaris translucent cloud ([4], [5]),
and all appear to share a common (FWHM) width $\sim 0.1$pc ([6]).
Moreover,  in active star-forming regions such as the Aquila Rift cloud ([7], [8]),  
most of the prestellar cores identified with $Herschel$ are located within gravitationally unstable filaments for which the mass 
per unit length exceeds the critical value ([9]) 
$M_{\rm line, crit} = 2\, c_s^2/G \sim 15\, M_\odot$/pc,  
where $c_{\rm s} \sim 0.2$~km/s is the isothermal sound speed for $T \sim 10$~K. 
These findings led us ([2])
to favor a scenario according to which core formation 
occurs in two main steps. First, large-scale magneto-hydrodynamic (MHD) turbulence generates an intricate network of 
filaments in the ISM  (cf. [10]);
second, the densest filaments fragment into 
prestellar cores by gravitational instability (cf. [11]). 
%
This scenario provides an {\it explanation} of the star formation threshold 
at a gas surface density  $\Sigma_{\rm gas}^{\rm th} \sim $~130~$M_\odot \, {\rm pc}^{-2} $ 
found in recent infrared studies of the star formation rate in both Galactic and extragalactic cloud complexes  (e.g. [12]): 
Given the typical  $\sim $~0.1~pc width of interstellar filaments ([6]), 
the threshold $\Sigma_{\rm gas}^{\rm th}$ corresponds to within a factor of $< 2$ to 
the critical mass per unit length  $M_{\rm line, crit}$ above which gas filaments at $T \sim 10$~K are gravitationally unstable.

\begin{center}
\vspace*{-0.1 cm}
\includegraphics[width=175mm,height=95mm]{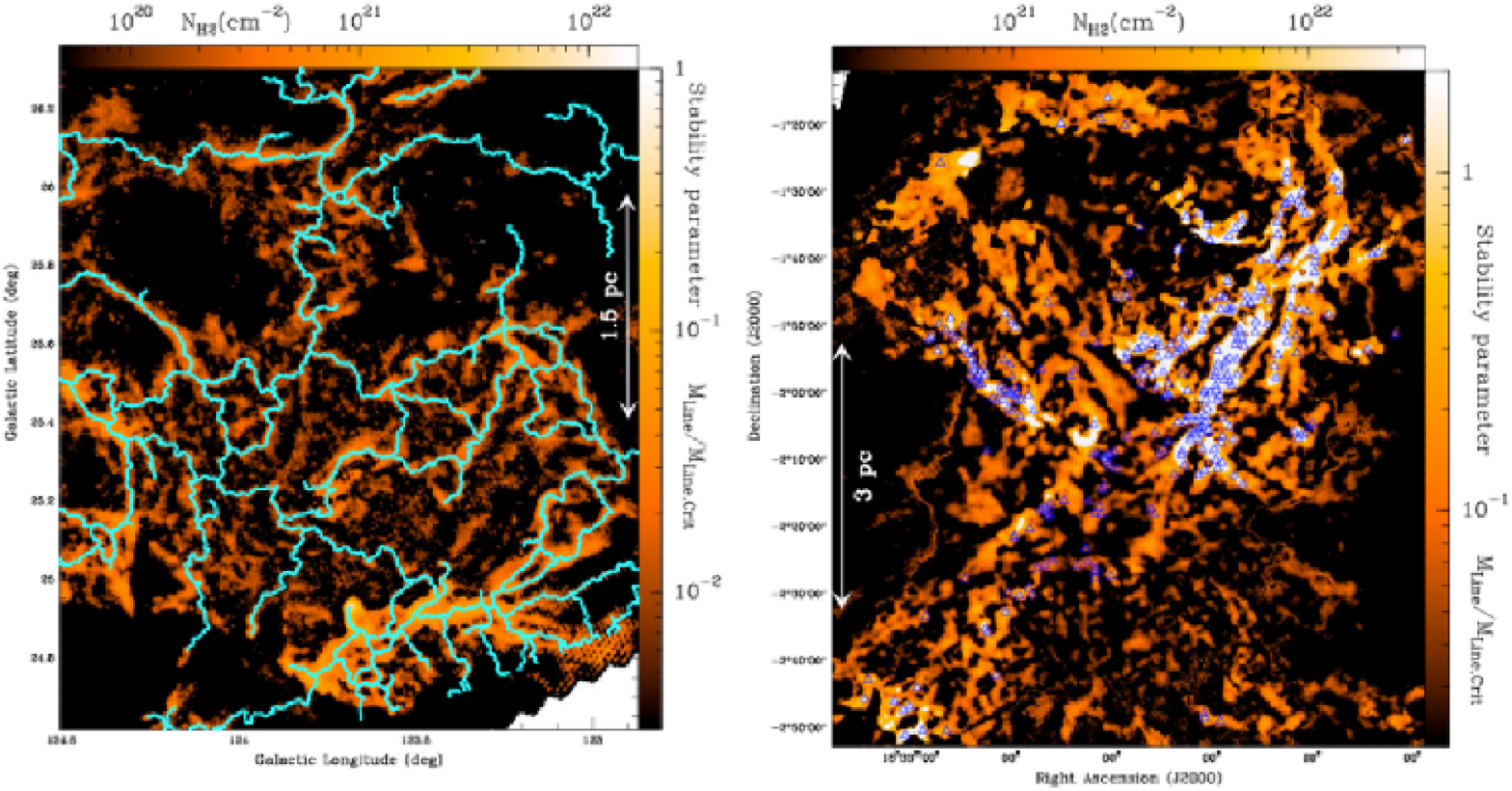}
\end{center}

\vspace*{-0.3 cm}
{\bf Fig. 1: }  {
Column density maps of two fields in Polaris (left) and Aquila (right) derived from $Herschel$ data ([2]). 
The contrast of the filaments has been enhanced using a curvelet transform (cf. [3]). 
The skeleton of the filament network identified in Polaris with the DisPerSE algorithm ([13]) 
is shown in light blue in the left panel.
Given the typical width $\sim $~0.1~pc of the filaments ([6]), 
these maps are equivalent to {\it maps of the mass per unit length along the filaments}. 
The areas where the filaments have a mass per unit length larger than half the critical value $2\, c_s^2/G$ and are thus likely gravitationally unstable 
have been highlighted in white. 
The bound prestellar cores identified by [7] 
in Aquila are shown as small triangles in the right panel; there are no bound cores in Polaris. 
Note the good correspondence between 
the spatial distribution of prestellar cores and the regions where the filaments are unstable to gravitational collapse.
(Adapted from [2].)
}

\medskip

Our $Herschel$ first results also confirm the existence of a close relationship between the 
prestellar core mass function (CMF) and the stellar IMF ([2], [7] 
-- see Fig.~2).
Interestingly, the peak of the prestellar CMF at $ \sim 0.6\, M_\odot $ as observed in the Aquila complex (cf. Fig.~2b) 
corresponds to the Jeans or Bonnor-Ebert mass $ M_{\rm BE} \sim 0.6\, M_\odot\, \times \left(T/10\, {\rm K}\right)^2 \times  \left(\Sigma/150\, M_\odot\, {\rm pc}^{-2}\right)^{-1} $
within marginally critical filaments with $ M_{\rm line} \approx M_{\rm line, crit} \sim 15\, M_\odot$/pc and surface densities 
$\Sigma \approx \Sigma_{\rm gas}^{\rm crit} \sim 150\, M_\odot \, {\rm pc}^{-2} $. 
Likewise,  the median spacing $\sim 0.08$~pc observed between the prestellar cores of Aquila roughly matches the 
thermal Jeans length within marginally critical filaments.
%
This is consistent with the idea that gravitational fragmentation is the dominant physical mechanism generating prestellar cores 
within the filaments. Furthermore, a typical prestellar core mass of $\sim 0.6\, M_\odot$  translates into a characteristic star 
or stellar system mass of $\sim 0.2\, M_\odot$, assuming a local star formation efficiency $ \epsilon_{\rm core} \sim 30\%$ 
at the level of individual cores ([14]). 
Therefore, our $Herschel$ findings strongly support Larson's interpretation ([15]) of the peak of the IMF in terms 
of the typical Jeans mass in star-forming clouds. 
Overall, our results suggest that the gravitational fragmentation of supercritical filaments produces the prestellar CMF which, 
in turn, accounts for the log-normal ``base'' 
of the IMF. It remains to be seen, however, whether the bottom end of the IMF 
and the Salpeter power-law slope at the high-mass end can be explained by the same mechanism.  


\begin{center}
\includegraphics[width=85mm,height=75mm]{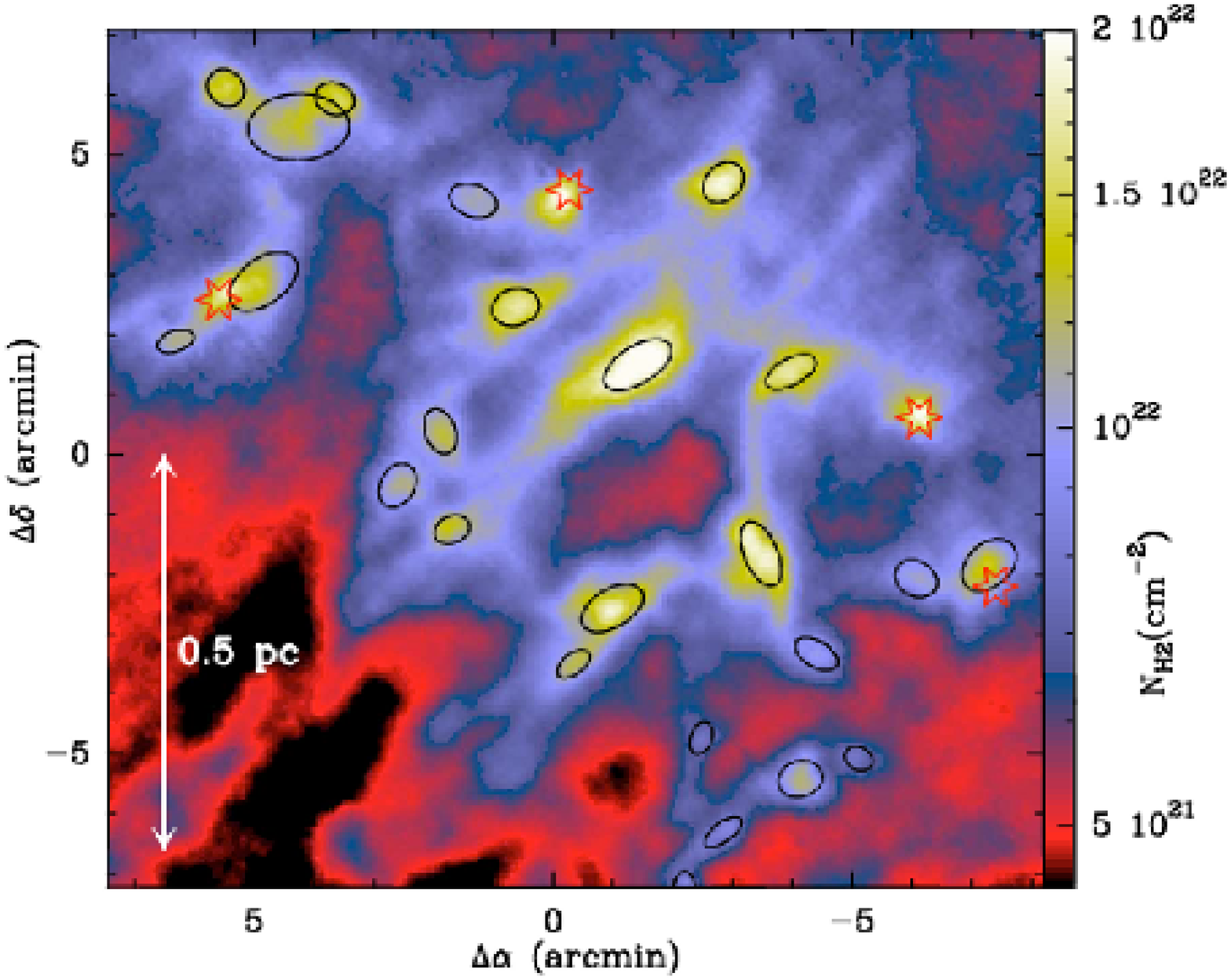} 
\includegraphics[width=75mm,height=75mm]{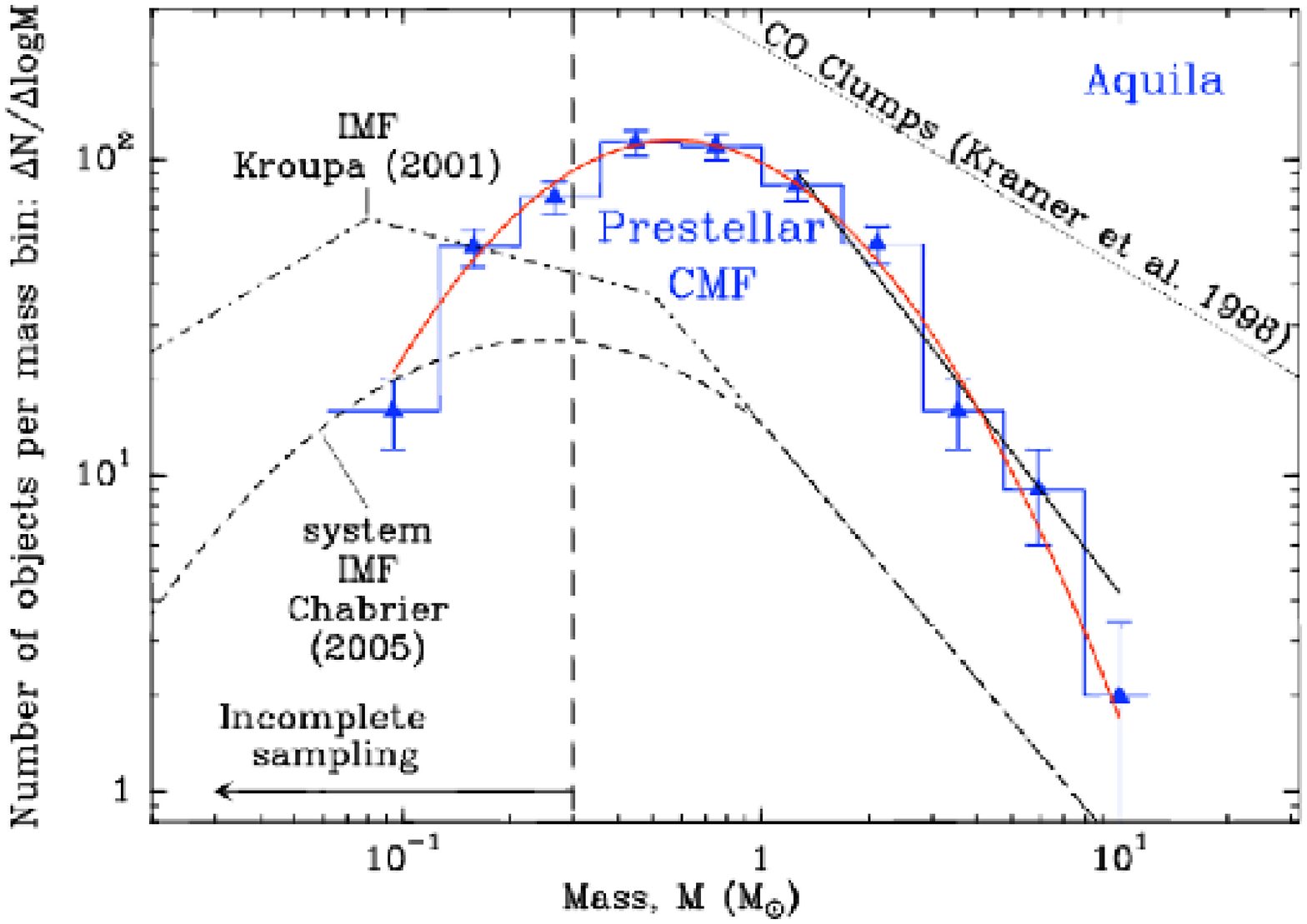} 
\end{center}

\vspace*{-0.1 cm}
{\bf Fig. 2: }  {
{\bf Left:} Close-up column density image of a small subfield in the Aquila Rift complex 
showing several candidate prestellar cores identified with $Herschel$ (adpated from [7]). 
The black ellipses mark the major and minor FWHM sizes determined for these cores at $\lambda = 250\, \mu$m 
by the source extraction algorithm {\it getsources} ([3]). 
Four protostellar cores are also shown by red stars.
{\bf Right:} Core mass function (histogram with error bars) of the 541 candidate prestellar cores identified with $Herschel$ 
in Aquila ([2], [7]). 
The IMF of single stars (corrected for binaries -- e.g. [16]), 
the IMF of multiple systems (e.g. [17]), 
and the typical mass spectrum of CO clumps (e.g. [18]) 
are shown for comparison.
A log-normal fit to the observed CMF is superimposed; it peaks at $\sim 0.6\, M_\odot $, which corresponds to the Jeans 
mass within marginally critical filaments at $T \sim 10$~K.
}

\bigskip

{\bf References}

\begin{description}

\item[1] Andr\'e, P. , \& Saraceno, P., in The Dusty and Molecular Universe
ESA SP-577,  p. 179  (2005).

\vspace{-0.3cm}

\item[2] Andr\'e, Ph., Men'shchikov, A., Bontemps, S. et al., A\&A, 518, L102 (2010).

\vspace{-0.3cm}

\item[3] Men'shchikov, A., Andr\'e, Ph., Didelon, P. et al., A\&A, 518, L103  (2010).

\vspace{-0.3cm}

\item[4] Ward-Thompson, D., Kirk, J.M., Andr\'e, Ph. et al., A\&A, 518, L92  (2010).

\vspace{-0.3cm}

\item[5] Miville-Desch\^enes, M.-A., Martin, P.G., Abergel, A. et al., A\&A, 518, L104 (2010).

\vspace{-0.3cm}

\item[6] Arzoumanian, D., Andr\'e, Ph., Didelon, P.  et al., A\&A, 529, L6 (2011).

\vspace{-0.3cm}

\item[7] K\"onyves, V., Andr\'e, Ph., Men'shchikov, A. et al., A\&A, 518, L106 (2010).

\vspace{-0.3cm}

\item[8] Bontemps, S., Andr\'e, Ph., K\"onyves, V. et al., A\&A, 518, L85 (2010).

\vspace{-0.3cm}

\item[9]  Ostriker, J., ApJ, 140, 1056 (1964).

\vspace{-0.3cm}

\item[10]  Padoan, P., Juvela, M., Goodman, A.A., Nordlund, A., ApJ, 553, 227  (2001).

\vspace{-0.3cm}

\item[11] Inutsuka, S-I, \& Miyama, S.M., ApJ, 480, 681 (1997).

\vspace{-0.3cm}

\item[12] Heiderman, A., Evans, N.J., Allen, L.E. et al., ApJ, 723, 1019 (2010).

\vspace{-0.3cm}

\item[13]  Sousbie, T.,  MNRAS,  414, 350 (2011).

\vspace{-0.3cm}

\item[14] Matzner, C.D., \& McKee, C.F., ApJ, 545, 364 (2000).

\vspace{-0.3cm}

\item[15]  Larson, R.B., MNRAS, 214, 379 (1985).

\vspace{-0.3cm}

\item[16]  Kroupa, P., MNRAS, 322, 231 (2001).

\vspace{-0.3cm}

\item[17]  Chabrier, G., in The Initial Mass Function 50 years later, Eds. E. Corbelli et al., p.41 (2005).

\vspace{-0.3cm}

\item[18]  Kramer, C., Stutzki, J., Rohrig, R., Corneliussen, U., A\&A, 329, 249 (1998).

\end{description}

\newpage

\subsection{Peter L. Biermann, with help from Julia K. Becker, Lauren\c{t}iu  I. Caramete, Lou Clavelli, L\'{a}szl\'{o} \'{A}. Gergely, Ben Harms, Gopal Krishna, Athina Meli, Biman Nath, Eun-Suk Seo, Vitor de Souza, Paul Wiita, \& Todor Stanev}

\vskip -0.3cm

\begin{center}

P.L.B, LIC: MPI for Radioastronomy, Bonn, Germany; 
P.L.B: Dept. of Phys., Karlsruher Institut f{\"u}r Technologie KIT, Germany, 
P.L.B., L.C., B.H.: Dept. of Phys. \& Astr., Univ. of Alabama, Tuscaloosa, AL, USA; 
P.L.B.: Dept. of Phys., Univ. of Alabama at Huntsville, AL, USA; 
P.L.B.: Dept. of Phys. \& Astron., Univ. of Bonn, Germany ; 
J.K.B.: Dept. of Phys., Univ. Bochum, Bochum, Germany; 
L.I.C.: Institute for Space Sciences, Bucharest, Romania; 
L.A.G.: Department of Theoretical Physics, University of Szeged, Szeged, Hungary; 
G.K.: NCRA, Tata Institute, Pune, India; 
A.M.: IFPA, Department of Physics, University of Li{\`e}ge, Belgium; 
B.M.: Raman Research Institute, Bangalore, India; 
E.-S.S.: Dept. of Physics, Univ. of Maryland, College Park, MD, USA; 
V.dS.: Universidade de S$\tilde{a}$o Paulo, Instituto de
        F\'{\i}sica de S$\tilde{a}$o Carlos, Brazil; 
P.W.: Dept.\ of Physics, The College of New Jersey, Ewing, New Jersey, USA; 
T.St.: Bartol Research Inst., Univ. of Delaware, Newark, DE, USA\\

\bigskip

{\bf ASTROPHYSICS OF WARM DARK MATTER} 

\end{center}

\bigskip

Dark matter has been first detected 1933 (Zwicky [1]) and basically behaves like a non-EM-interacting gravitational gas of particles.  From particle physics Supersymmetry suggests with an elegant argument that there should be a lightest supersymmetric particle, which is a dark matter candidate, possibly visible via decay in odd properties of energetic particles and photons:  We have discovered i) an upturn in the CR-positron fraction, ii) an upturn in the CR-electron spectrum, iii) a flat radio emission component near the Galactic Center (WMAP haze), iv) a corresponding IC component in gamma rays (Fermi haze and Fermi bubble), v) the 511 keV annihilation line also near the Galactic Center (Integral), and most recently, vi) an upturn in the CR-spectra of all elements from Helium (CREAM), with a hint of an upturn for Hydrogen, vii)  A flat $\gamma$-spectrum at the Galactic Center (Fermi), and viii) have the complete cosmic ray spectrum available through $10^{15}$ to $10^{18}$ eV (KASCADE-Grande).  

\medskip

All the above features can be quantitatively explained with the action of cosmic rays accelerated in the magnetic winds of very massive stars, when they explode (Biermann et al., [2 - 6]): this work is based on predictions from 1993 ([7 - 11]); this approach is older and simpler than adding WR-star supernova CR-contributions with pulsar wind nebula CR-contributions, and also simpler than using the decay of a postulated particle.  This concept gives an explanation for the cosmic ray spectrum as Galactic plus one extragalactic source, Cen A ([4,6]).  The data do not require any extra source population below the MWBG induced turnoff - commonly referred to as the GZK-limit: [12,13].  This is possible, since the magnetic horizon appears to be quite small (consistent with the cosmological MHD simulations of Ryu et al. [14,15]).  It also entails that Cen A is our highest energy physics laboratory accessible to direct observations of charged particles.  All this allows to go back to galaxy data to derive the key properties of the dark matter particle: Work by [16 - 23] clearly points to a keV particle.  A right-handed neutrino is a Fermion candidate to be this particle (e.g. [24 - 31]; also see [32,33]: This particle has the advantage to allow star formation very early, near redshift 80, and so also allows the formation of supermassive black holes: they possibly formed out of agglomerating massive stars, in the gravitational potential well of the first DM clumps, whose mass in turn is determined by the properties of the DM particle.  Black holes in turn also merge, but in this manner start their mergers at masses of a few million solar masses, about ten percent of the baryonic mass inside the initial dark matter clumps.  This readily explains the supermassive black hole mass function ([34]).  The formation of the first super-massive stars might be detectable among the point-source contributions to the fluctuations of the MWBG at very high wave-number (Atacama); their contribution is independent of redshift.  The corresponding gravitational waves are not constrained by any existing limit, and could have given a substantial energy contribution at high redshift.  Our conclusion is that a right-handed neutrino of a mass of a few keV is the most interesting candidate to constitute dark matter.

\bigskip

{\bf References}

\begin{description}

\item[1]  Zwicky, F., {\it Helvetica Physica Acta} {\bf 6}, 110 (1933)
\vspace{-0.3cm}
\item[2]  Biermann, P. L., Becker, J. K., Meli, A., Rhode, W., 		
	Seo, E.-	S., Stanev, T., \PRL {\bf 103}, 061101 (2009); arXiv:0903.4048
\vspace{-0.3cm}
\item[3]  Biermann, P.L., Becker, J.K., Caceres, G., Meli, A., Seo, E.-S.,
\& Stanev, T., \ApJL {\bf 710}, L53 - L57 (2010); arXiv:0910.1197
\vspace{-0.3cm}
\item[4]  Gopal-Krishna, Biermann, P.L., de Souza, V., Wiita, P.J., 
\ApJL {\bf 720}, L155 - L158 (2010); arXiv:1006.5022
\vspace{-0.3cm}
\item[5]  Biermann, P.L., Becker, J.K., Dreyer, J., Meli, A., Seo, E.-S., \& Stanev, T., \ApJ  {\bf 725}  184 - 187 (2010); arXiv: 1009.5592
\vspace{-0.3cm}
\item[6]  Biermann, P.L., \& de Souza, V., eprint arXiv: 1106.0625 (2011)
\vspace{-0.3cm}
\item[7] Biermann, P.L., \AA {\bf 271}, 649 (1993); astro-ph/9301008
\vspace{-0.3cm}
\item[8] Biermann, P.L., \& Cassinelli, J.P.,  \AA {\bf 277},
  691 (1993); astro-ph/9305003
\vspace{-0.3cm}
\item[9] Biermann, P.L., \& Strom, R.G., \AA {\bf 275}, 659 (1993); astro-ph/9303013
   $10^{4}$ GeV and the radio emission from supernova remnants
\vspace{-0.3cm}   
\item[10]  Stanev, T., Biermann, P.L., \& Gaisser, T.K., 
  \AA {\bf 274}, 902 (1993); astro-ph/9303006
\vspace{-0.3cm}
\item[11]  Biermann, P.L., invited plenary lecture at 23rd Internat. Conf. 
on Cosmic Rays, in Proc. ``Invited, Rapporteur and Highlight papers''; Eds. 
D. A. Leahy et al., World Scientific, Singapore, p. 45 (1995)
\vspace{-0.3cm}   
\item[12]  Greisen, K., \PRL  {\bf 16}, 748 (1966)  
\vspace{-0.3cm}
\item[13]  Zatsepin, G. T., Kuz'min, V. A., {\it Zh. E.T.F. Pis'ma Redaktsiiu} 
{\bf 4}, p.114 (1966); transl. {\it J. of Exp. and Theor. Physics Lett.} 
{\bf 4}, 78 (1966)
\vspace{-0.3cm}
\item[14]  Ryu, D., Kang, H., Cho, J., Das, S.,\Science {\bf 320}, 909 (2008)
\vspace{-0.3cm}
\item[15]  Das, S., Kang, H., Ryu, D., Cho, J., \ApJ  {\bf 682}, 29 (2008)
\vspace{-0.3cm}
\item[16]  Dalcanton, J. J., Hogan, C. J., \ApJ {\bf 561}, 35 - 45 (2001); 
arXiv:astro-ph/0004381
\vspace{-0.3cm}
\item[17]  Gilmore, G.,  et al., eprint  astro-ph/0703308 (2007)
\vspace{-0.3cm}
\item[18]  Wyse, R., \& Gilmore, G.,  eprint arXiv/0708.1492 (2007)
\vspace{-0.3cm}
\item[19]  Strigari, L. E. \etal., eprint astro-ph/0603775 (2006)
\vspace{-0.3cm}
\item[20] Boyanovsky, D., de Vega, H. J., Sanchez, N. G., \PRD  {\bf 77}, id. 043518 (2008)
\vspace{-0.3cm}
\item[21]  Gentile, G., Famaey, B., Zhao, H., Salucci, P., \Nature {\bf 461},  627  (2009)
\vspace{-0.3cm}
\item[22]  de Vega, H. J., \& Sanchez, N. G., \MNRAS {\bf 404}, 885 (2010); arXiv:0901.0922 (2009)
%
\vspace{-0.3cm}
\item[23]  de Vega, H. J., \& Sanchez, N. G., eprint arXiv:0907.0006 (2009)
 
\vspace{-0.3cm}
\item[24] Kusenko, A., Segre, G., \PLB  {\bf 396}, 197 (1997)
\vspace{-0.3cm}
\item[25] Fuller, G. M., Kusenko, A., Mocioiu, I., Pascoli, S., \PRD {\bf 68}, id. 103002 (2003)
\vspace{-0.3cm}
\item[26] Kusenko, A., \IJMP {\bf D 13}, 2065 (2004)
\vspace{-0.3cm}
\item[27]  Kusenko, A., {\it Phys. Rep.} {\bf 481}, 1 (2009)
\vspace{-0.3cm}
\item[28] Biermann, P. L., \& Kusenko, A., \PRL {\bf 96}, 091301 (2006); astro-ph/0601004
\vspace{-0.3cm}
\item[29]  Stasielak, J., Biermann, P.L:, \& Kusenko, A., \ApJ {\bf 654}, 290-303 (2007); 
   astro-ph/0606435
\vspace{-0.3cm}
\item[30]  Loewenstein, M., Kusenko, A., Biermann, P.L., \ApJ {\bf 700}, 426 - 435 (2009); arXiv:0812.2710
\vspace{-0.3cm}
\item[31]  Kusenko, A., Int.\ J.\ Mod.\ Phys.\ D {\bf 13}, 2065 (2004); astro-ph/0409521
\vspace{-0.3cm}
\item[32]  Kusenko, A., Takahashi, F., Yanagida, T. T., \PLB {\bf 693}, 144 (2010)
\vspace{-0.3cm}
\item[33] Adulpravitchai, A., Gu, P.-H., Lindner, M., \PRD  {\bf 82}, id. 073013 (2010)
\vspace{-0.3cm}
\item[34]  Caramete, L.I., Biermann, P.L., \AA {\bf 521}, id.A55 (2010); arXiv:0908.2764
vspace{-0.3cm}
\item[35]  Biermann, P. L., Becker, J. K., Caramete, A. Curutiu, L., Engel, R., Falcke, H., 
Gergely, L. A., Isar, P. G., Maris, I. C., Meli, A., Kampert, K. -H., Stanev, T., 
Tascau, O., Zier, C., invited review for the conference CRIS2008, Malfa, 
Salina Island, Italy, Ed. A. Insolia,{\it Nucl. Phys. B, Proc. Suppl.} 
{\bf 190}, 61 - 78 (2009); arXiv: 0811.1848v3
\vspace{-0.3cm}
\item[36] Caramete, L.I., Tascau, O., Biermann. P.L., \& Stanev, T., submitted \AA (2022); arXiv:1106.5109
\vspace{-0.3cm}
\item[37]  Caramete, L.I., Biermann, P.L., submitted (2011); arXiv:1107.2244

\end {description}
 
\newpage
 
\subsection{Daniel Boyanovsky}
 
\begin{center}
Department of Physics and Astronomy, \\
University of Pittsburgh, Pittsburgh, PA 15260
\end{center}

\begin{center}
\textbf{Warm Dark Matter at small scales.}
\end{center}

The current `standard model of cosmology' $\Lambda\mathrm{CDM}$
correctly describes large scale structure but evidence has been emerging
that suggest several problems with this
paradigm at \emph{small scales} [1]. The small scale problems
may be ameliorated by invoking a Warm Dark Matter candidate in the form
of a particle with a mass in the
  $\sim \,\mathrm{keV}$ range [2]. A sterile neutrino, namely a
neutrino that does not feature standard model weak interactions is a
suitable candidate [3].
Warm Dark Matter particles feature a non-vanishing velocity dispersion
which leads to a cutoff in the matter power spectrum as a consequence of
free streaming. The essential
ingredient is the distribution function of the WDM candidate:
$f_0(p)+F_1(\vec{p},\vec{k},\eta)$ where $f_0(p)$ is the unperturbed
distribution function which is a solution of the
  collisionless Boltzmann equation in absence of gravitational
perturbations and $F_1$ is a solution of the (linearized) Boltzmann
equation with gravitational perturbations.
Moments of $f_0$ determine the WDM abundance and the phase space density
which provide upper and lower bounds on the mass [2]
Reference [4]
provide a semi-analytic method to obtain the transfer function and power
spectra for arbitrary $f_0$. A WDM particle that decouples from the
cosmological plasma when it is
  ultrarelativistic has three distinct stages of evolution: i) during
the radiation dominated stage when the particle is ultrarelativistic,
ii) radiation dominated stage when the particle
  is non-relativistic, iii) matter dominated stage. In stages i) and ii)
gravitational perturbations are dominated by the radiation fluid, while
in stage iii) the Boltzmann equation
  becomes a self-consistent integral-differential equation via Poisson's
equation. The solution of the Boltzmann equation during stages i) and
ii) yield the initial conditions on the
  distribution function for stage iii).  Sterile neutrinos produced
non-resonantly via two different mechanisms: mixing with active
neutrinos and from the decay of scalar or vector
bosons near the electroweak scale yield very different distribution
functions. The power spectra for these species is obtained and compared
in ref. [4], two noteworthy
features emerge: i) a quasidegeneracy between the distribution function
and the mass of the particle: a less massive WDM candidate with a
distribution function that favors
low momenta features a similar power spectrum as a more massive particle
but with a near thermal distribution function, ii) the emergence of warm
dark matter oscillations at
scales of the order of the free streaming scale. The first feature
suggests that constraints on the mass from the Lyman-$\alpha$ must be
taken with a caveat since a reliable
assessment requires knowledge of the distribution function, and the
second feature suggests that typical fits to the power spectra in terms
of power laws cannot reliable describe
small scales. Non-vanishing velocity dispersion and free streaming lead
to a redshift dependence of the matter and velocity power spectra which
affects peculiar velocities and
suggests that using the $z=0$ power spectra for N-body simulations with
initial conditions at large $z$ incur in substantial errors
underestimating the power spectrum of peculiar
velocities. For details see [4]

\bigskip

{\bf References}

\begin{description}

\item[1] See the contribution by G. Gilmore to these proceedings.

\item[2] See D. Boyanovsky, H. J. de Vega, N. Sanchez,
Phys.Rev.D77:043518,(2008); 

H. J. de Vega, N. G. Sanchez, Mon. Not. Roy. Astron. Soc.404:885 (2010); 

H. J. de Vega, N. G. Sanchez,
Int. J. Mod. Phys. A26:1057-1072 (2011), and references therein.
 
\item[3] A. Kusenko, Phys.Rept.481:1-28, (2009); 

J. Stasielak, P. L. Biermann, A. Kusenko, Acta Phys. Polon. B38: 3869-3878, (2007).

\item[4] D. Boyanovsky, Jun Wu, Phys. Rev. D83:043524, (2011); 
D. Boyanovsky, Phys. Rev. D83:103504, (2011).

\end{description}

\newpage

\subsection{Carlo Burigana$^{1,2}$, Alessandro Gruppuso$^{1}$, Nazzareno Mandolesi$^{1}$, Pietro Procopio$^{1}$, and Paolo Natoli$^{2,1}$\\
On behalf of of the {\it Planck} Collaboration}

\vskip -0.3cm

\begin{center}

$^{1}$INAF-IASF Bologna, Via Piero Gobetti 101, I-40129, Bologna, Italy\\
$^{2}$Dipartimento di Fisica, Universit\`a degli Studi di Ferrara, Via Giuseppe Saragat 1, I-44100 Ferrara, Italy

\bigskip

{\bf The {\it Planck} satellite: from the first astrophysical results to cosmological promises} 

\end{center}

\medskip

At the date of this Conference, the {\it Planck} cosmic microwave background (CMB) anisotropy probe, launched into space
on 14 May 2009, 
accumulated $\simeq 23.5$ months of data, corresponding to about four complete sky surveys. 
The spacecraft will continue to operate until the consumption of the cryogenic liquids on January 2012 with its two instruments, 
the High Frequency Instrument (HFI), based on bolometers working between 100 and 857 GHz,  
and the Low Frequency Instrument (LFI), based on radiometers working between 30 and 70 GHz.  
A further 12 months extension has been approved for observations with LFI only, cooled down with the cryogenic system provided by HFI. 
A summary of the {\it Planck} performance is provided in Table \ref{table:sens}. 
{\it Planck\/} is sensitive to linear polarisation up to $353\,$GHz.

\begin{table}[!ht]
  \caption{
{\it Planck\/} performances. 
The average sensitivity, $\delta$T/T,
per  FWHM$^2$ resolution element (FWHM is reported in arcmin) 
is given in CMB temperature units (i.e. equivalent thermodynamic temperature) for 28 
months of integration. The white noise (per frequency channel for LFI and 
per detector for HFI) in 1~sec of 
integration (NET,  in  $\mu$K $\cdot \sqrt{{\rm s}}$) is also given in CMB temperature units. The other used acronyms are: 
DT = detector technology, N of R (or B) = 
number of radiometers (or bolometers), EB = effective bandwidth (in GHz).
Adapted from [1,2].}
 	\begin{tabular}{l c c c}
\hline
LFI & & &\\
\hline
	Frequency (GHz) &	$30\,$	& $44\,$	& $70\,$ \\
\hline
	InP DT	& MIC	& MIC	& MMIC \\
	FWHM 	& 33.34 &	26.81 &	13.03 \\
	N of R (or feeds)	& 4 (2)	& 6 (3)	& 12 (6) \\
	EB	& 6	& 8.8 	& 14 \\
	NET & 159 & 197 & 158 \\	
	$\delta$T/T [$\mu$K/K] (in $T$)	& 2.48    & 3.82 & 6.30 \\
	$\delta$T/T [$\mu$K/K] (in $P$)	& 3.51    & 5.40 & 8.91 \\
\end{tabular}
	\begin{tabular}{l c c}
\hline
HFI & & \\
\hline
	Frequency (GHz) 	& $100\,$	& $143\,$	 \\
\hline
         FWHM  in $T$ ($P$) 	& (9.6) & 7.1 (6.9)  \\
         	N of B in $T$ ($P$) & (8) & 4 (8)  \\
	EB	in $T$ ($P$) & (33) & 43 (46) 	\\
	NET in $T$ ($P$) & 100  (100) & 62 (82)  \\
	$\delta$T/T [$\mu$K/K] in $T$ ($P$) &  2.1 (3.4)  & 1.6 (2.9) \\
\hline
\end{tabular}
	\begin{tabular}{l c c}
	Frequency (GHz) &	 $217\,$  & $ 353 \,$ \\
\hline
	FWHM 	 in $T$ ($P$) & 4.6 (4.6) &  4.7 (4.6) \\
	N of B in $T$ ($P$) 	& 4 (8) & 4 (8)  \\
	EB	in $T$ ($P$) &  72 (63) 	& 99 (102)  \\
	NET  in $T$ ($P$) & 91 (132) & 277 (404) \\
	$\delta$T/T [$\mu$K/K] in $T$ ($P$)	& 3.4 (6.4)  & 14.1 (26.9) \\
\hline
\end{tabular}
	\begin{tabular}{l c c}
	Frequency (GHz) &	 $545\,$  & $ 857 \,$ \\
\hline
	FWHM 	 in $T$ & 4.7 &  4.3 \\
	N of B in $T$	& 4 & 4  \\
	EB 	in $T$ & 169 	& 257 \\
	NET in $T$ & 2000 & 91000 \\
	$\delta$T/T [$\mu$K/K] in $T$	& 106    & 4243 \\
\hline
\end{tabular}
\label{table:sens}
\end{table}

\noindent
The first scientific results (http://www.sciops.esa.int/index.php?project=PLANCK\&page=Planck\_Published\_Papers)
have been released on January 2011, while
few other papers have been released very recently.
They describe the instrument performance in flight including thermal behaviour (papers I--IV),
the LFI and HFI data analysis pipelines (papers V--VI),
the main astrophysical results about Galactic science (papers XIX--XXV),
extragalactic sources and far-IR background (papers XIII--XVIII and [3]),
and Sunyaev-Zel'dovich effects and cluster properties  (papers VIII--XII and [4]),
providing to the scientific community the {\it Planck\/} Early Release Compact Source Catalog (ERCSC) 
(paper VII and {\it The Explanatory Supplement to the Planck Early Release Compact Source Catalogue}).

The presentation of a next set of (some tens of) astrophysical papers is planned by the first half of 2012.
Astrophysical papers will be roughly divided into two wide categories: those mainly based only on total intensity data and
those requiring well established polarization data. 
Some of the {\it Planck\/} updated astrophysical results will be presented in occasion of the Conference 
{\it Astrophysics from radio to sub-millimeter wavelengths: the {\it Planck\/} view and other experiments} to be held in Bologna on 13-17 February 2012.
The first publications of the main cosmological implications are expected in early 2013, together with the delivery of a first set of {\it Planck\/} maps
and cosmological products based on the first 15 months of data.
{\it Planck\/}  will open a new era in our understanding of the universe and of its astrophysical structures (see [5]
for descriptions of the {\it Planck\/} scientific programme). 
In this lecture we review the current status of {\it Planck\/} data analysis focussing on the production of the first astrophysical results. 
Moreover, we provide a general overview of the {\it Planck\/} fundamental cosmological promises. {\it Planck\/}  will improve the accuracy of current measures of a wide set of cosmological parameters by a factor from $\sim 3$ to $\sim 10$ and will characterize the geometry of the universe with unprecedented accuracy thanks to its excellent mapping and removal of all astrophysical emissions achievable through its so wide frequency coverage.  {\it Planck\/} will put light on many of the open issues in the connection between the early stages of the universe and the evolution of the cosmic structures, from the characterization of primordial conditions and perturbations to the late phases of cosmological reionization. 
We discuss also the {\it Planck\/}  perspectives for some crucial selected topics linking cosmology to fundamental physics (the neutrino masses and effective species number, the primordial helium abundance, the parity property of CMB maps and its connection with CPT symmetry with emphasis to the Cosmic Birefringence, the detection of the stochastic field of gravitational waves through the identification of the so-called B-mode angular power spectrum of the CMB anisotropies -- see Fig. 1), showing how  
{\it Planck\/} represents also an extremely powerful fundamental and particle physics laboratory.

\begin{figure}
    \centering
    \includegraphics[width=0.6\hsize]{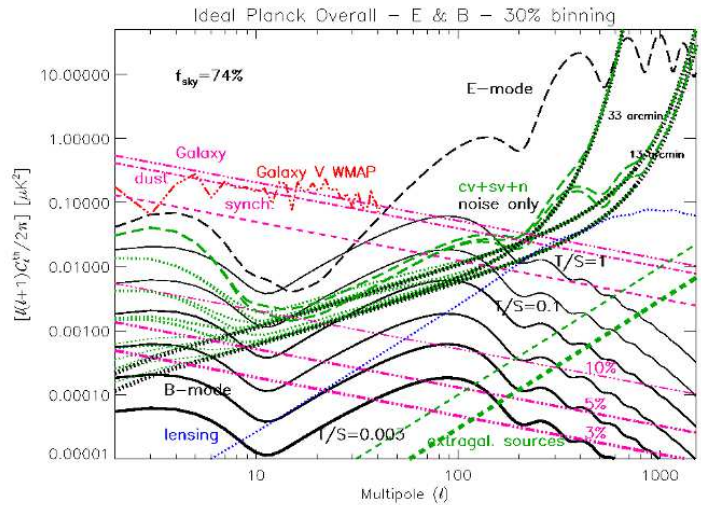}
    \caption{CMB E polarisation modes (black long dashes) compatible with {\it WMAP\/} data and CMB B polarisation modes (black solid lines) for different tensor-to-scalar 
ratios $T/S=r$ of primordial perturbations) are compared to {\it Planck\/} overall sensitivity to the power spectrum for two surveys and whole mission,
assuming the noise expectation has been subtracted, and two different FWHM angular resolutions. 
The plots include cosmic and sampling variance plus instrumental noise 
(green dots for B modes, green long dashes for E modes, 
labeled with cv+sv+n; black thick dots, noise only) 
assuming a multipole binning of 30\%. 
Note that the cosmic and sampling (74\% sky coverage
excluding the sky regions mostly affected by Galactic emission) 
variance implies a dependence of the overall sensitivity on $r$ at low multipoles,
relevant to the parameter estimation; 
instrumental noise only determines the capability of detecting the B mode. 
The B mode induced by lensing (blue dots) is also shown.
Galactic synchrotron
(purple dashes) and dust (purple dot-dashes) polarised emissions produce the
overall Galactic foreground (purple three dot-dashes). {\it WMAP\/} 3-yr
power-law fits for uncorrelated dust and synchrotron have been used. For
comparison, {\it WMAP\/} 3-yr results (http://lambda.gsfc.nasa.gov/) derived from the foreground
maps using HEALPix tools
(http://healpix.jpl.nasa.gov/) [6], which use is acknowledged,
are shown: 
power-law
fits provide (generous) upper limits to the power at low multipoles.
Residual contamination levels by Galactic
foregrounds (purple three dot-dashes) are shown for 10\%, 5\%, and 3\% of the
map level, at increasing thickness. 
We plot also as thick and thin green dashes 
realistic estimates of 
the residual
contribution of un-subtracted extragalactic sources, $C_\ell^{\rm res,PS}$ and
the corresponding uncertainty, $\delta C_\ell^{\rm res,PS}$.
}
\end{figure}


{{\small{
{\bf Acknowledgements --} {{\it Planck} (http://www.esa.int/Planck) 
is a project of the European Space Agency - ESA - with instruments provided by two scientific Consortia
funded by ESA member states (in particular the lead countries: 
France and Italy) with contributions from NASA (USA), and telescope
reflectors provided in a collaboration between ESA and a scientific
Consortium led and funded by Denmark.
We acknowledge the support by the ASI/INAF Agreement I/072/09/0 for the {\it Planck\/}  LFI Activity of Phase E2. 
}}}


{\bf References}

\begin{description}

\item[1] N. Mandolesi, et al., 2010, A\&A 520, A3

\vspace{-0.3cm}

\item[2] J.-M. Lamarre, et al., 2010, A\&A 520, A9

\vspace{-0.3cm}

\item[3] P. Giommi, et al., 2011, A\&A, submitted, arXiv:1108.1114

\vspace{-0.3cm}

\item[4] {\it Planck\/} Collaboration, 2011, A\&A, submitted, arXiv:1106.1376

\vspace{-0.3cm}

\item[5] {\it Planck\/} Collaboration, 2005, ESA publ. ESA-SCI(2005)/01, arXiv:0604069; J. Tauber, et al., 2010, A\&A 520, A1

\vspace{-0.3cm}

\item[6] K.M. G\'orski, et al. 2005, ApJ, 622, 759

\end{description}

\newpage

\subsection{Hector J. de Vega and Norma G. Sanchez}

\vskip -0.3cm

\begin{center}

H.J.dV: LPTHE, CNRS/Universit\'e Paris VI-P. \& M. Curie \& Observatoire de Paris, Paris, France\\
N.G.S: LERMA, CNRS/Observatoire de Paris, Paris, France

\medskip

{\bf Predictions of the Effective Theory of Inflation in the Standard Model of the 
Universe and the CMB+LSS data analysis}

\end {center}

\medskip

Inflation is today a part of the Standard Model of the Universe supported by
the cosmic microwave background (CMB) and large scale structure (LSS)
datasets. Inflation solves the horizon and flatness problems and
naturally generates  density fluctuations that seed LSS and CMB anisotropies,
and tensor perturbations (primordial gravitational waves). 
Inflation theory is
based on a scalar field  $ \varphi $ (the inflaton) whose potential
is fairly flat leading to a slow-roll evolution. 

\medskip

We focuse here on the following new aspects of inflation. We present the
effective theory of inflation \`a la {\bf Ginsburg-Landau} in which
the inflaton potential is a polynomial in the field $ \varphi $ and has
the universal form $ V(\varphi) = N \; M^4 \;
w(\varphi/[\sqrt{N}\; M_{Pl}]) $, where $ w = {\cal O}(1) , \;
M \ll M_{Pl} $ is the scale of inflation and  $ N \sim 60 $ is the number
of efolds since the cosmologically relevant modes exit the horizon till
inflation ends. The slow-roll expansion becomes a systematic $ 1/N $ expansion and
the inflaton couplings become {\bf naturally small} as powers of the ratio
$ (M / M_{Pl})^2 $. The spectral index and the ratio of tensor/scalar
fluctuations are $ n_s - 1 = {\cal O}(1/N), \; r = {\cal O}(1/N) $ while
the running index turns to be $ d n_s/d \ln k =  {\cal O}(1/N^2) $
and therefore can be neglected. The {\bf energy scale of inflation }$ M \sim 0.7
\times 10^{16}$ GeV turns to be completely determined by the amplitude of the scalar 
adiabatic fluctuations [1-2]. 

\bigskip

A complete analytic study plus the
Monte Carlo Markov Chains (MCMC) analysis of the available
CMB+LSS data (including WMAP5) with fourth degree trinomial potentials
showed [1-3]: 

\begin {itemize}

\item {{\bf(a)} the {\bf spontaneous breaking} of the
$ \varphi \to - \varphi $ symmetry of the inflaton potential.} 

\medskip

\item {{\bf(b)} a {\bf lower bound} for $ r $ in new inflation:
$ r > 0.023 \; (95\% \; {\rm CL}) $ and $ r > 0.046 \;  (68\% \;
{\rm CL}) $. }

\medskip

\item {{\bf(c)} The preferred inflation potential is a {\bf double
well}, even function of the field with a moderate quartic coupling
yielding as most probable values: $ n_s \simeq 0.964 ,\; r\simeq
0.051 $. This value for $ r $ is within reach of forthcoming CMB
observations. }

\medskip

\item {{\bf(d)} The present data in the effective theory of
inflation clearly {\bf prefer new inflation}. }

\medskip

\item { {\bf(e)} Study of higher degree
inflaton potentials show that terms of degree higher than four do not
affect the fit in a significant way. In addition, horizon exit happens for
$ \varphi/[\sqrt{N} \; M_{Pl}] \sim 0.9 $ making higher order terms
in the potential $ w $ negligible [4]. }

\item { {\bf(f)} Within the Ginsburg-Landau potentials in new inflation,
$ n_s$ and  $r$ in the $(n_s, r)$ plane are within the universal banana region 
fig. \ref{banana} and $ r $ is in the range $ 0.021 < r < 0.053 $ [4].}

\end {itemize}


We summarize the physical effects of
{\bf generic} initial conditions (different from Bunch-Davies) on the
scalar and tensor perturbations during slow-roll and
introduce the transfer function $ D(k) $ which encodes the observable
initial conditions effects on the power spectra.
These effects are more prominent in the \emph{low}
CMB multipoles: a change in the initial conditions during slow roll can
account for the observed CMB {\bf quadrupole suppression} [1].

\medskip

Slow-roll inflation is generically preceded by a
short {\bf fast-roll} stage. Bunch-Davies initial conditions are the
natural initial conditions for the fast-roll perturbations. 

\medskip

The characteristic time scale of the fast-roll era turns to be 
$ t_1= (1/m) \; \sqrt{V(0)/[3 \; M^4] } \sim 10^4 \;  t_{Planck} $.
The {\bf whole} evolution of the fluctuations along the 
decelerated and inflationary 
fast-roll and slow-roll eras is computed in ref. [5]. 

\medskip

The Bunch-Davies initial conditions 
(BDic) are generalized for the fast-roll case in which the potential felt by the 
fluctuations can never be neglected. The fluctuations feel 
a {\bf singular attractive} potential near the $ t = t_* $ singularity 
(as in the case of a particle in a central singular potential) with {\bf exactly} the 
{\bf critical} strength ($ -1/4 $) allowing the fall to the centre.

The power spectrum gets {\bf dynamically modified} by the effect of the 
fast-roll eras and the choice of BDic at a finite time, 
through the transfer function $ D(k) $. The power spectrum
vanishes at $ k = 0 . \; D(k) $ presents a first peak for $ k \sim 2/\eta_0 $
($ \eta_0 $ being the conformal initial time), 
then oscillates with decreasing amplitude and vanishes asymptotically for $ k \to \infty $.
The transfer function $ D(k) $ affects the {\bf low} CMB multipoles $ C_{\ell} $:
the change $ \Delta C_{\ell}/ C_{\ell} $ for $ 1 \leq \ell \leq 5 $ is computed in [5] as a
function of the starting instant of the fluctuations $ t_0 $.
CMB quadrupole observations indicate large {\bf suppressions} which
are well reproduced for the range  $ t_0 - t_\ast \gtrsim 0.05/m \simeq 10100 \; t_{Planck} $.

\medskip

A MCMC analysis of the WMAP+SDSS data {\bf including fast-roll} shows that the quadrupole
mode exits the horizon about 0.2 efold before fast-roll ends and its
amplitude gets suppressed. In addition, fast-roll fixes the {\bf initial
inflation redshift} to be $ z_{init} = 0.9 \times 10^{56} $ and
the {\bf total number} of efolds of inflation to be $ N_{tot} \simeq 64 $ [1,3].
Fast-roll fits the TT, the TE and the EE
modes well reproducing the quadrupole supression. 

\medskip

A thorough study of the
{\bf quantum loop corrections} reveals that they are very small and controlled by
powers of $(H /M_{Pl})^2 \sim {10}^{-9} $, {\bf a conclusion that validates the
reliability of the effective theory of inflation [1].} 

\medskip

This work [1-4] shows how powerful is
the Ginsburg-Landau effective theory of inflation in predicting
observables that are being or will soon be contrasted to observations.

\medskip

The Planck satellite is right now measuring with unprecedented accuracy 
the primary CMB anisotropies.
The Standard Model of the Universe (including inflation) provides the context to analyze the CMB 
and other data. The Planck performance for $ r $ related to the primordial $ B $ 
mode polarization, will depend on the quality of the data analysis.

\medskip

The Ginsburg Landau approach to inflation allows to take high benefit of the CMB data.
We evaluate the Planck precision to the recovery of cosmological
parameters within a reasonable toy model for residuals of systematic
effects of instrumental and astrophysical origin based on publicly
available information.
We use and test two relevant models: the $\Lambda$CDM$r$ model,
i. e. the standard $\Lambda$CDM model augmented by  $ r $, and the
$\Lambda$CDM$r$T model, where the scalar spectral index, $ n_s $, and $ r $
are related through the theoretical `banana-shaped' curve $ r = r(n_s) $
coming from the double-well inflaton potential (upper boundary of the banana region
fig. \ref{banana}. In the latter case,
$ r = r(n_s) $ is imposed as a hard constraint in the MCMC data
analysis. We take into account the white noise sensitivity of Planck in the 70,
100 and 143 GHz channels as well as the residuals from systematics
errors and foregrounds. Foreground residuals turn to affect
only the cosmological parameters sensitive to the B modes [6].

\medskip

In the Ginsburg-Landau inflation approach, better measurements on $ n_s $, 
as well as on TE and EE modes will improve 
the prediction on $ r $ even if a detection of B modes is still
lacking [6].

\medskip

Forecasted B mode detection probability by
the most sensitive HFI-143 channel:
At the level of foreground residual equal to
30\% of our toy model, only a 68\% CL detectiof $ r $ is very likely.
For a 95\% CL detection the level of
foreground residual should be reduced to 10\%
or lower of the adopted toy model. The possibility to detect
$ r $ is borderline [6].

\begin{figure}[ht]
\includegraphics[height=6cm,width=10cm]{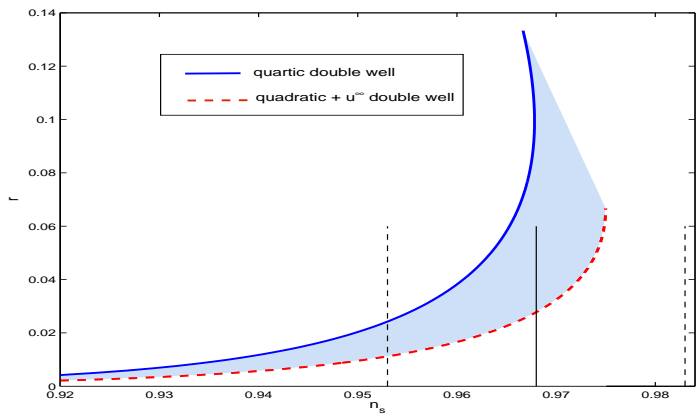}
\caption{The universal banana region $ \cal B $ in the $ (n_s, r) $-plane
  setting $ N = 60 $. The upper border of the region $ \cal B $ corresponds to
  the fourth order double--well potential (new inflation). The lower border is
  described by the potential $ V(\varphi) = \frac12{ m^2} \,
  \left(\frac{m^2}{\lambda} - \varphi^2\right) $ for $ \varphi^2 < m^2/\lambda $
  and $ V(\varphi) = \infty $ for $ \varphi^2 > m^2/\lambda $ [4].  We
  display in the vertical full line the observed value $ n_s = 0.968 \pm
  0.015 $ using the WMAP+BAO+SN data set.  The broken vertical lines delimit
  the $ \pm 1 \, \sigma$ region.}
\label{banana}
\end{figure}

\begin{figure}[ht]
  \includegraphics[height=7cm]{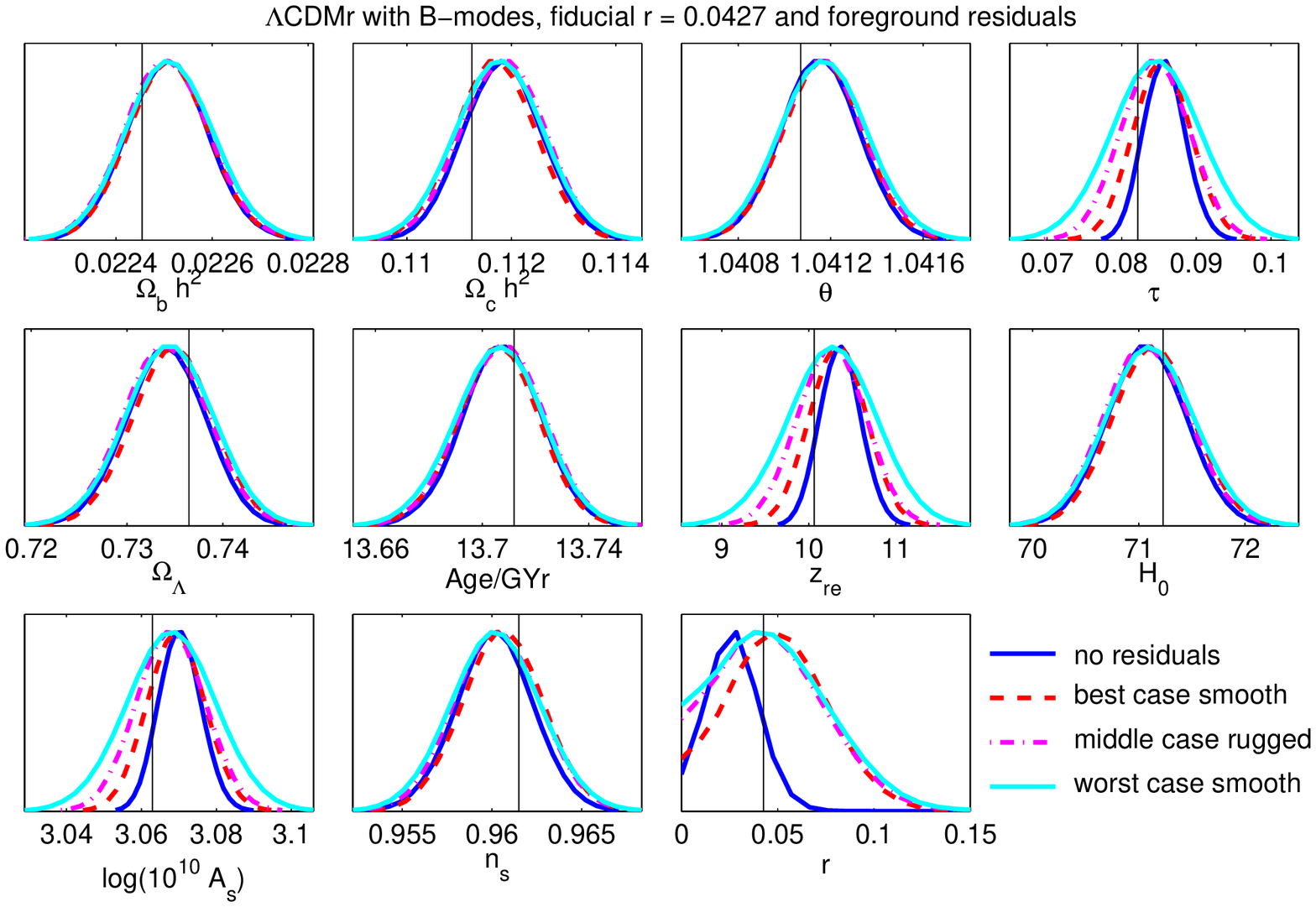}\\
  \includegraphics[height=7cm]{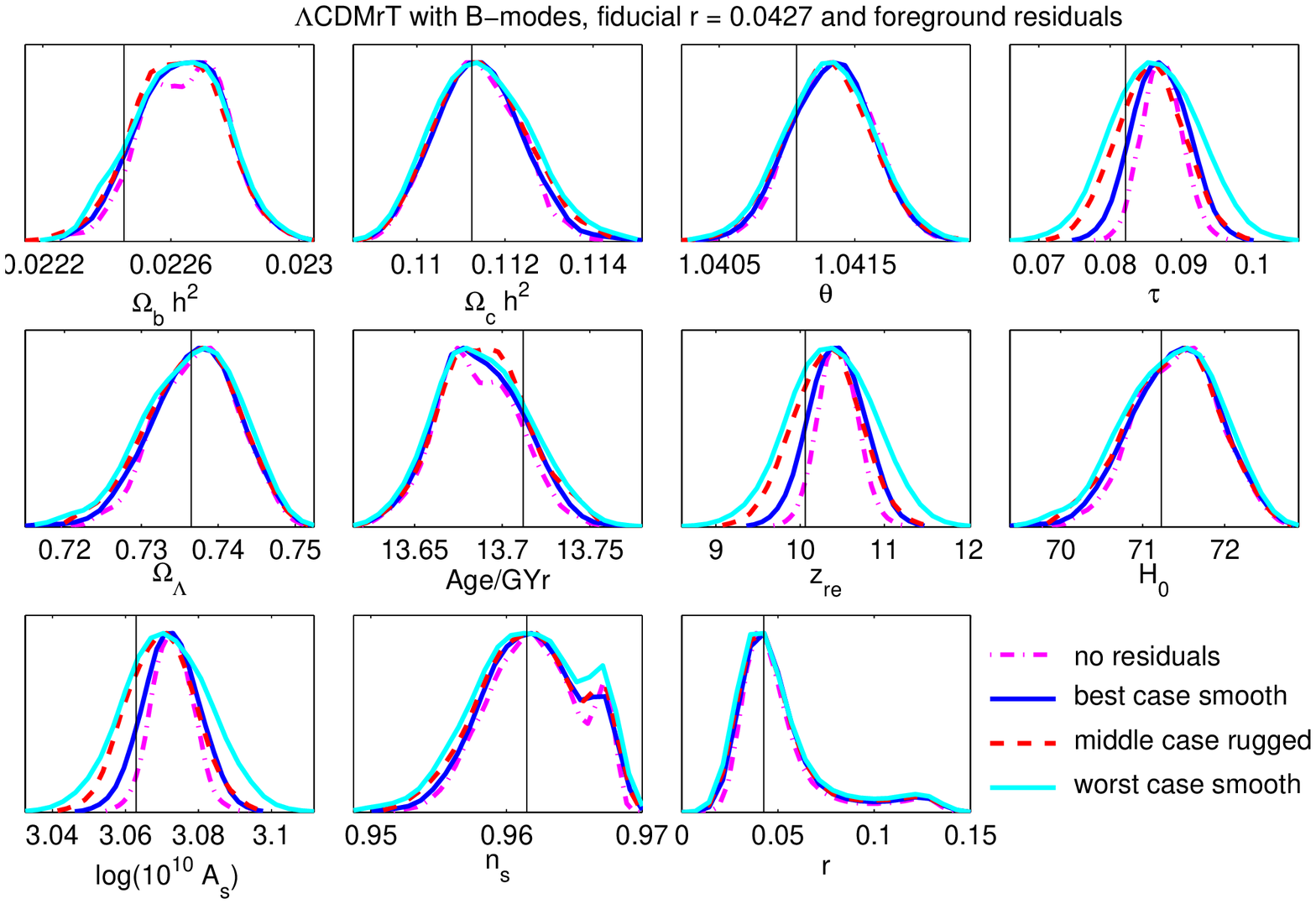}
  \caption{Forecasts for Planck [6]. 
Upper panel: Cumulative $3-$channel marginalized likelihood distributions,
    including $ B $ modes and foreground residuals, of the cosmological
parameters for the $\Lambda$CDM$r$ model.  The fiducial ratio is 
$ r = 0.0427 $. Lower panel:
Cumulative marginalized likelihoods from the three channels for the
    cosmological parameters for the $\Lambda$CDM$r$T model including $ B $ modes
    and fiducial ratio $ r = 0.0427 $ and the foreground residuals. 
We plot the
distributions  in four cases: (a) without residuals, (b) best case smooth:
with 30\% of the toy model residuals in the $TE$ and $E$ modes.
(c) worst case smooth: with the toy model residuals in the $TE$ and $E$ modes.
(d) with 65\% of the toy model 
residuals in the $TE$ and $E$ modes 
and $ 88 \mu K^2 $ in the $T$ modes rugged by 
Gaussian fluctuations of $ 30 \% $ relative strength. }
  \label{ufa7}
\end{figure}


{\bf References}

\medskip
 
[1]  Review article: D. Boyanovsky, C. Destri, H. J. de Vega, N. G. Sanchez

Int. J. Mod. Phys. A24, 3669-3864 (2009) and  author's references therein.

\medskip

[2] C. Destri, H. J. de Vega, N. Sanchez,
Phys. Rev. D77, 043509 (2008), astro-ph/0703417.

\medskip

[3] C. Destri, H. J. de Vega, N. G. Sanchez, arXiv:0804.2387.
Phys. Rev. D 78, 023013 (2008).

\medskip

[4] C. Destri, H. J. de Vega, N. G. Sanchez, arXiv:0906.4102, 
Annals of Physics, {\bf 326}, 578 (2011).\\
\noindent D. Boyanovsky, H. J. de Vega, C. M. Ho et N. G. Sanchez, 
Phys. Rev.  D75, 123504 (2007).

\medskip

[5] C. Destri, H. J. de Vega, N. G. Sanchez, 
Phys. Rev. D81, 063520 (2010).

\medskip

[6] C. Burigana, C. Destri, H. J. de Vega, A. Gruppuso, N. Mandolesi, P. Natoli, N. G. Sanchez,
arXiv:1003.6108, Astrophysical Journal, 724, 588 (2010).

\newpage



\subsection{Hector J. de Vega and Norma G. Sanchez}

\vskip -0.3cm

\begin{center}

HJdV: LPTHE, CNRS/Universit\'e Paris VI-P. \& M. Curie \& Observatoire de Paris.\\
NGS: Observatoire de Paris, LERMA \& CNRS

\bigskip

{\bf keV scale dark matter from theory and observations and 
galaxy properties from linear primordial fluctuations} 

\end{center}

In the context of the standard Cosmological model the nature of dark matter (DM)
is unknown. Only the DM gravitational effects are noticed and they are necessary
to explain the present structure of the Universe. DM particle candidates are not present in the standard model (SM) of particle physics. Particle model independent theoretical analysis combined with astrophysical data from
galaxy observations points towards a DM particle mass in
the {\bf keV scale} (keV = 1/511 electron mass) [1-4]. Many extensions of the SM can be envisaged to include a DM particle with mass in the keV scale and
weakly enough coupled to the Standard Model particles to fulfill
all particle physics experimental constraints.

DM particles can decouple being ultrarelativistic (UR) at 
$ \; T_d \gg m $ or non-relativistic $ \; T_d \ll m $.
They may  decouple at or out of local thermal equilibrium (LTE).
The DM distribution function: $ F_d[p_c] $ freezes out at decoupling
becoming a function of the  comoving momentum $ p_c = $.
$ P_f(t) = p_c/a(t) = $ is the physical momentum. 
Basic physical quantities can be expressed in terms of the
distribution function as the velocity fluctuations,
$ \langle \vec{V}^2(t) \rangle = \langle \vec{P}^2_f(t) \rangle/m^2 $ 
and the DM energy density $ \rho_{DM}(t) $
where $ y = P_f(t)/T_d(t) = p_c/T_d $ is the integration variable and
$ g $  is the number of internal degrees of freedom of the DM 
particle; typically $ 1 \leq g \leq 4 $.

\medskip

{\bf Two} basic quantities characterize DM: its particle mass $ m $
and the temperature $ T_d $ at which DM decouples.  $ T_d $
is related by entropy conservation to the number of
ultrarelativistic degrees of freedom $ g_d $ at decoupling by
$ \quad T_d = \left(2/g_d \right)^\frac13 \; T_{cmb} \; ,
\; T_{cmb} = 0.2348 \; 10^{-3} \; $ eV.
One therefore needs {\bf two} constraints to determine the values of
$ m $ and $ T_d $ (or $ g_d $).

\medskip

One constraint is to reproduce the known cosmological DM density today.
$\rho_{DM}({\rm today})= 1.107 \; {\rm keV/cm}^3 $.

Two independent further constraints are considered in refs. [1-4].
First, the phase-space density $ Q=\rho/\sigma^3 $ [1-2] and second the
surface acceleration of gravity (surface density) in DM dominated galaxies [3-4].
We therefore provide {\bf two} quantitative ways to derive the value $ m $ 
and $ g_d $ in refs. [1-4].

\medskip

The phase-space density $ Q $ is invariant under the
cosmological expansion and can {\bf only decrease} 
under self-gravity interactions 
(gravitational clustering). The value of $ Q $ today follows
observing dwarf spheroidal satellite galaxies of the Milky Way (dSphs):
$ Q_{today} = (0.18 \;  \mathrm{keV})^4 $ (Gilmore et al. 07 and 08).
We compute explicitly $ Q_{prim} $ (in the primordial universe) and it turns
to be proportional to $ m^4 $ [1-4].

\medskip

During structure formation $ Q $
{\bf decreases} by a factor that we call $ Z $. Namely, 
$ Q_{today} = Q_{prim}/Z $. The value of $ Z $ is galaxy-dependent.
The spherical model gives $ Z \simeq 41000 $
and $N$-body simulations indicate: $ 10000 >  Z > 1 $ (see [1]).
Combining the value of $ Q_{today} $ and $\rho_{DM}({\rm today}) $ with 
the theoretical analysis yields that $ m $ must be in the keV scale and 
$ T_d $ can be larger than 100 GeV. More explicitly, we get 
general formulas for $ m $ and $ g_d $ [1]:
$$ 
m = \frac{2^\frac14 \; \sqrt{\pi}}{3^\frac38 \; g^\frac14 } \; 
Q_{prim}^\frac14
\; I_4^{\frac38} \; I_2^{-\frac58} \; , \quad
g_d = \frac{2^\frac14 \; g^\frac34}{3^\frac38 \; 
\pi^\frac32 \; \Omega_{DM}} \; 
 \; \frac{T_{\gamma}^3}{\rho_c} \; Q_{prim}^\frac14 \; 
\left[I_2 \; I_4\right]^{\frac38}
$$
where $ I_{2 \, n} = \int_0^\infty y^{2 \, n} \; F_d(y) \; dy 
\quad , \quad n=1, 2 $ 
and $ Q_{prim}^\frac14 = Z^\frac14 \; \; 0.18 $ keV using the dSphs data,
$T_{\gamma} = 0.2348 \; {\rm meV } 
\; , \; \Omega_{DM} = 0.228 $ and $ \rho_c = (2.518 \; {\rm meV})^4$.
These formulas yield for relics decoupling UR at LTE:
$$ 
m = \left(\frac{Z}{g}\right)^\frac14 \; \mathrm{keV} \; 
\left\{\begin{array}{l}
         0.568 \\
              0.484      \end{array} \right. \; , \;
 g_d = g^\frac34 \; Z^\frac14 \; \left\{\begin{array}{l}
         155~~~\mathrm{Fermions} \\
              180~~~\mathrm{Bosons}      \end{array} \right. \; . 
$$
Since $ g = 1-4 $, we see that 
$ g_d \gtrsim 100 \Rightarrow  T_d \gtrsim 100 $ GeV.
Moreover, $ 1 < Z^\frac14 < 10 $ for $ 1 < Z < 10000 $.
For example for DM Majorana fermions $ (g=2) \; m \simeq 0.85 $ keV.

\medskip

We get results for $ m $ and $ g_d $ on the same scales for DM particles 
decoupling UR out of thermal equilibrium [1]. For a specific model of 
sterile neutrinos where decoupling is out of thermal equilibrium:
$$
0.56 \; \mathrm{keV} \lesssim m_{\nu} \;  
Z^{-\frac14} \lesssim 1.0 \; \mathrm{keV}
\quad ,  \quad 15 \lesssim g_d  \;  Z^{-\frac14}\lesssim 84
$$
For relics decoupling non-relativistic
we obtain similar results for the DM particle mass: keV 
$ \lesssim m \lesssim $ MeV [1].

\medskip

Notice that the dark matter particle mass $ m $ and decoupling temperature 
$ T_d $ are {\bf mildly} affected by the uncertainty in the factor $ Z $ through a 
power factor 
$ 1/4 $ of this uncertainty, namely, by a factor $ 10^{\frac14} \simeq 1.8 $

\medskip

The comoving free-streaming) wavelength, and the Jeans' mass are obtained in the range
$$
\frac{0.76}{\sqrt{1+z}} \; {\rm kpc} <\lambda_{fs}(z) <
\frac{16.3}{\sqrt{1+z}} \; {\rm kpc} \; , \; 0.45 \; 10^3 \; M_{\odot} 
< \frac{M_J(z)}{(1+z)^{+\frac32}} < 0.45 \; 10^7  \; 
\; M_{\odot} \; .
$$

These values at $ z = 0 $ are consistent with the $N$-body simulations 
and are of the order of the small dark matter structures observed today .
By the beginning of the matter dominated era $ z \sim 3200 $, the masses are of the 
order of galactic masses $ \sim 10^{12} \; M_{\odot} $ and the comoving free-streaming 
wavelength scale turns to be of the order of the galaxy sizes today 
$ \sim 100 \; {\rm kpc}$.

\medskip

 Lower and upper bounds for the dark matter annihilation cross-section $ \sigma_0 $ 
are derived: $ \sigma_0 > (0.239-0.956) \; 10^{-9} \; \mathrm{GeV}^{-2} $ and 
$ \sigma_0 < 3200 \; m \; \mathrm{GeV}^{-3} \; . $ There is at least five orders of 
magnitude between them, the dark matter non-gravitational self-interaction is 
therefore negligible (consistent with structure formation and observations, as 
well as by comparing X-ray, optical and lensing observations of the merging of 
galaxy clusters with $N$-body simulations).

\medskip

Typical `wimps' (weakly interacting massive particles) with mass $ m = 100 $ GeV 
and $ T_d = 5 $ GeV  would require a huge $ Z \sim 10^{23} $, well above
the upper bounds obtained and cannot reproduce the observed galaxy properties. 
They produce an extremely short free-streaming or Jeans length $ \lambda_{fs} $ today $ 
\lambda_{fs}(0) \sim 3.51 \; 10^{-4} \; {\rm pc} = 72.4  \; {\rm AU} \; $ that would
correspond to unobserved structures much smaller than the galaxy structure.
Wimps result are strongly disfavoured. 

\medskip

Galaxies are described by a variety of physical quantities:

\medskip

(a) {\bf Non-universal} quantities: mass, size, luminosity, fraction of DM,
DM core radius $ r_0 $, central DM density $ \rho_0 $.

(b) {\bf Universal} quantities: surface density $ \mu_0 \equiv r_0 \; \rho_0 $
and DM density profiles. $M_{BH}/M_{halo}$ (or halo binding energy).
The galaxy variables are related by
{\bf universal} empirical relations. Only one variable remains free. 
That is, the galaxies are a one parameter family of objects.
The existence of such universal quantities may be explained by the
presence of attractors in the dynamical evolution. 
The quantities linked to the attractor always reach the same value
for a large variety of initial conditions. This is analogous
to the universal quantities linked to fixed points in critical
phenomena of phase transitions.The universal DM density profile in Galaxies has the scaling property: 
$$
\rho(r) = \rho_0 \; F\left(\displaystyle\frac{r}{r_0}\right) \quad , \quad  
F(0) = 1 \quad , \quad x \equiv  \displaystyle\frac{r}{r_0} \; ,
$$
where $ r_0 $ is the DM core radius.
As empirical form of cored profiles one can take Burkert's form for $ F(x) $.
Cored profiles {\bf do reproduce} the astronomical observations.

\medskip

The surface density for dark matter (DM) halos 
and for luminous matter galaxies is defined as: 
$ \mu_{0 D} \equiv r_0 \; \rho_0, $ 
$ r_0 = $ halo core radius, $ \rho_0 = $ central density for DM galaxies.
For luminous galaxies  $ \rho_0 = \rho(r_0) $
(Donato et al. 09, Gentile et al. 09).
Observations show an Universal value for $ \mu_{0  D} $: independent of 
the galaxy luminosity for a large number of galactic systems 
(spirals, dwarf irregular and spheroidals, elliptics) 
spanning over $14$ magnitudes in luminosity and of different Hubble types.
Observed values:
$$
\mu_{0  D} \simeq 120 \; \frac{M_{\odot}}{{\rm pc}^2} = 
5500 \; ({\rm MeV})^3 = (17.6 \; {\rm Mev})^3 \quad , \quad
5 {\rm kpc}  < r_0 <  100 {\rm kpc} \; .
$$
Similar values $ \mu_{0  D} \simeq 80  \; \frac{M_{\odot}}{{\rm pc}^2} $ are observed in 
interstellar molecular clouds of size $ r_0 $ of different type and composition over 
scales $ 0.001 \, {\rm pc} < r_0 < 100 $ pc (Larson laws, 1981).
Notice that the surface gravity acceleration is 
given by $\mu_{0  D}$ times Newton's constant.

\bigskip

We combine in refs.[3-4] the theoretical evolution of density fluctuations 
computed from first principles since decoupling till today to the observed properties of
galaxies as the surface density and core radius. We obtain that 
(i) the dark matter particle mass must be in the keV scale both for in and out
of thermal equilibrium decoupling (ii) the density profiles are cored for 
keV scale DM particles and cusped for GeV scale  DM particles (wimps). 

\bigskip

{\bf References}

\begin{description}

\item[1]  H. J. de Vega, N. G. Sanchez,  arXiv:0901.0922, 
Mon. Not. R. Astron. Soc. 404, 885 (2010).

\item[2] D. Boyanovsky, H. J. de Vega, N. G. Sanchez, 	
arXiv:0710.5180, Phys. Rev. {\bf D 77}, 043518 (2008).

\item[3] H. J. de Vega, N. G. Sanchez, Int. J. Mod. Phys. A26: 1057 (2011),
arXiv:0907.0006.

\item[4] H. J. de Vega, P. Salucci, N. G. Sanchez, arXiv:1004.1908.

\end{description}

\newpage

\subsection{Joanna Dunkley for the ACT Collaboration}

\vskip -0.3cm

\begin{center}

Oxford Astrophysics, Denys Wilkinson Building, Keble Road, Oxford OX1 3RH, UK\

\bigskip

{\bf Cosmology from ACT: the small-scale CMB} 

\end{center}

\medskip

{\bf Introduction:} The Cosmic Microwave Background has been measured over the whole sky by three generations of satellites at increasingly high resolution: COBE, WMAP, and currently Planck. The Atacama Cosmology Telescope (ACT) measures a small fraction of the sky, but at even higher resolution and sensitivity [1]. In doing so it measures not only the primordial light from the time of recombination at $z\sim1100$, but also secondary distortions to the light from intervening cosmic structures, and additional microwave emission from individual galaxies.

ACT is located at an elevation of 5190m in the Atacama desert in Chile, one of the driest locations on the planet. It has a 6m primary mirror and over 3000 detectors at three frequencies in the range 148-270~GHz. It has observed the microwave sky for four seasons from 2007-2010, covering about 800 square degrees at arcminute resolution, in a Southern region at -55$^\circ$ declination, and a celestial Equatorial region overlapping with the SDSS Stripe 82. Maps of the microwave sky are generated using a maximum likelihood map-maker, solving for the sky signal simultaneously with the atmospheric signal, for the detector arrays at each frequency.

\medskip

{\bf The CMB spectrum from ACT:} The power spectrum was estimated from the 2008 Southern data  at 148 and 220~GHz using the flat-sky approximation, and is reported in [2] in the angular range $500<\ell<10000$. At large scales the signal is dominated by the primordial CMB, and the Silk damping tail is clearly seen. At smaller scales the extragalactic point source power dominates, with a larger contribution at 220~GHz due to thermal dust emission from star-forming galaxies. As reported in [3], the total spectrum is well fit by the sum of a lensed CMB component, a Sunyaev-Zel'dovich (SZ) contribution, and both radio and infrared emission from individual galaxies. Focusing on the CMB-dominated region, seven acoustic peaks can now be seen, as shown in Figure \ref{fig:spectrum} from [3]. The spectrum is consistent with the $\Lambda$CDM model estimated from WMAP data [4]. The broader range of angular scales now make it possible to rule out certain cosmological models that were previously degenerate with $\Lambda$CDM at larger scales, but that have different degrees of Silk damping and peak positions at smaller scales.

\begin{figure}[b]
\includegraphics[width=0.5\hsize]{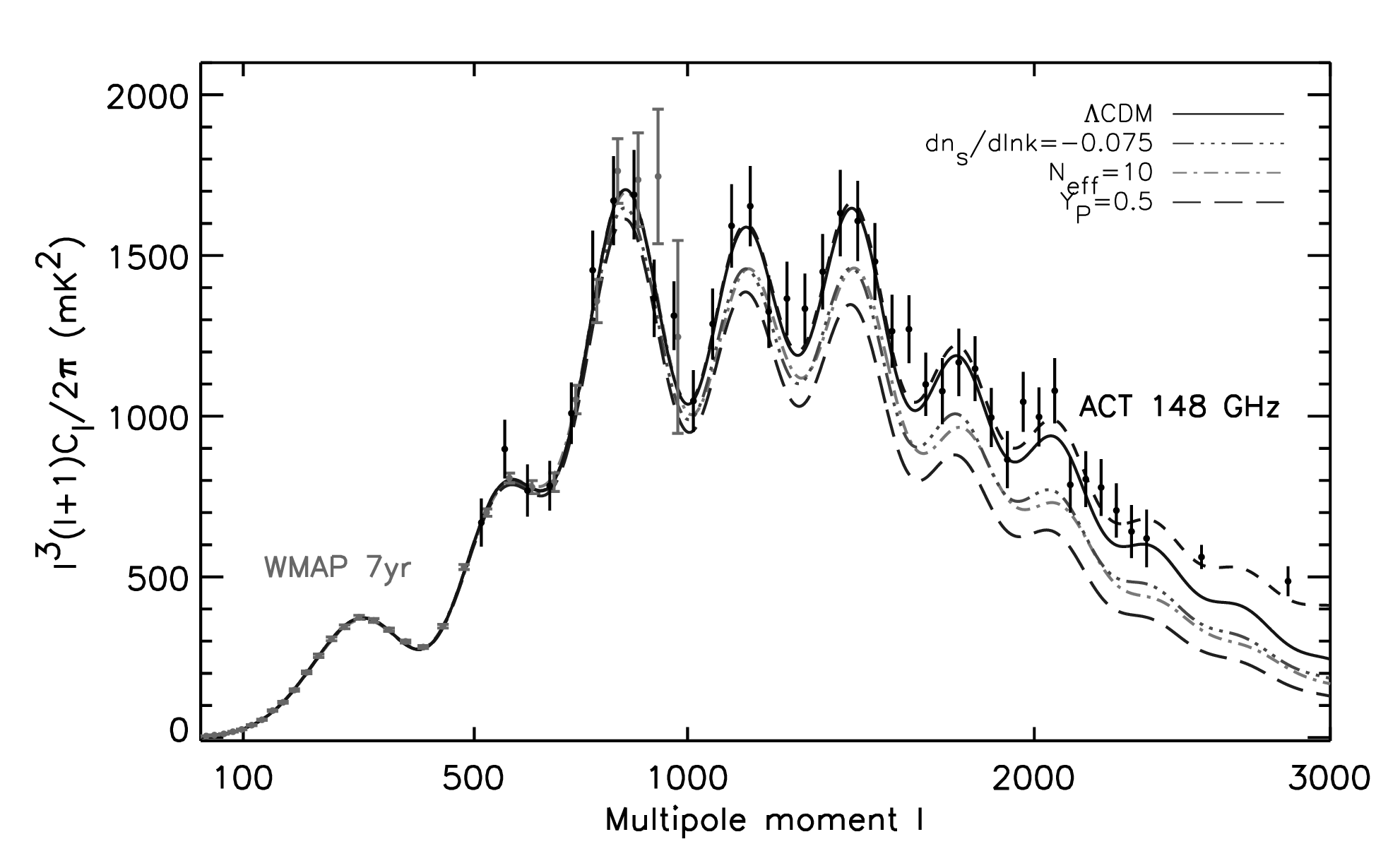}
\caption{{\it Figure from Dunkley et al. 2011 [3]:} Four cosmological models that are degenerate at large scales can be distinguished at smaller scales using ACT, constraining the number of neutrino species, helium fraction, and running of the spectral index. 
\label{fig:spectrum}
}
\end{figure}
\medskip
{\it Inflation and early universe physics:} We have estimated cosmological parameters from ACT combined with the WMAP 7-year data [4], and late-time Baryon Acoustic Oscillation and Hubble constant data. This leads to improved limits on inflationary parameters including the primordial spectral index, running of the spectral index with scale, and tensor-to-scalar ratio, all reported in [3]. A measure of the primordial spectrum as a function of scale is reported in [5], finding no significant deviations from power law fluctuations. We also better constrain the effective number of neutrino species, detect primordial helium at high significance, and constrain the possible string tension of an additional cosmic string component. The final ACT spectrum, together with the recently published spectrum from the South Pole Telescope [6], promises to further test early universe physics.

\medskip

{\bf Late-time effects:} The primary CMB provides an excellent probe of inflationary and early universe parameters, and allows a measure of the relative densities of components in the universe, if a flat geometry is assumed. However, the well-known geometric degeneracy prevents the expansion rate and geometry from both being determined from the CMB. To measure the geometry and behaviour of dark energy as a function of cosmic time, lower-redshift probes are necessary. However, the small-scale CMB can now also be used to probe late-time physics, due to secondary effects.

\medskip
{\it Lensing:} CMB photons are gravitationally deflected by large-scale structure potentials, with deflection angles typically a few arcminutes. A measurement of the deflection field can provide a measurement of the matter fluctuations and the geometry of the universe. Lensing directly affects the temperature power spectrum by smoothing the acoustic peaks and increasing small-scale power, and is observed at almost 3$\sigma$ in the ACT power spectrum [2]. This detection is improved by reconstructing the lensing field directly, using the property that lensing couples modes of different scales, generating a non-zero 4-point function in excess of that expected for a Gaussian field. Measuring the lensing 4-point function has led to a 4$\sigma$ detection of the lensing power spectrum in the Equatorial ACT data, reported in [7] as the first direct detection of CMB lensing. The deflection power is consistent with the $\Lambda$CDM model, and helps constrain cosmological models. A universe with no dark energy, and positive spatial curvature, fits the CMB temperature spectrum almost as well as a flat $\Lambda$CDM  model, as shown in Figure \ref{fig:lensing}. However, this model has enhanced lensing deflection power, and is disfavored at more than 3$\sigma$ when the lensing data is included [8]. The CMB alone can now provide evidence for a dark energy component in the universe.

\begin{figure}[t]
\includegraphics[width=0.48\hsize]{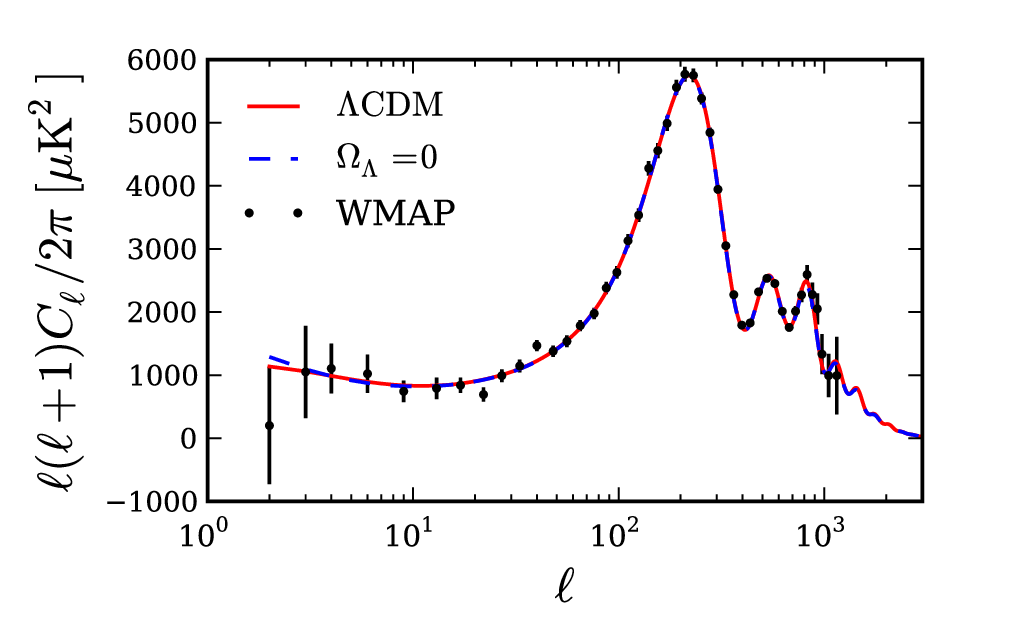}
\hskip -0.2in
\includegraphics[width=0.48\hsize]{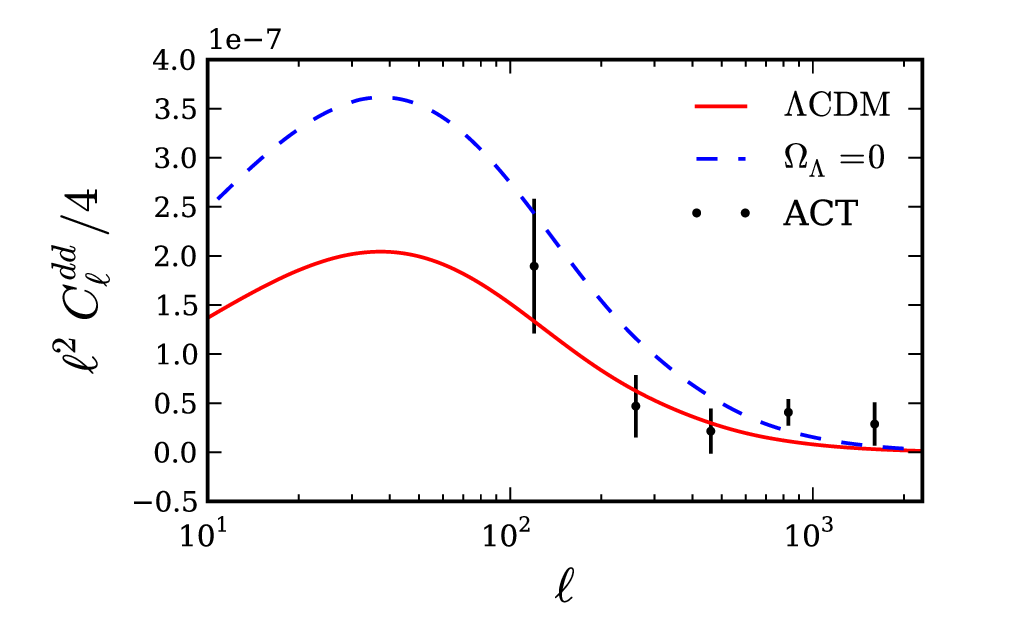}
\vskip-0.2in
\caption{{\it Figure from Sherwin et al. 2011 [8]:} Two models that are degenerate in CMB temperature power (left) can be distinguished using CMB lensing (right): the curved model with no Dark Energy gives more lensing than the $\Lambda$CDM model.
\label{fig:lensing}
}
\end{figure}
\medskip
{\it Sunyaev-Zel'dovich:} The CMB is inverse Compton scattered by hot electrons in galaxy clusters, shifting the black-body spectrum such that clusters can be identified. Their number as a function of mass and redshift is sensitive to the evolution of structure formation, and can be used to probe properties of dark energy and structure growth. In the Southern 2008 ACT data, 23 clusters were detected and optically confirmed at high significance [9,10], and a further sample have been detected in the Equatorial data. Constraints on cosmological parameters from cluster counts are consistent with the $\Lambda$CDM model [11], although constraints are limited by uncertainties in the cluster masses. The mass is expected to scale with the SZ signal, and such scaling relations have been observed in the ACT data [12]; further multi-wavelength observations are expected to better determine masses and improve  constraints.

\medskip

{\bf Conclusion:} By considering the primordial CMB power spectrum, lensing deflection estimates, and cluster number counts, all measured with ACT, the $\Lambda$CDM cosmological model is found to be consistent over a wide range of scales and redshifts. The ACT receiver was removed from the telescope early 2011, and the upgraded receiver, ACTPol, is due to begin observations in 2012. With improved sensitivity and polarization capabilities, it will measure the primordial power spectrum in polarization, and allow a significantly improved measurement of the lensing signal.

\medskip

\bigskip

{\bf References}

\begin{description}

\item[1] http://www.physics.princeton.edu/act/; D. Swetz et al., arXiv:1007.0290 (2010)

\vspace{-0.3cm}
\item[2] S. Das et al., arXiv:1009.0847, ApJ 729, 62 (2011)

\vspace{-0.3cm}
\item[3] J. Dunkley et al., ApJ accepted, arXiv:1009.0866 (2011)

\vspace{-0.3cm}
\item[4] D. Larson et al., arXiv:1105.4887, ApJS 192, 16 (2011)

\vspace{-0.3cm}
\item[5] R. Hlozek et al., ApJ submitted, arXiv:1105.4887 (2011)

\vspace{-0.3cm}
\item[6] R. Keisler et al., ApJ submitted, arXiv:1105.3182 (2011)

\vspace{-0.3cm}
\item[7] S. Das et al, arXiv:1103.2124, PRL 107, 021301 (2011)

\vspace{-0.3cm}
\item[8] B. Sherwin et al., arXiv:1105.0419, PRL 107, 021302 (2011)

\vspace{-0.3cm}
\item[9] T. Marriage et al., arXiv:1010.1065, ApJ 737, 61 (2011)

\vspace{-0.3cm}
\item[10] F. Menanteau et al., arXiv:1006.5126, ApJ 723, 1523 (2010)

\vspace{-0.3cm}

\item[11] N. Sehgal et al., arXiv:1010.1025, ApJ 732, 44 (2011)

\vspace{-0.3cm}
\item[12] N. Hand et al., arXiv:1101.1951, ApJ 736, 39 (2011)

\vspace{-0.3cm}

\end{description}

\newpage

\subsection{Gerard Gilmore}

\vskip -0.3cm

\begin{center}

G. Gilmore, Institute of Astronomy, Cambridge, UK

\bigskip

{\bf 
Observational Imprints of Dark Matter on Small Astrophysical Scales}

\end{center}

\medskip

If elementary particles make up (most of) Dark Matter, the properties
of the (probably several) different particles will leave an imprint on
the power spectrum at small scales. These imprints will in turn affect the 
smallest gravitational potential wells, and will in turn affect the
observable properties of the galaxies which today occupy the smallest
potential wells. On similar scales, astrophysical effects - such as
the minimum potential in which metal-free primordial gas can readily
cool to form stars- are known to be important. Distinguishing between
astrophysics and particle physics in the histories of small systems
today is thus both challenging and important. The smallest galaxies,
those with masses in the range where particle properties will be
observable if they exist, are the dSph galaxies. These low-luminosity
systems are studied in detail only in the Local Group, with extant facilities.

\bigskip

Substantial efforts are being made to quantify both the statistical
(number counts, correlations, spatial distributions..) and the internal
properties (star formation histories, dark matter density profiles,
...) of these low-mass systems. 

\bigskip

Recent discovery surveys have accentuated the {\bf Satellite 
Problem}. Although some 25 dSph are now known assciated with each of
the Milky way and M31, their numbers remain orders of magnitude too
low compared to simple LCDM predictions. Further, the spatial
distributions of these satellites are concentrated in sheets/groups,
more so than is anticipated. Early hints that the dSph have a radically
different {\bf Half-Light Size Distribuion} than do baryonic (star
cluster) systems, and are - when sufficiently far from a deep tidal
field -- all larger than about 100-pc [1], are still
apparent. Does this reflect an underlying scale in the dark matter?
Why are such low-luminosity galaxies so large?

\bigskip

A key requirement is to quantify the role of {\bf Baryonic Feedback}
on galaxy formation. Dominant feedback is required in LCDM to solve
the Satellite Problem, and to reproduce something like the observed
galaxy luminosity function. Feedback is not a free parameter, to be
sub-grid fixed. It is a consequence of the star formation rate, which
is determinable from the {\bf chemical abundance distribution function
of the earliest stars}. Substantial progress is being made in
quantifying the early histories of the local dSph. In all cases, very
low star formation rates are required by observation. This implies
{\bf very low baryonic feedback.} It also implies that the field stars
in the Milky Way {\bf were not formed in now tidally destroyed} dSph.
Appropriate high-resolution simulation efforts [2.3] conclude
that {\bf baryon feedback does not affect DM structure} and that CDM
still cannot produce realistic galaxies with realistic feedback and
star formation recipes.

\bigskip

{\bf Direct kinematic probing of dSph density profiles} is making
significant progress. Substantially improved techniques have been
developed and implemented to derive very precise kinematics for faint
stars in dSph [4]. Sophisticated distribution-function methodes for
deriving potentials from projected kinematics continue to be
developed, albeit slowly. Serious effort is being invested to find
ways to test the effect on the host galaxy of the tens of thousands of
star-less CDM halos predicted. Most interestingly, several authors
have recently confirmed that the total mass enclosed within a
half-light radius is a robust parameter. Clever new work [5] is
using the existence of multiple populations inside a single dSph
galaxy to determine the mass enclosed with the half-light radius of
each population, thus providing several integrated mass determinations
inside a single profile. 

\bigskip

This new kinematic and chemical abundance work has considerable potential to
provide direct determinations of what are apparently primordial
DM-dominated density profiles.

\bigskip

{\bf References}

\begin{description}

\item[1] Gilmore etal 2007  ApJ...663..948

\vspace{-0.3cm}

\item[2] Sawala etal 2011 MNRAS.413..659

\vspace{-0.3cm}

\item[3] Parry etal 2011 arXiv 1105.3474

\vspace{-0.3cm}

\item[4] Koposov, Gilmore etal 2011 arXiv 1105.4102

\vspace{-0.3cm}

\item[4] Walker, M.G. \& Penarrubia, J. arXiv:1108.2404 
The Astrophysical Journal, Volume 742, Issue 1, article id. 20 (2011)

\end{description}

\newpage

\subsection{A. Kashlinsky, F. Atrio-Barandela, H. Ebeling, D. Kocevski and A. Edge}

\vskip -0.3cm

\begin{center}
A.K.: SSAI and Observational Cosmology Laboratory, Code
665, Goddard Space Flight Center, Greenbelt MD 20771 (alexander.kashlinsky@nasa.gov)\\
F. A-B.: Fisica Teorica, University of Salamanca, 37008 Salamanca, Spain\\
H.E.: Institute for Astronomy, University of Hawaii, 2680,  Woodlawn Drive, Honolulu, HI 96822\\
D.K.: University of California Observatories/Lick Observatory, University of California, Santa Cruz, CA 95064\\
A. E.: Department of Physics, University of Durham, South Road,
Durham DH1 3LE, UK\\
\bigskip

{\bf Large-scale peculiar flows of clusters of galaxies: measurements and implications}

\end{center}

\medskip

Peculiar velocities of cosmological objects provide an important test of the physics and 
initial conditions in the earliest moments of the Universe's evolution. In standard cosmological 
paradigm, large-scale peculiar velocities arise from gravitational instability due to mass 
inhomogeneities seeded during inflationary expansion. On sufficiently large scales, ${>\atop{\sim}} 100$ Mpc, 
this leads to a robust prediction of the amplitude and coherence length of these velocities independently of 
cosmological parameters or evolution of the Universe. A variety of methods have been employed over the years 
to measure peculiar velocities starting with several galaxy distance indicators. For clusters of galaxies, 
their peculiar velocities can be measured from the kinematic component of the Sunyaev-Zeldovich (KSZ) 
effect produced by the Compton scattering of the cosmic microwave background (CMB) photons off the hot 
intracluster gas. 

Here we present the results of the measurements of the large-scale peculiar velocities we have conducted 
over the past several years. The measurements utilize the method proposed by Kashlinsky \& 
Atrio-Barandela [1] which uses the cumulative KSZ effect from an all-sky catalog of X-ray clusters. 
The method identified a statistic (the dipole of the CMB temperature field evaluated over cluster 
catalog pixels) which can isolate a signal remaining from the KSZ effect produced by coherently moving 
clusters. The X-ray cluster catalog was assembled for this purpose from the ROSAT all-sky survey data 
and, as of now, contains about $\sim 1,500$ clusters with spectroscopically measured redshifts. It is 
then applied to all-sky CMB maps from the WMAP satellite from the 3-, 5-, and 7-yr total integrations. 
The results have been been presented in multiple refereed publications [2-7] and a comprehensive summary 
is now being prepared for Physics Reports [8]. 

Our results first established that the cluster hot gas distribution is, on average, well described by
 the NFW density profiles [2]. Because the hydrostatic equilibrium temperature of the NFW-like distributed 
gas decreases towards outer radii, the thermal SZ (TSZ) component from such clusters decreases with 
increasing aperture size. This empirically established property was then used by us in [3,4] for 3-yr 
WMAP data to identify a statistically significant dipole remaining at apertures containing zero CMB 
monopole, $\langle \delta T\rangle$. The signal was present, at high statistical significance, 
exclusively at the cluster pixels associated with the apertures containing hot gas (within which 
$\langle \delta T\rangle$ was still systematically negative) and its amplitude remained constant 
(within the statistical uncertainties) out to the largest depths probed. After assembling and using 
a much larger (X-ray luminosity limited) cluster catalog with 5-yr WMAP data, coupled with a revised 
statistical treatment accounting for correlations between primary CMB remaining in different WMAP 
channels, we have confirmed the signal and demonstrated that the amplitude of the dipole remaining at 
zero monopole aperture increases systematically with the cluster $L_X$-bin. This confirms that the 
signal must originate from the SZ component and is inconsistent with it arising from some putative 
systematics originating in primary CMB or instruments. 

In [6] we have developed an analytical formalism to understand the errors in our measurement - which has been further verified with numerical/empirical data analysis - and have shown that our filtering scheme removes primary CMB down to the fundamental limit  imposed by the cosmic variance; any alternative filtering scheme must satisfy the formalism developed there. Because the dipole is measured at zero monopole and its amplitude correlates well with the cluster $L_X$, we believe that KSZ is the only viable explanation of the detected signal and its properties; no other explanation has been proposed in the literature as of now. The larger cluster catalog has enabled binning by the cluster $L_X$ resulting in a more accurate measurement extending to significantly larger scales. The measured dipole implies a flow of about 600-1,000 km/sec remaining coherent out to at least $\sim 750$ Mpc. 

In [7] we have shown that the signal can be readily confirmed with the existing {\it public} cluster data in 5- and 7-yr WMAP data; to facilitate verification and enable better testing of the result we have made the data publicly available from \url{http://www.kashlinsky.info/bulkflows/data_public}. 

We further address consistency of these large scale measurements with peculiar velocities on smaller scales derived independently by different techniques. The results cast doubt that the gravitational instability from the observed mass distribution is the sole - or even dominant  - cause of the detected motions. Instead it appears that the flow extends across the observable Universe and may be indicative of either the globally different world structure (such as the primeval preinflationary structure of space-time and its landscape), or require modifications of known physics (e.g. arising from higher-dimensional structure of gravity). We review these possibilities in light of the measurements discussed here. 

In order to improve the results, still better eliminate any possible systematics and to measure the 
`dark flow' - and its properties - with higher accuracy and to still large scales, we have designed 
an experiment dubbed {\it SCOUT} ({\bf S}unyaev-Zeldovich {\bf C}luster {\bf O}bservations as probes of the {\bf U}niverse's {\bf T}ilt) which we are currently conducting. The {\it SCOUT} experiment will catalog up to $\sim2,000$ clusters of galaxies extending to $z\sim 0.7$ and double the depth of the measured flow and its accuracy, while improving the current calibration uncertainties, upon application to forthcoming CMB data releases.


\bigskip

{\bf References}

\begin{description}

\item[1] Kashlinsky, A. \& Atrio-Barandela, F.  2000, Astrophys. J., 536, L67

\vspace{-0.3cm}

\item [2] Atrio-Barandela, F.,
Kashlinsky, A., Kocevski, D. \& Ebeling, H.  2008, Ap.J.
(Letters), 675, L57

\vspace{-0.3cm}

\item[3] Kashlinsky, A., Atrio-Barandela, F., Kocevski, D. \& Ebeling, H. 2008, Ap.J., 686,
L49

\vspace{-0.3cm}

\item[4] Kashlinsky, A., Atrio-Barandela, F., Kocevski, D. \& Ebeling, H. 2009, Ap.J., 691,
1479

\vspace{-0.3cm}

\item[5] Kashlinsky, A., Atrio-Barandela, F., Ebeling, H., Edge, A. \& Kocevski, D.  2010, Ap.J., 712, L81

\vspace{-0.3cm}

\item[6] Atrio-Barandela, F., Kashlinsky, A., Ebeling, H. \& Kocevski, D. 2010, Ap.J., 719, 77

\vspace{-0.3cm}

\item[7] Kashlinsky, A., Atrio-Barandela, F. \& Ebeling, H. 2011, 731, 1

\vspace{-0.3cm}

\item[8] Kashlinsky, A. et al 2011, Physics Reports, in preparation

\end{description}

\newpage

\subsection{A. Kogut}

\vskip -0.3cm
\begin{center}

NASA Goddard Space Flight Center, Greenbelt, MD USA

\bigskip
{\bf Testing the Standard Model with the Primordial Inflation Explorer} 

\end{center}
\medskip


Inflation has a central place in modern cosmology.
The many $e$-foldings of the scale size during inflation
force the geometry of space-time to asymptotic flatness
while dilating quantum fluctuations in the inflaton potential
to the macroscopic scales responsible for seeding 
large-scale structure in the universe.
Inflation provides a simple, elegant solution
to multiple problems in cosmology,
but it relies on extrapolation of physics
to energies greatly exceeding direct experimentation
in particle accelerators.

Linear polarization of the cosmic microwave background (CMB)
provides a direct test of inflationary physics.
Gravity waves generated during inflation
later interact with the CMB to impart a characteristic
curl pattern (B-mode) in the linear polarization.
Detecting the innflationary signature
in polarization will be difficult.
As recognized in multiple reports [1--3],
there are three fundamental challenges:
sensitivity, 
foreground emission from within the Galaxy,
and rigorous control of systematic errors from instrumental effects.
Satisfying the simultaneous requirements of sensitivity, 
foreground discrimination, and immunity to systematic errors
presents a technological challenge. 

The Primordial Inflation Explorer (PIXIE)
is an Explorer-class mission to detect and characterize 
the polarization signal from an inflationary epoch
in the early Universe.
Figure 1 shows the instrument concept.
PIXIE combines multi-moded optics with a 
Fourier Transform Spectrometer (FTS)
to provide breakthrough sensitivity for CMB polarimetry
using only four semiconductor detectors.
The design addresses each of the principal challenges 
for CMB polarimetry.
A multi-moded ``light bucket'' provides nK sensitivity
using only four detectors. 
A polarizing Fourier Transform Spectrometer (FTS)
synthesizes 400 channels across 2.5 decades in frequency 
to provide unparalleled separation of CMB from Galactic foregrounds. 
PIXIE's highly symmetric design
enables operation as a nulling polarimeter to provide
the necessary control of instrumental effects.

\begin{figure}[b]
\centerline{
\includegraphics[height=4.0in]{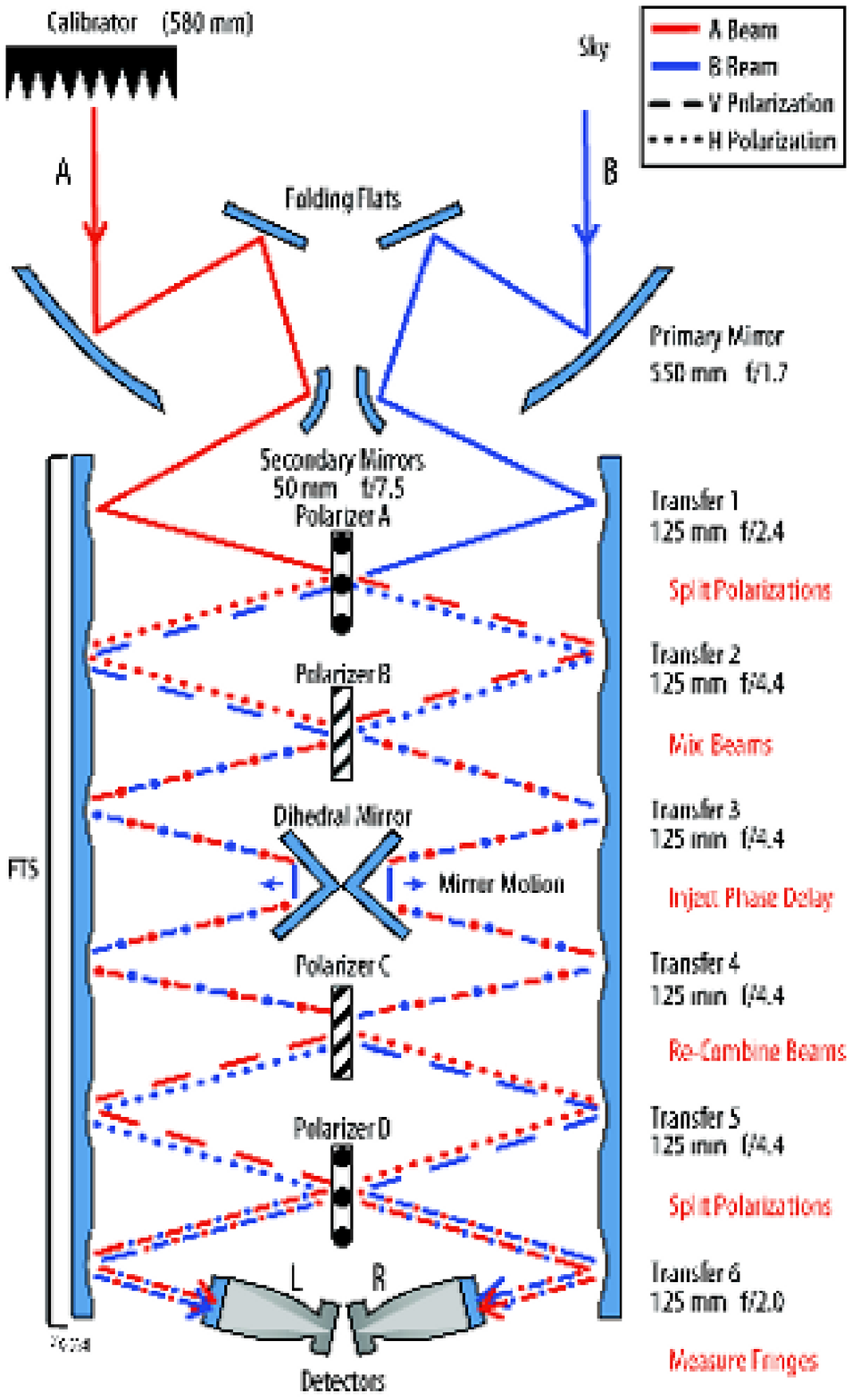}
\includegraphics[height=4.0in]{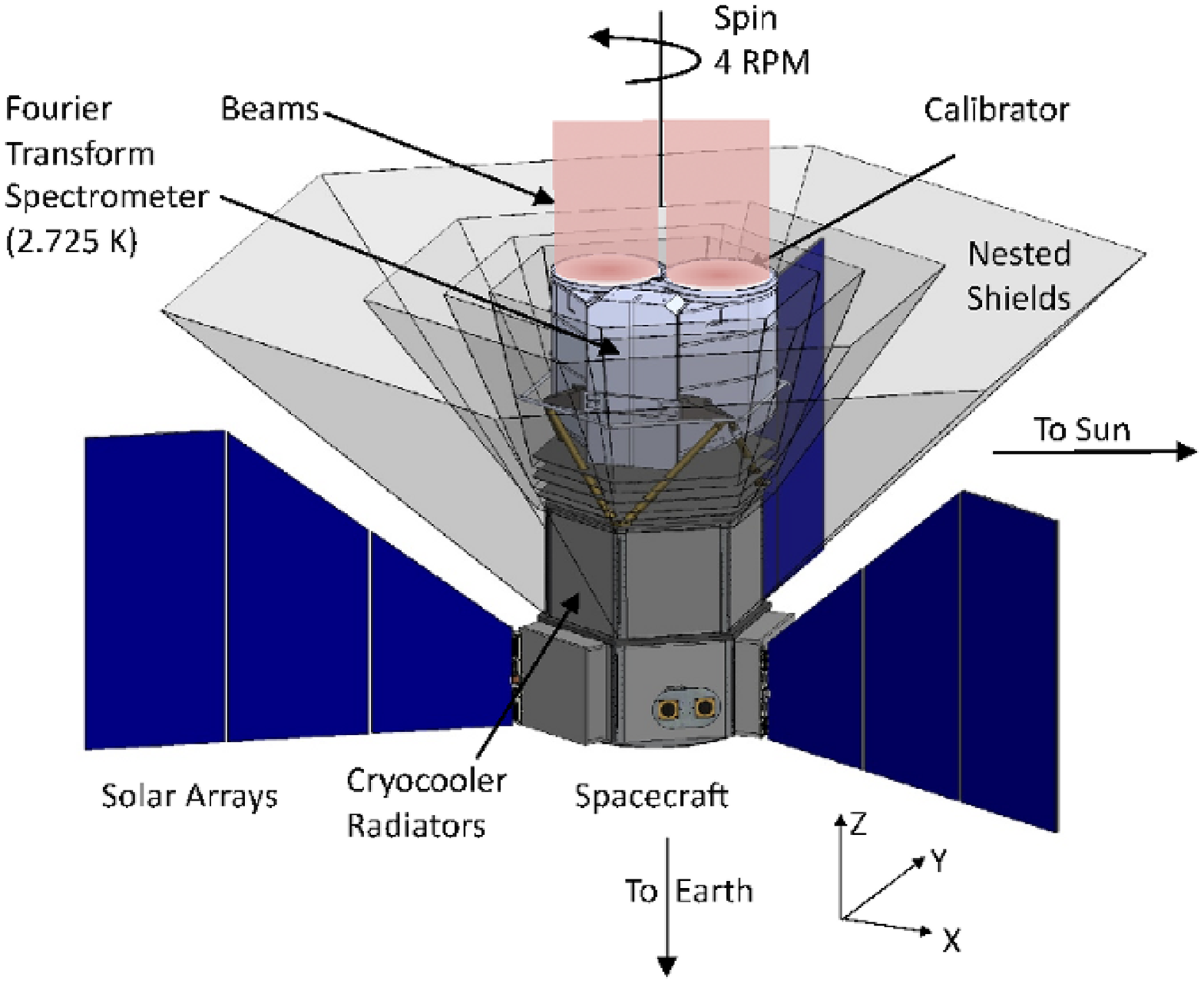}
}
\caption
{(Left) PIXIE optical signal path.
As the dihedral mirrors move,
the detectors measure a fringe pattern proportional to the
Fourier transform of the difference spectrum
between orthogonal polarization states from the two input beams
(Stokes Q in instrument coordinates).
A full-aperture blackbody calibrator
can move to block either input beam,
or be stowed to allow both beams to view the same patch of sky.
(Right) PIXIE observatory with calibrator stowed.  
PIXIE spins at 4 RPM and
will observe from a 660 km polar sun-synchronous orbit. 
}
\end{figure}

A dihedral mirror within the FTS translates $\pm$2.5mm
to provide $\pm$10 mm optical phase delay between the two input beams.
The power at the detectors
as a function of the dihedral position $z$
may be written
\begin{eqnarray}
P_{Lx} &=& \frac{1}{2} ~\int(E_{Ax}^2+E_{By}^2)+(E_{Ax}^2-E_{By}^2) \cos(4z\omega /c)d\omega    \nonumber \\
P_{Ly} &=& \frac{1}{2} ~\int(E_{Ay}^2+E_{Bx}^2)+(E_{Ay}^2-E_{Bx}^2) \cos(4z\omega /c)d\omega	\nonumber \\
P_{Rx} &=& \frac{1}{2} ~\int(E_{Ay}^2+E_{Bx}^2)+(E_{Bx}^2-E_{Ay}^2) \cos(4z\omega /c)d\omega    \nonumber \\
P_{Ry} &=& \frac{1}{2} ~\int(E_{Ax}^2+E_{By}^2)+(E_{By}^2-E_{Ax}^2) \cos(4z\omega /c)d\omega~,
\label{full_p_eq}
\end{eqnarray}
where
$\omega$ is the angular frequency of incident radiation,
$x$ and $y$ are orthogonal polarizations on the sky,
L and R refer to the detectors in the left and right concentrators,
and A and B refer to the two input beams.
PIXIE operates as a nulling polarimeter:
when both beams view the sky,
the instrument nulls all unpolarized emission
so that the FTS fringe pattern responds only to the sky polarization.
The resulting null operation greatly reduces
sensitivity to systematic errors
from unpolarized sources.
Normally the instrument collects light
from both co-aligned telescopes.
A full-aperture blackbody calibrator
can move to block either beam,
replacing the sky signal in that beam
with an absolute reference source near 2.7 K,
or be stowed to allow both beams to view the same sky patch.
When the calibrator blocks either beam,
the fringe pattern encodes information
on both the temperature distribution on the sky (Stokes I)
as well as the linear polarization.
Interleaving observations with and without the calibrator
allows straightforward transfer of the absolute calibration scale
to linear polarization,
while providing a valuable cross-check of the polarization solutions
obtained in each mode.

\begin{figure}[b]
\centerline{
\includegraphics[height=3.0in]{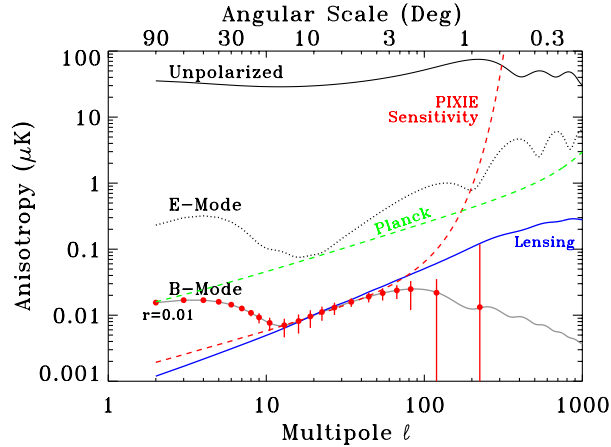}}
\caption{
Angular power spectra for 
unpolarized,
E-mode,
and B-mode polarization
in the cosmic microwave background.
The dashed red line shows the PIXIE sensitivity
to B-mode polarization 
at each multipole moment $\ell \sim 180\deg/\theta$.
The sensitivity estimate assumes a 4-year mission
and includes the effects of foreground subtraction
within the cleanest 75\% of the sky
combining PIXIE data at frequencies $\nu < 600$ GHz.
Red points and error bars show the response 
within broader $\ell$ bins
to a B-mode power spectrum with amplitude $r = 0.01$.
PIXIE will reach the confusion noise (blue curve)
from the gravitational lensing of the E-mode signal
by cosmic shear along each line of sight,
and has the sensitivity and angular response
to measure even the minimum predicted B-mode power spectrum
at high statistical confidence.
}
\end{figure}

PIXIE will map the full sky in both absolute intensity 
and linear polarization 
(Stokes $I$, $Q$, and $U$ parameters)
with angular resolution $2{{\rlap.}^\circ}$
~in each of 400 frequency channels
15 GHz wide
from 30 GHz to 6 THz.
Typical sensitivities within each 
$1^\circ \times 1^\circ$ pixel are
\begin{equation}
\delta I_\nu^{I} = 4 \times 10^{-24}~
{\rm W~m}^{-2}~{\rm s}^{-1}~{\rm sr}^{-1}
\nonumber
\end{equation}
for Stokes $I$
and
\begin{equation}
\delta I_\nu^{QU} = 6 \times 10^{-25}~
{\rm W~m}^{-2}~{\rm s}^{-1}~{\rm sr}^{-1}
\nonumber
\end{equation}
for Stokes $Q$ or $U$.
The resulting data set supports a broad range of science goals [4].

The primary science goal is the characterization
of primordial gravity waves from an inflationary epoch
through measurement of the CMB B-mode power spectrum.
PIXIE will measure the CMB linear polarization
to sensitivity of 70 nK per $1\deg \times 1\deg$ pixel,
including the penalty for foreground subtraction.
Averaged over the cleanest 75\% of the sky,
PIXIE can detect B-mode polarization
to 3 nK sensitivity,
well below the 30 nK predicted from large-field inflation models.
The sensitivity is comparable to the ``noise floor'' 
imposed by gravitational lensing,
and allows robust detection of primordial gravity waves
to limit $r < 10^{-3}$ at more than 5 standard deviations
(Figure 2).

PIXIE provides a critical test for light dark-matter candidates.
Neutralinos are an attractive candidate for dark matter;
the annihilation of $\chi \bar{\chi}$ pairs
in the early universe 
releases energy to the CMB
and distorts the spectrum away from the blackbody shape.
The resulting chemical potential 
can be estimated as
\begin{equation}
\mu \sim 3 \times 10^{-4} 
~f
~\left( \frac{\sigma v}{6 \times 10^{-26}~{\rm cm}^3~{\rm s}^{-1}} \right)
~\left( \frac{m_\chi}{1 ~{\rm MeV}} \right)^{-1}
~\left( {\Omega_{\chi}}h^2 \right)^2 .
\label{wimp_mu_eq}
\end{equation}
where
$f$ is the fraction of the total mass energy
released to charged particles,
$\langle \sigma v \rangle$ 
is the velocity-averaged annihilation cross section,
$\Omega_{\chi}$ is the dark matter density,
and
$h = H_0 / 100~{\rm km~s}^{-1}~{\rm Mpc}^{-1}$
is the Hubble constant [5,6].
The dark matter annihilation rate varies as 
the square of the number density.
For a fixed $\Omega_{\chi}$
the number density is inversely proportional 
to the particle mass.
The chemical potential distortion
is thus primarily sensitive to lower-mass particles.
PIXIE will probe neutralino mass range $m_\chi < 80$ keV
to provide a definitive test
for light dark matter models [7].

\begin{figure}[b]
\centerline{
\includegraphics[height=3.0in]{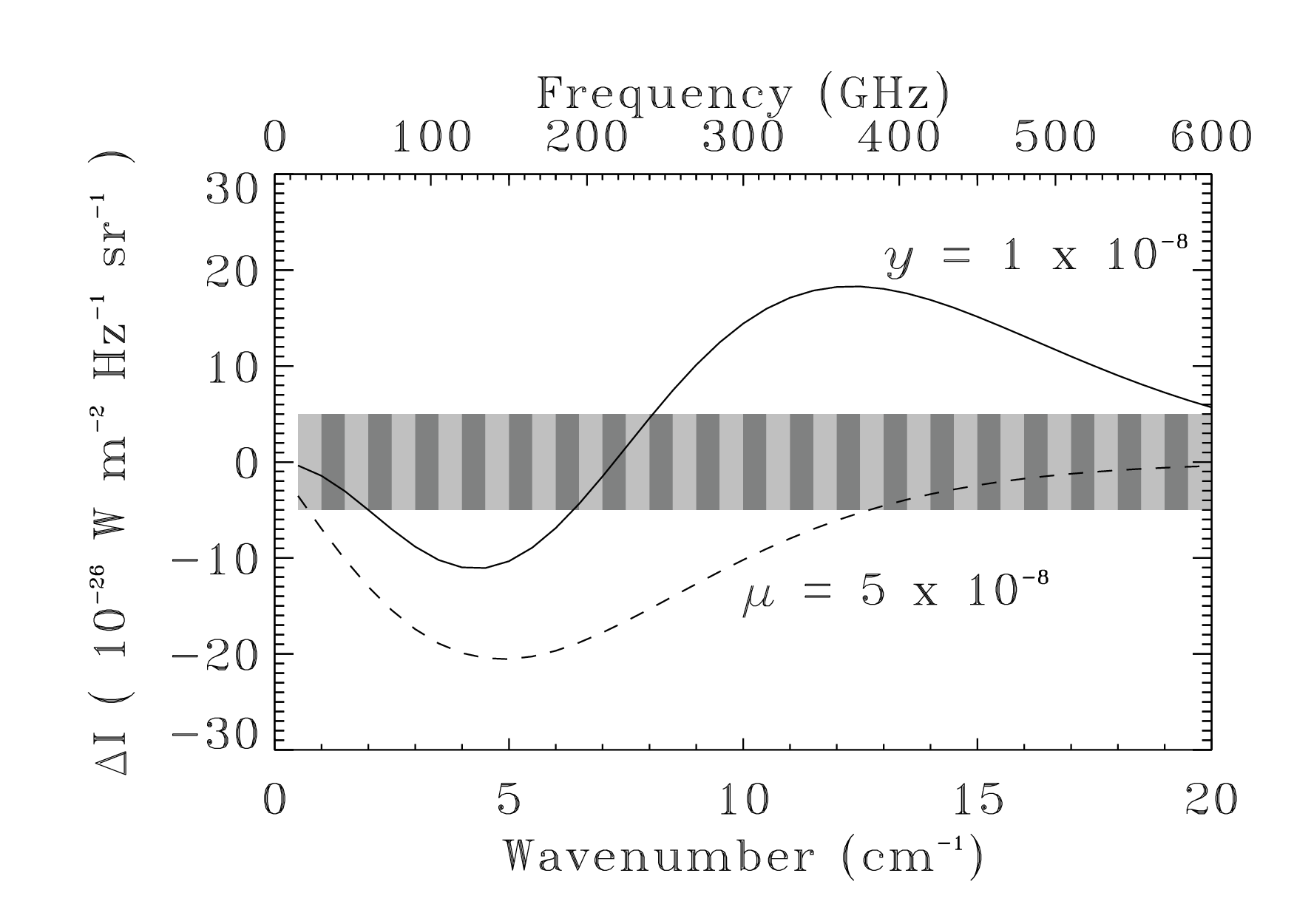}}
\caption{
Distortions to the CMB blackbody spectrum
compared to the PIXIE instrument noise in each synthesized frequency channel.
The curves show 5$\sigma$ detections
of Compton ($y$) 
and chemical potential ($\mu$)
distortions.
PIXIE measurements of the $y$ distortion
determine the temperature of the intergalactic medium
at reionization,
while the $\mu$ distortion
probes early energy release
from dark matter annihilation
or Silk damping of primordial density perturbations.
}
\end{figure}


\bigskip
{\bf References}

\begin{description}

\item[1] J.~Bock, et al.,
astro-ph/0604101,
Task Force on Cosmic Microwave Background Research (2006)

\vspace{-0.3cm}

\item[2] J.~Dunkley, et al.,
arXiv:0811.3915,
AIP Conference Proceedings {\bf 1141} 222 (2009) 

\vspace{-0.3cm}

\item[3] S.~Dodelson, et al.,
arXiv:0902.3796,
Astro2010: The Astronomy and Astrophysics Decadal Survey White Paper 67 (2009) 

\vspace{-0.3cm}

\item[4] A.~Kogut, et al.,
arXiv:1105.2044,
Journal of Cosmology and Astroparticle Physics, in press (2011)

\vspace{-0.3cm}

\item[5] J.~Silk and A.~Stebbins,
ApJ, {\bf 269}, 1 (1983)

\vspace{-0.3cm}

\item[6]P.~McDonald, et al.,
Physical Review D {\bf 63}, 023001 (2001) 

\vspace{-0.3cm}

\item[7]
H.~J.~de Vega and N.~G.~Sanchez,
Mon. Not. R. Astron. Soc. {\bf 404}, 885 (2010)

\end{description}

\newpage

\subsection{Anthony Lasenby}

\vskip -0.3cm

\begin{center}

Astrophysics Group, Cavendish Laboratory, J.J. Thomson Avenue,
Cambridge CB3 0HE, U.K.\\ and Kavli Institute for Cosmology, c/o Institute of Astronomy, Madingley Road, Cambridge, CB3 0HA, U.K.

\bigskip

{\bf CMB Observations: Current Status and Implications for Theory}

\end{center}

\medskip

The aim of this talk was to give an overview of the current state of CMB observations and their scientific 
implications. This last year has been exciting for this area, mainly as regards CMB observations on small 
angular scales, i.e.\ at the high-$\ell$ part of the CMB power spectrum. Also the Sunyaev-Zeldovich story 
continues to be very interesting, and the first Planck results have now become available on this.

\medskip
 
About 90\% of the photons in the CMB reach us directly from decoupling, telling us about the physics at 
the time of recombination, whilst the rest carry imprints of what happened on the way. Also, when emitted 
at recombination, the CMB has encoded within it information dating from about $10^{-36}$ seconds after 
the big bang.
Huge advances in technology over the past few years, are enabling us to measure all three of these 
aspects with rapidly increasing precision, with the key modern frontiers being {\em polarization} 
and {\em the high resolution temperature power spectrum}.

\medskip
 
On the polarization front, the main new results during the past year have come from the QUIET 
experiment, as reported in [1]. Unlike most other current experiments this uses 
{\em coherent} rather than bolometric techniques
The feeds look at a 1.4 m primary, and the whole is mounted on the CBI mount in Chile. The results 
given in [1] are able to confirm the first peak in the $EE$ spectrum first 
seen in the BICEP experiment [2]. The current direct limit on the primordial 
tensor to scalar ratio from QUIET is $r<0.9$ (95\%), which is larger than that of $r<0.73$ given 
in [2], but they stress that their systematic error in $r$ determination 
($<0.1$) is the smallest yet, thus promising well for the future measurements (with more data 
and a further frequency) from QUIET.

\medskip
 
The successor to BICEP, BICEP2 was deployed to the South Pole in November 2009. This has 512 
detectors at $150 {\rm \, GHz}$ and 8 times the mapping speed of BICEP1 has been achieved, and 
with similar angular scales and $\ell$-range coverage. The latest step is that the {\em KECK} 
array is being deployed, which has effectively $5\times$ BICEP2 cryostats [3].
Results from both this and BICEP2 are eagerly awaited.

\medskip
 
Results on $r$ and the slope of the power law spectrum of primordial perturbations, $n_s$, are crucial
 discriminators of the dynamics of inflationary models. Constraints from WMAP7 [4]
in the $(r,n_s)$ plane have already effectively ruled out a purely quartic inflation potential, 
$V(\phi)\propto \phi^4$, and even $V(\phi)\propto \phi^2$ could be considered to be under some pressure. 
In these circumstances it is interesting to note that potentials of the form $V(\phi)\propto\phi^p$ 
with $p\leq 1$ are now starting to be of interest in string cosmology-inspired inflation scenarios. 
For example, [5] discusses three different types of {\em monodromy}-type theory 
(this is where a {\em shift symmetry} in the theory is able to suppress higher-order corrections to the 
potential) within string theory, each leading to a $p\leq 1$. A further way in which the looming 
constraints on monomial potentials can be avoided is via {\em mixtures} of terms, such as would occur 
if using the spontaneous symmetry breaking-type potential, of the form $(\phi^2-v^2)^2$, 
for $v$ a constant. 
Terms of this form are thought to arise within a supergravity-type approach to inflation, and their 
observational consequences are discussed in [6].

\medskip
 
Returning to experiments, the prospects for the balloon-borne Spider experiment are discussed in [7].
This experiment has now returned to the idea of medium duration flights around Antarctica, rather 
than ultra-long flights along a line of latitude through Australia. The first full flight is scheduled 
for December 2012, and will provide a sky coverage fraction of $f_{\rm sky} \sim 0.1$ and probe the 
multipole range $10 \leq \ell \leq 300$ with the greatest sensitivity coming at $150 {\rm \, GHz}$. 
The maps from Spider could be of great interest as providing $B$-mode observations which can be 
combined with those from the Planck Satellite. The latter experiment is still on course to provide 
first cosmological information in January 2013, with the first release of primordial CMB polarization 
data likely in January 2014.

\medskip
 
A major highlight over the past year has been the release of high-$\ell$ power spectrum results from both 
the Atacama Cosmology Telescope (ACT), and the South Pole Telescope (SPT). 
The talk by Jo Dunkley described the ACT results, and we concentrate here on those from the SPT, 
as described in the impressive recent paper 
from Keisler {\it et al.} [8]. This described results from 790 square degrees of 
sky measured at $150 {\rm \, GHz}$, and displays a CMB power spectrum in which 9 peaks can now be clearly 
discerned. The high-$\ell$ part of the CMB spectrum now gives a lever arm in which (as for results for the 
ACT experiment) for the first time various degeneracies in cosmic parameters 
can be resolved using CMB-alone 
data. Additionally, various parameters sensitive to the high-$\ell$ part of the spectrum, such as (a) the 
tilt of the primordial spectrum, $n_s$; (b) a possible running in the primordial spectrum, 
parameterised by 
$n_{\rm run}$; (c) the primordial Helium abundance $Y_p$; and (d) the effective number of neutrino species 
at decoupling, $N_{\rm eff}$, are all starting to be constrained by the data. 
The basic observational result [8],
seems to be that there is a preference for models with slightly more damping at smaller scales than would 
be predicted with currently favoured values of these parameters, although with a strong degeneracy 
affecting what it is possible to say individually about each of $ n_{\rm run}, \; Y_p $ and $ N_{\rm eff} $ 
alone.

\medskip
 
A further highlight of the year has been the first release of Planck results for the 
Sunyaev-Zeldovich (SZ) 
effect in clusters of galaxies. The resolution of Planck for SZ studies (at best $\sim 5'$) is lower than 
for most ground-based observations, but is  compensated for by all-sky coverage, plus a good frequency 
discrimination of the effect due to multiple frequency channels spanning the whole range over which the 
effect is important. The `Early Release SZ Catalogue' (ESZ) [9] contained a sample 
of 189 clusters of which 169 were previously known. Amongst the new clusters was a candidate which 
turned out, following XMM-Newton confirmation, to be the first {\em super}cluster to be detected via the 
blank-field SZ effect. Other telescopes have now confirmed some of the other candidates, including the 
observations by the Arcminute MicroKelvin Imager in Cambridge, which was able to provide confirmation 
within the ESZ paper for a previously unknown cluster, and has done the same for another subsequently
[10], as well as providing a refined position estimate.

\medskip
 
The story on the SZ `deficit' discussed in last year's summary, which drew attention to the factor of 
between 0.5 and 0.7 by which the SZ decrements measured by WMAP7 were below what was expected for the 
given clusters using standard X-ray models for temperature and profile, [4],
has taken an interesting turn in that for the much larger sample surveyed by Planck, no such effect has 
been found. 
This is discussed in [11], where additionally to emphasising that no deficit has 
been found, it is pointed out that even when the analysis is redone in two bins, with pressure profiles 
corresponding to `cool cores' or `morphologically disturbed' (respectively) then the results are still 
robust to this, with maximum deviations at only the few percent level. Since this division was thought 
to be important in explanation of the WMAP7 results, [4], it is unclear currently 
how to explain the apparent discrepancy in findings, and suggests that there is still quite a bit to 
learn in this area.

\medskip

Finally, returning to early universe cosmology, the Lasenby \& Doran model [12]
is still doing surprisingly well in comparison to observations. This model has a slightly closed universe, 
in which constraints on both the initial and final development of the universe have to be obeyed, 
and which links the two via an overall constraint on the elapse of total conformal time. 
This link is able to provide 
a numerical estimate of the value of the current cosmological constant in terms of an exponential of the 
number of e-folds of inflation, thus obtaining a very small number ($10^{-122}$) in natural units, from 
one not so far from unity ($\sim 50$). In [13], the model is compared with other 
current models, such as $\Lambda$CDM with a power law initial spectrum and the addition of either spatial 
curvature or a running spectral index, and shown to outperform these in terms of Bayesian evidence, 
probably through having both a naturally occurring dip in the predicted primordial spectrum at large 
scales, and also fitting the high-$\ell$ CMB spectrum better.

\bigskip

{\bf References}

\begin{description}

\item[1] QUIET Collaboration, C. {Bischoff} \etal\
\newblock {First Season QUIET Observations: Measurements of CMB Polarization
  Power Spectra at 43 GHz in the Multipole Range 25 $\leq \ell \leq$ 475}.
\newblock {\em ArXiv e-prints}, December 2010.
\newblock arXiv:1012.3191.

\item[2] 
H. C. {Chiang} \etal\
\newblock {Measurement of Cosmic Microwave Background Polarization Power
  Spectra from Two Years of BICEP Data}.
\newblock {\em \apj}, 711:1123--1140, March 2010.

\item[3] 
C.D. {Sheehy} \etal\
\newblock {The Keck Array: a pulse tube cooled CMB polarimeter}.
\newblock {\em ArXiv e-prints}, April 2011.
\newblock arXiv:1104.5516.

\item[4] 
E.{Komatsu} \etal\
\newblock {Seven-year Wilkinson Microwave Anisotropy Probe (WMAP) Observations:
  Cosmological Interpretation}.
\newblock {\em \apjs}, 192:18--+, February 2011.

\item[5] 
X.{Dong}, B.{Horn}, E.{Silverstein}, and A.{Westphal}.
\newblock {Simple exercises to flatten your potential}.
\newblock {\em \prd}, 84(2):026011--+, July 2011.

\item[6] 
A.{Linde}, M.{Noorbala}, and A.{Westphal}.
\newblock {Observational consequences of chaotic inflation with nonminimal
  coupling to gravity}.
\newblock {\em jcap}, 3:13--+, March 2011.

\item[7] 
A.A. {Fraisse} \etal\
\newblock {SPIDER: Probing the Early Universe with a Suborbital Polarimeter}.
\newblock {\em ArXiv e-prints}, June 2011.
\newblock arXiv:1106.3087.

\item[8] 
R.{Keisler} \etal\
\newblock {A Measurement of the Damping Tail of the Cosmic Microwave Background
  Power Spectrum with the South Pole Telescope}.
\newblock {\em ArXiv e-prints}, May 2011.
\newblock arXiv:1105.3182.

\item[9] 
{Planck Collaboration} \etal\
\newblock {Planck Early Results VIII: The all-sky Early Sunyaev-Zeldovich
  cluster sample}.
\newblock {\em ArXiv e-prints}, January 2011.
\newblock arXiv:1101.2024.

\item[10] 
{AMI Consortium}, N.{Hurley-Walker} \etal\
\newblock {Further Sunyaev-Zel'dovich observations of two Planck ERCSC clusters
  with the Arcminute Microkelvin Imager}.
\newblock {\em \mnras}, 414:L75--L79, June 2011.

\item[11] 
{Planck Collaboration} \etal\
\newblock {Planck early results: Statistical analysis of Sunyaev-Zeldovich
  scaling relations for X-ray galaxy clusters}.
\newblock {\em ArXiv e-prints}, January 2011.
\newblock arXiv:1101.2043.

\item[12] 
A.{Lasenby} and C.{Doran}.
\newblock {Closed universes, de Sitter space, and inflation}.
\newblock {\em \prd}, 71(6):063502--+, March 2005.

\item[13] 
J.A. {Vazquez}, A.N. {Lasenby}, M.{Bridges}, and M.P. {Hobson}.
\newblock {A Bayesian study of the primordial power spectrum from a novel
  closed universe model}.
\newblock {\em ArXiv e-prints}, March 2011.
\newblock arXiv:1103.4619.
\end{description}
\newpage

\subsection{John C. Mather}

\vskip -0.3cm

\begin{center}

NASA Goddard Space Flight Center, Greenbelt, MD 20771 USA

\bigskip

{\bf James Webb Space Telescope and the Origins of Everything - Progress and Promise} 

\end{center}

\medskip

{\bf Abstract.}
James E. Webb built the Apollo program and led NASA to a successful moon landing.  We honor his leadership with the most powerful space telescope ever designed, capable of observing the early universe within a few hundred million years of the Big Bang, revealing the formation of galaxies, stars, and planets, and showing the evolution of solar systems like ours. Under study since 1995, it is a project led by the United States National Aeronautics and Space Administration (NASA), with major contributions from the European and Canadian Space Agencies (ESA and CSA). It will have a 6.6 m diameter aperture (corner to corner), will be passively cooled to below 50 K, and will carry four scientific instruments: a Near-IR Camera (NIRCam), a Near-IR Spectrograph (NIRSpec), a near-IR Tunable Filter Imager (TFI), and a Mid-IR Instrument (MIRI). It is planned for launch in 2018 on an Ariane 5 rocket to a deep space orbit around the Sun - Earth Lagrange point L$_2$, about $1.5 \times 10^6$ km from Earth. The spacecraft will carry enough fuel for a 10 yr mission.

\medskip
{\bf International Partnership.}
Goddard Space Flight Center leads the NASA team, and is supported by a  prime contract to Northrop Grumman Aerospace Systems and their subcontractors, including ATK, Ball Aerospace, and ITT. The NIRCam comes from the University of Arizona with Lockheed Martin.  The NIRSpec comes from ESA with Astrium, and its microshutter array is provided by Goddard Space Flight Center. The Fine Guidance Sensor and the Tunable Filter Imager come from CSA with Comdev. All of the near IR detectors come from Teledyne. The mid IR instrument is built by a European consortium led by the UK ATC, in partnership with Jet Propulsion Laboratory, and its detectors come from Raytheon. Europe is also providing the Ariane 5 rocket.

\medskip
{\bf Scientific Objectives.}
A project summary has been published by Gardner et al. [1]. Additional documents about JWST are available here: http://www.jwst.nasa.gov/ and here: http://www.stsci.edu/jwst/doc-archive. A recent meeting ``Frontier Science Opportunities with the James Webb Space Telescope'' was held at the Space Telescope Science Institute and the webcast archive is available online: http://www.stsci.edu/institute/conference/jwst2011.
Four key topics were used to guide the design of the observatory:

{\bf The end of the dark ages: first light and reionization.} This theme requires the largest feasible infrared telescope, since the first objects of the universe are faint, rare, and highly redshifted. It requires a wide wavelength range, to distinguish  high-redshift objects from cool local objects, and to estimate the ages and photometric redshifts of the stellar contents from colors.  It also requires powerful multi-object infrared spectroscopy, to determine the physical conditions and redshifts of the earliest objects.

{\bf The assembly of galaxies.} It is now thought that galaxies form around dark matter concentrations, 
as products of extensive merger trees, but it is difficult to tell whether simulations match observations.  
The interaction between dark matter, ordinary matter, black holes (when and how did they form?), stars, 
winds, and magnetic fields is extraordinarily complex ``gastrophysics''.  It is currently impossible to 
simulate the full dynamic range from stars and planetary systems to magnetized jets, outflows, dust 
formation, etc., so the formation and assembly of galaxies is still an observational science. With its 
infrared imaging and spectroscopy, JWST will reveal details of galactic mergers and show changes of 
properties with distance (time).

The JWST will also address some aspects of dark energy and dark matter. It can extend the Hubble measurements of distant supernovae, achieving greater precision by using rest-frame IR photometry, where the candles are more standard and the dust obscuration is less than at visible wavelengths.  It can also improve the calibration of the Hubble constant, by extending the range of each step of the distance ladder and  eliminating some steps.  It can also extend maps of the dark matter distribution to higher redshift, because it can observe many more faint and higher redshift background galaxies than current observatories.

{\bf The birth of stars and protoplanetary systems.}  Local star and planet formation is uniquely observable in the infrared, because of the low temperatures of young objects, and their typical location within obscuring dust clouds. Infrared observations complement radio observations very well.

{\bf Planetary systems and the origins of life.} Traces of the formation of the Solar System are found everywhere from the bright planets to the faint comets, asteroids, and dwarf planets of the outer solar system. JWST's image quality, field of view, and ability to track moving targets are essential to detect and analyze many such targets,  In addition, transit spectroscopy and direct imaging of extra-solar planetary systems are feasible, especially now that the Kepler observatory has cataloged 1235 candidate transiting planets.  With the discovery of a favorable nearby target (small star, large Earth), it may be possible to measure the atmospheric composition of an Earth-like planet and to detect the presence of liquid water.

\medskip
{\bf Observatory Design.} The observatory is composed of a deployable telescope, a scientific instrument package, and a spacecraft bus, separated by a deployable sun shield to enable the telescope and instruments to cool to $<$ 50 K.

{\bf Telescope.} The telescope uses a three-mirror anastigmat design to provide a wide field of view with diffraction-limited imaging at 2 $\mu$m.  The primary is nearly parabolic with an aperture of 6.6 m, and is composed of 18 hexagonal segments.  These segments are made of beryllium, machined to a few mm thickness with stiffening ribs, and polished at room temperature to a shape that will be nearly ideal (20 nm rms error) at the operating temperature.  All the segments have been polished and coated with gold.  The primary mirror segments are deployed after launch and adjusted to the right position and curvature by actuators having step sizes of a few nm. The secondary mirror is also deployed after launch. The final focus is determined using wavefront sensing algorithms, based on star images taken in and out of focus. These algorithms were derived from those used for the Hubble Space Telescope repair. The predicted image quality (point spread function) is available online at http://www.stsci.edu/jwst/software/webbpsf. A small movable flat mirror located at an image of the primary mirror (a fine steering mirror) is used to null the image motion as sensed by the fine guidance sensor in the instrument package; a signal from its control loop also goes to help maintain the pointing of the whole spacecraft,

{\bf Instrument Package.} All of the instruments are mounted to a carbon-fiber structure attached to the back of the telescope. The near IR camera covers 0.6 to 5 $\mu$m in two bands observed simultaneously, with 0.034 and 0.068 arcsec pixels respectively.  The camera  includes the wavefront sensing equipment, and a coronagraph. The near IR spectrometer provides low (R 100) and medium (R 1000 and 3000) spectroscopy.  It can observe 100 simultaneous targets with a microshutter array selector, and also provides fixed slits and an image slicing integral field configuration. The fine guidance sensor is a two-field camera that provides error signals to the pointing control system. The tunable filter imager provides high contrast imaging and spectroscopy in the near IR band.  All of the near IR instruments use HgCdTe detectors. The mid IR instrument provides imaging, coronography, and integral field spectroscopy from 5 to 28 $\mu$m, using Si:As detectors.  The exposure time calculator is available on line at http://jwstetc.stsci.edu/etc. With long exposures, nJy sensitivities are possible (AB magnitude $>$ 31.4).

{\bf Spacecraft.} 
The spacecraft bus includes the command and telemetry system, a deployable telemetry antenna, a pointing control system, a solar power system, and a deployable 5-layer sunshield the size of a singles tennis court. Behind the shield, the instrument package and the telescope radiate heat to the dark sky and reach temperatures of 40-50 K, cold enough for operating all the near IR ($<5$ $ \mu$m) detectors.  A helium compressor provides active cooling for the mid IR instrument to operate below 7 K, without stored cryogens. Pointing control is provided by reaction wheels and thrusters controlled by sun sensors, gyros (no moving parts!) and star trackers.

{\bf Orbit and Orientation.}
The Ariane 5 launch vehicle will ascend from Kourou, French Guiana, to place the observatory in orbit around the Sun-Earth Lagrange point L$_2$, approximately 1.5$\times10^6$ km from Earth. Bi-propellant thrusters will adjust the orbit to stay within several hundred thousand km of the L$_2$ point (which is near the end of the Earth's umbra), and to avoid shadows from the Moon or the Earth.  The orbit is unstable and small adjustments will be required every few weeks.  Fuel is also required to maintain the orientation of the observatory; small torque unbalances from the solar radiation pressure build up and must be compensated with thruster firings.  The fuel tanks are sized to last for 10 years of operation after initial observatory checkout.  

{\bf Operations.}
The scientific operation of the JWST will be very similar to that of the Hubble Space Telescope, based on observing proposals evaluated by time allocation committees.  Approximately 1/2 year of observations are allocated to the instrument teams and the interdisciplinary scientists. The European share of general observation time is to be 15\% and the Canadian share 5\% based on their contributions to the mission. It will take about 2 months to reach the L$_2$ orbit and cool down to operating temperature, and 4 more months are allocated for initial setup, focussing, characterization, and optimization of the observatory.

\medskip
{\bf Project Status.}
Two of the flight instruments (MIRI and NIRSpec) are completed and tested in Europe, and all four are  to be delivered to Goddard Space Flight Center within a year, for integration into the Integrated Science Instrument Module, and combined testing with a telescope simulator. Also, all the flight telescope mirrors (18 primary mirror hexagons, and the secondary, tertiary, and fine steering mirror) have been polished and coated with thin layers of  IR-reflecting gold. The near IR detectors (HgCdTe made by Teledyne) were degrading while they were kept warm, the cause has been determined, and new part designs are being made.

{\bf Replan.}
Following two major reports (TAT and ICRP) chaired by John Casani a year ago, the JWST management has been replaced at both Goddard Space Flight Center and NASA Headquarters, and the project has been replanned for a launch date in 2018.  The new plan provides additional funds for risk reduction and testing and raises JWST to one of NASA's top three priorities, along with commercial crew and the space launch system (SLS).  The NASA budget for 2012 is currently under discussion by the US Congress.

\bigskip

{\bf Reference}

\begin{description}

\item[1] J. Gardner et al., Space Science Reviews, 123 (\#4), 2006; arXiv:astro-ph/0606175

\end{description}

\newpage

\subsection{A. Yu. Smirnov}

\vskip -0.3cm

\begin{center}

International Centre for Theoretical Physics, Trieste, Italy

\bigskip

{\bf Status of the theory of neutrino mass and mixing} 

\end{center}

\medskip

Essentially, the theory of neutrino mass and mixing does not exist yet. 
We even do not know what is nature of neutrino mass:  Dirac or Majorana, 
soft  (enviroment-dependent) or hard; 
we do not know the absolute mass scale; number of neutrinos, {\it etc.}.     
There are direct and possibly indirect connections between  
neutrinos and Dark matter (DM  - the main subject of this school).  
Known active neutrinos compose hot component of the dark matter 
influencing structure formation in the Universe.   
New neutrino states (sterile neutrinos), if exist,  
can form, depending on mass, hot or warm component of DM. 
Concerning indirect connection,
the same flavor symmetries which explain the mixing pattern of light neutrinos 
can also ensure stability of DM particles. 
Mechanisms of neutrino mass generation 
imply existence of new particles, e.g. RH neutrinos,  
which can play the role of DM. Again, symmetry 
responcible for smallness of neutrino mass can 
lead to stability of the DM particles. 

\medskip


The most important observational features which play role in 
uncovering of the origin of neutrino mass include the following.  
(i) Cosmological and oscillation data show that at least one neutrino mass 
should be in the interval (0.04 - 0.30) eV. Smallness of neutrino mass 
can be characterized by ratio  
$m_\nu/m_\tau \sim 10^{-10}$ which can be connected to some other mass hierarchy. 
(ii)  Neutrinos have the weakest mass hierarchy (if any) which may be 
related to large lepton mixing.   
(iii)  The lepton mixing pattern is described aproximately by the Tri-Bimaximal (TBM) scheme [1] according to which $\sin^2 \theta_{23} = 1/2$, $\sin^2 \theta_{12} = 1/3$ and $\sin^2 \theta_{13} = 0$. 
(iv) Global fit of the oscillation data [2] shows substantial deviations from TBM:  $1/2 - \sin^2 \theta_{23} = 0.08$  (which is about $2\sigma$ off); $1/3 - \sin^2 \theta_{12} = 0.02$  ($2\sigma$ off) and $\sin^2 \theta_{13} = 0.020 - 0.025$   
($3\sigma$ off).  

\medskip

Understanding neutrino mass and mixing is on cross-roads. 
Smallness of neutrino mass can be due to existence of
  
-  New large mass scale,  which is realized in the seesaw mechanism.

-  Extra spatial dimensions;   
here the smallness is explained by the overlap mechanism and different localization of the left and right handed components of neutrinos. It is given by ratio of the width of the 3D brane and size of extra dimension or by large warp factor.  These two mechanisms are related to  properties 
of the RH neutrinos, they explain simultaneously 
suppression of the usual Dirac mass terms and finite mass. 
 
 -  New symmetries which forbid the neutrino Dirac mass term 
 (this also can be due to absence of the RH components). 
Here finite neutrino masses can be generated by radiative corrections   
or high dimensional operators. 


\medskip

There are three different approaches to explain the lepton mixing:
  
1. Tri-bimaximal mixing [2] as the first order approximation. 
If not accidental, TBM implies certain flavor (fundamental) symmetry [3]. 
The symmetry should be broken differently in the 
charged lepton and neutrino sectors. Different residual symmetries 
are responcible for the TBM mixing. 

2. Quark-lepton complementarity [4] is 
based  on observation that the lepton mixing equals 
approximately bi-maximal mixing 
({\it i.e.} maximal 1-2 and 2-3 mixings) minus quark mixing. 
This implies the quark-lepton unification 
or quark-lepton symmetry and existence of some new structures or interactions which 
generate the bi-miximal mixing. The latter can be the see-saw 
machanism and specifically certain pattern of the 
Majorana mass matrix of RH neutrinos. 

3. Quark-lepton  universality means that  
the lepton mixing is organized according to the same principle as 
quark mixing. Probably mixing angles are related to the mass hierachies.  Large lepton mixing is due to the smallness of neutrino 
mass and weak mass hierarchy of neutrinos. 
The same mechanism which explains the smallness 
of neutrino masses is responsible for enhancement of the lepton mixing.  
Again usual see-saw can be behind this picture. 

\medskip

Probably correct approach is  go as far as possible 
along the quark and lepton unification and reduce to minimum 
features/principles which distinguish quarks and leptons. Neutrality of neutrinos and therefore possibility to have  Majorana mass are the key points. Following this line one can chose  GUT based on $SO(10)$ with all the known fermions 
(plus RH neutrinos) in the same 16-plet. The  singlets of SO(10) 
may exist and mix with neutrinos only, and it is this mixing which produces difference between lepton and quark mixings and explains smallness of the neutrino mass. 

\medskip

TBM can be accidental without any 
fundamental symmetry and principle behind; e.g. it can be an interplay of different  independent  contributions [5]. Due to observed deviations of mixing from TBM the  symmetry relations between the elements of neutrino mass matrix can be broken maximally. No simple and convincing model for TBM has been proposed so far. Models have  complicated structure and  large number of assumptions 
 and free parameters.  
Often  no connection between  masses and mixing exists and 
additional symmetries are introduced to explain mass hierarchies. 
Inclusion of quarks leads to further complication.   
Grand unification imposes additional requirements. 

\medskip

The latest measurements of 1-3 mixing [6] further support this point of view, favouring the possibilities 2 and 3.  
Indeed, too small value of this mixing,  $\sin \theta_{13}  \sim  \sin^2 \theta_C$,  is typical prediction of the  flavor-symmetry models of TBM [3].   
On the other hand according to the  
Quark-lepton complementarity   
$\sin^2 \theta_{13}  \approx  2\sin^2 \theta_C$, which is close to the observed value [4].  Typical GUT prediction 
without special symmetry in the leptonic sector is 
$\sin^2 \theta_{13} \sim \Delta m^2_{21}/ \Delta m^2_{31}$ [7], again in a good agreement with data.

\medskip

For further progress in theory of neutrino mass and mixing it is 
important to settle down the issue of existence of sterile neutrinos. Variety of sterile neutrinos has been proposed with masses in different energy ranges from $10^{-3}$ eV to $10^2$ MeV. Some new hints in favour of the eV-scale (LSND) $\nu_S$ follow from MiniBooNE results, reactor anomaly  
and Gallium calibration experiments [8], [9].   
Furthermore, Cosmology (mainly CMB)  indicates existence 
of additional radiation in the Universe. Actually Cosmology 
``likes'' additional but light neutrinos [10]:   
for single  $\nu_S$  it gives the bound $\Delta m^2 < 0.25$ eV$^2$. 
Toghether with new MINOS bound on mixing [11] this essentially 
excludes the oscillation interpretation of the results from  
LSND and MiniBooNE.   

Mixing of the eV-scale neutrino with light neutrinos with mixing angles 
required by LSND is not a small perturbation. 
The corresponding contribution to the light neutrino mass matrix,  
$\delta m$, can be larger than the original elements $m_0$. 
$\delta m$ can change structure (symmetries) of the original 
mass matrix completely. It can produce the dominant $\mu - \tau$  block 
with small determinant, enhance the lepton mixing, generate TBM,  
be origin of difference of the quark and lepton mixings. 

\medskip

IceCube  can perform very sensitive search  for the LSND-type  sterile neutrinos. 
The $\nu_\mu - \nu_s$ oscillations with $\Delta m^2 \sim 1$ eV$^2$   are enhanced in matter of the Earth in the energy range $0.5 - 3 $ TeV [12]. This distorts the energy spectrum 
and zenith angle distribution of the 
atmospheric muon neutrinos. 
Effect  on the zenith angle distribution integrated over the neutrino energy is about $(10 - 20) \%$ [13]. Statistics is not the problem for IceCube: it has already about $10^5$ neutrino events and the main goal is to understand  systematics. 
The effect in IceCube depends not only on admixture of $\nu_\mu$ in the heavy, mainly sterile state, $U_{\mu 4}$, but also on admixture of $\nu_\tau$,  $U_{\tau 4}$.  
Part of the parameter-space formed by $m_S$,  $U_{\mu 4}$ and 
$U_{\tau 4}$ can be excluded with already existing data [13]. 

\medskip

Another interesting scenario is sterile neutrino with very small mass: 
$m_S \sim 10^{-3}$ eV [14]. Being mixed with $\nu_e$ this neutrino can explain the absence of expected upturn of the energy spectrum of solar neutrinos at low  energies. If it mixes with the heaviest mass eigenstate, $\nu_3$,  
the equilibrium concentration of $\nu_S$ can be generated in the Early Universe thus producing additional radiation without creating problem for Large scale structure of the Universe. 
Existence of such a  sterile neutrino can be tested in the DeepCore experiment studying the  
zenith angle and energy distributions of the atmospheric neutrinos [14].

\bigskip

{\bf References}

\begin{description}

\item[1] P. F. Harrison, D. H. Perkins and W. G. Scott, Phys. Lett. B 530 (2002). 

\vspace{-0.3cm}

\item[2] G. L. Fogli et al., arXiv:1106.6028 [hep-ph].

\vspace{-0.3cm}

\item[3] G. Altarelli and F. Feruglio, Rev. Mod. Phys. 82 (2010) 2701.  

\vspace{-0.3cm}

\item[4]  A. Yu. Smirnov, arXiv: hep-ph/0402264,   
     M. Raidal, Phys. Rev. Lett. 93 (2004) 16180,  
     H. Minakata and A. Yu. Smirnov, Phys. Rev. D 70 (2004) 073009. 

\vspace{-0.3cm}

\item[5]   M. Abbas and A. Yu. Smirnov Phys. Rev. D82 9 (2010) 013008.  

\vspace{-0.3cm}

\item[6] K. Abe, {\it et al.} [The T2K Collaboration]
arXiv:1106.2822 [hep-ex]. 

\vspace{-0.3cm}

\item[7] H. S. Goh,  R.N. Mohapatra, S.-P.  Ng,  Phys. Lett. B570 (2003) 215, 
    B. Bajc et al., Phys. Rev.  D73 (2006)  055001.

\vspace{-0.3cm}

\item[8] C. Giunti and M. Laveder, arXiv:1107.1452  [hep-ph]. 

\vspace{-0.3cm}

\item[9] J. Kopp, M. Maltoni, T. Schwetz, arXiv:1103.4570 [hep-ph]. 

\vspace{-0.3cm}

\item[10] E. Giusarma et al., arXiv:1102.4774 [astro-ph]. 

\vspace{-0.3cm}

\item[11]  P. Adamson {\it et al} [ MINOS Collaboration ] arXiv:1104.3922 [hep-ex]. 

\vspace{-0.3cm}

\item[12] H. Nunokawa,  O. L. G. Peres and R. Zukanovich-Funchal, 
        Phys. Lett. B562 (2003) 279; S. Choubey HEP 0712 (2007) 014.

\vspace{-0.3cm}

\item[13] S. Razzaque and A.Yu. Smirnov,  arXiv:1104.1390, [hep-ph].

\vspace{-0.3cm}

\item[14] P. De Holanda and A. Yu. Smirnov,   
Phys. Rev. D83 113011 (2011), arXiv:1012.5627 [hep-ph].

\end{description}

\newpage
 
 \subsection{Sylvaine Turck-Chi\` eze}

\vskip -0.3cm

\begin{center}

CEA/IRFU/SAp, CE Saclay, 91191 Gif sur Yvette Cedex, France\\

\bigskip

{\bf  Helioseismology, Neutrinos and Dark Matter} 

\end{center}

\medskip

The Sun is certainly the best known star. But answers to crucial questions are still missing because its standard representation is too poor to reproduce all the present observations.  We ignore its initial mass and consequently the real history of the formation of Earth and Mars: we know already that its initial mass was larger than the present one to avoid what has been called the `Solar paradox'  [1]. We cannot predict its degree of activity for the next 20 years and its impact on the present Earth environment because its internal magnetic description is not yet under control. All these questions emerge once again because the knowledge of the solar interior has been improved tremendously this last 10 years thanks to the SoHO satellite that transforms it in a real laboratory of plasma physics. One needs to explain properly the present sound speed  profile and the present  internal rotation profile. They are now really under control and they show the limitations of the standard solar model [2,3]. See Figure 1.

A large success of the last decade is  that all the different informations: detected solar neutrinos of different flavors or energies,  acoustic modes, gravity modes altogether lead to a coherent and new picture of our active star [4].   

So, this star can also put some constraints on the properties of dark matter  due to the position of the Sun in our Galaxy and  the accuracy obtained in   the whole radiative zone that contains practically 98\% of the solar mass [5,6,7].

\medskip
In fact, the idea that the Sun could show an indirect evidence of dark matter appears in the eighties [8]  to solve the `solar neutrino problem' with a CDM scenario. In this contribution we show how this idea has evolved, how WIMPs could have imprinted the solar core and how the range of WIMPs properties are reduced by the recent detection of gravity modes associated to boron neutrino detection. We finally discuss quickly also the case of the potential presence of lighter particles like sterile neutrinos of the keV range in the solar interior. 

\vskip 0.5cm
\noindent
{\bf Neutrinos,  Acoustic modes and Gravity modes}

The accumulation of WIMPS in the center of the Sun  decreases the
central temperature and increases the density.  These  WIMPS
rapidly become in thermic equilibrium with baryons (Hydrogen, Helium
and other elements), leading  to the formation of an
isothermal core that modifies  the sound speed and density profiles below  5\% $R_\odot$ [5]. 

The best way to check this possibility is to use the boron neutrino flux that is extremely dependent on the central temperature (to the exponent 20-24) and to detect gravity modes that add some independent constrains on the central density and its profile in the core below 0.1 $R_\odot$.  Both are now available after 30 years of effort.

The SNO boron detection [9,10] has completed the Kamiokande results and totally clarified the fact that different flavors of neutrinos must be considered to understand all the detections of the solar boron neutrino emitted flux (the flux coming from the reaction p+Be7 -$>$ $B8^*$ -$>$ 2 He4 + $e^+ + \nu_e$ that acts in the first 10\% $R_\odot$. 

In parallel the use of a large number of  acoustic modes detected aboard the SOHO satellite [2a] and confirmed with ground instruments [2b] has allowed to build a seismic solar model that avoids the variability of the standard model  prediction of the boron flux  noticed along time (see Table 6 of [4]). This point is important because, as mentioned previously, we know today that the physics of the standard model is incomplete. The seismic model is built to get practically no difference in sound speed between model and observation (Fig 1a). The acoustic modes detection allows an observational estimate of the sound speed down to 0.06 $R_\odot$ with a vertical error bar extremely small. One sees on the figure that the difference with SSM sound speed integrating the recent composition of CNO is  larger at more than 25 $\sigma$ from the observational result. 

The main characteristics of this seismic model [2, 11,12] are summarized in Table 1. This model allows a prediction of the gravity mode frequencies that are in very good agreement with the first observation of dipole gravity modes ([3,13]  and Figure 1b). Their period differences are smaller than 0.3 minute. It also produces results in agreement with the detection of all the different neutrino installations (see table 9 of [4]). So the central temperature, central density and core density profile of the Sun must be very near from those of the seismic model predictions, even  the detailed physics of the radiative zone is not totally under control: possible existence of a fossil field, possible bad determination of the transfer of energy, possible effect of dark matter...

\vskip 0.5cm
\noindent
{\bf Constraints  on Cold Dark Matter or Warm Dark Matter}

We use these previous strong constraints to limit the range of parameters of the WIMPs properties  using the prescription of [14] that introduces a capture, annihilation and evaporation rates.  We show that the new solar detections cannot  constrain  dark matter annihilation models as the estimated cross section is too small to produce any effect on the density profile. But we exclude the presence of   non-annihilating WIMPs in the Sun for masses $\le$ 10 GeV  and spin dependent cross sections $> 5 \,10^{-36}$ cm$^2$ [7], see Figure 1b. No signature is visible in the obtained profile in the solar core.  We note also that the seismic model has a central temperature higher than the SSM (T$_C$= 15.51, SSM neutrino prediction is not compatible with boron neutrino detection)  but a density slightly smaller by 2 or 3\%. This remark could favor other kinds of processes like a different gravitational effect due to WDM (sterile neutrinos ?) acting on the whole radiative zone.  In fact the radiative zone is certainly more complex than thought previously and one needs to include  a good energetic balance [1] and all the potential dynamical effects before any conclusion on this subject.

\medskip
In the coming years helio- and asteroseismology will provide complementary  results  to extend the present study and to check these probes of dark matter.
 
\bigskip

  \begin{figure}
\includegraphics[width=12pc] {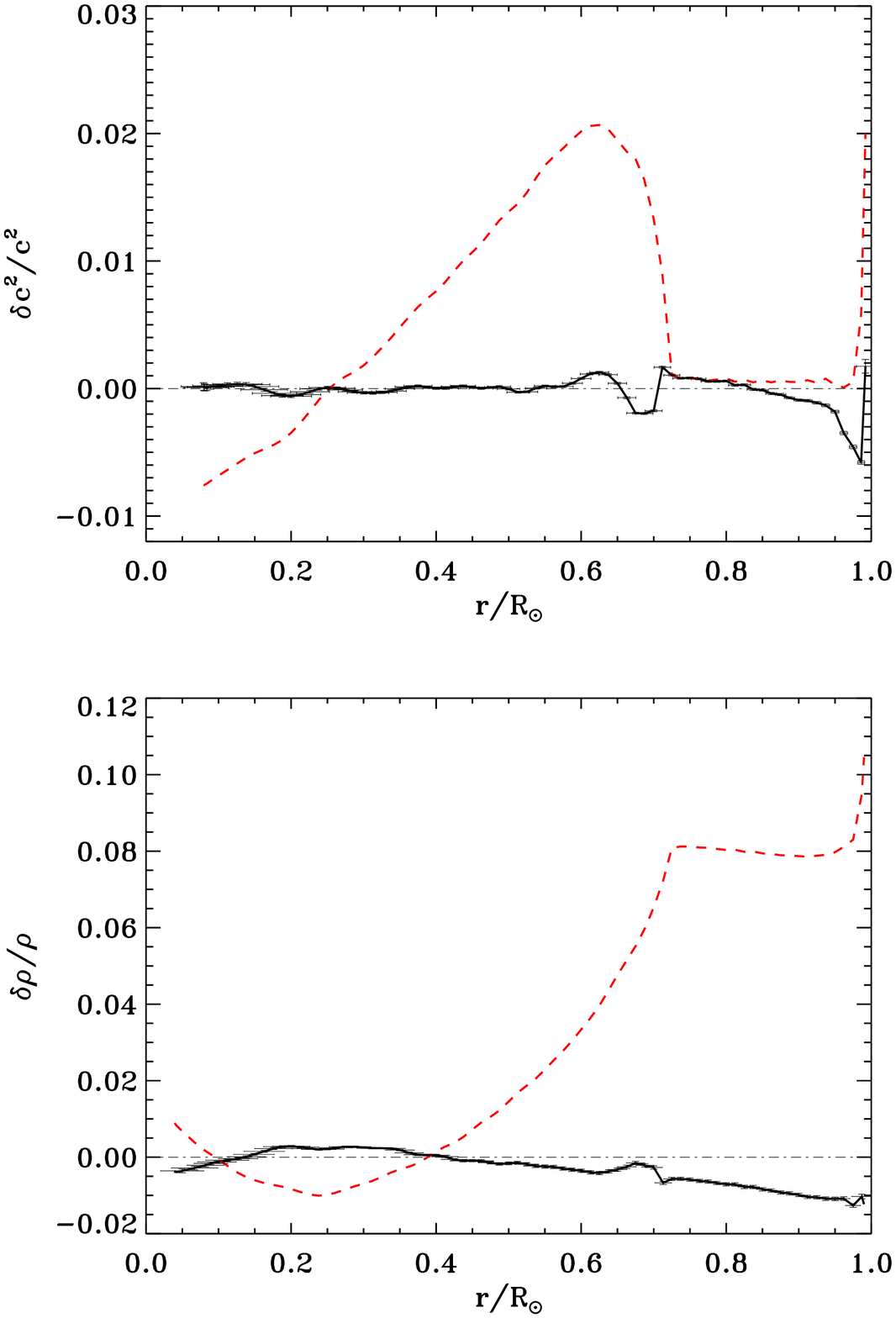}
\includegraphics[width=15pc] {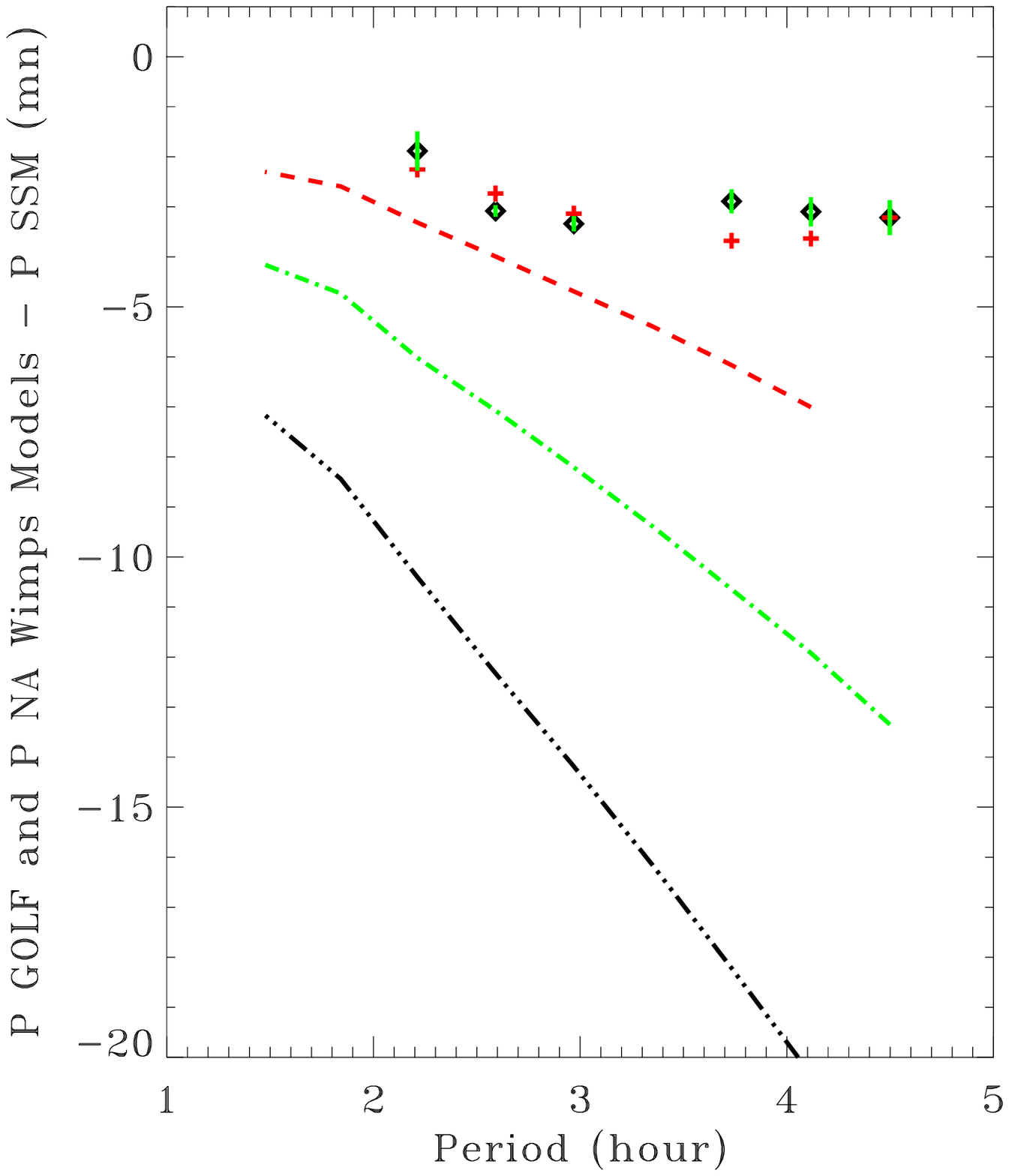}
\caption { \small {Left: Differences in squared sound speed between GOLF+MDI/SoHO and solar model predictions. Seismic model: full line + seismic error bars, blue ($ {- -}$) and red lines (small $ {- -}$) from SSM models without or with tachocline treatment. From [4]. Right: Differences in minutes between GOLF 
(green diamonds with error bars) or SSeM (crosses) and SSM gravity mode periods. Superimposed are the differences for solar model with DM using spin dependent cross section of 5 10$^{-36}$cm$^2$  for respectively 5, 7 and 10 Gev (black ${- ... -}$ line, green ${-.-}$ line and red small ${- -}$ line).  From [7]. }}
\normalsize
\label{fig:figure2}
\end{figure}
  \begin{table}
  \begin{center} 
   \caption{Central conditions of the seismic model}
\vspace{5mm}    
  \begin{tabular}{lccccc}
  \hline  
 &  $T_C$  & $ \rho_C\; \; $ & boron $\nu$ prediction \; \;	& gravity mode prediction l= 2, n = -10 \\
\hline      
     &  15.75 $10^6$K   \;\; &  153.6 g/ cm$^3$\; \; &  5.31 $\pm 0.6 \; 10^6 \nu / cm^3$  & 62.50   $\mu$Hz       \\
       \hline
 \end{tabular}
  \end{center}  
\label{tab:table1}
\end{table}

\bigskip

{\bf References}
\begin{description}

\item[1] S. Turck-Chi{\`e}ze, L. Piau and S. Couvidat, ApJ lett, 731, L29 (2011)

\vspace{-0.3cm}

\item[2]  a) S. Turck-Chi\`eze et al., ApJ, 555, L69 (2001); b) S. Basu et al., ApJ., 699, 1403 (2009) 

\vspace{-0.3cm}

\item[3] S. Turck-Chi\`eze, et al., ApJ, 715, 153 (2010)
R.~A. Garc\'ia,  et al.,  SOHO24, JPCS, 271, 12046 (2011)

\vspace{-0.3cm}

\item[4] S. Turck-Chi{\`e}ze and S. Couvidat, Report in Progress in Physics, 74, 086901 (2011)

\vspace{-0.3cm}

\item[5] I. Lopes \& J. Silk,  ApJ., 722, 95L and Science, 330, 462L (2010)

\vspace{-0.3cm}

\item[6] M. Taoso et al.,  Phys. Rev. D, 82, 3509 (2010)

\vspace{-0.3cm}

\item[7] S. Turck-Chi\`eze  et al., ApJ  submitted (2011)

\vspace{-0.3cm}
\item[8] D. N. Spergel  \& W. H. Press, ApJ., 294, 663 (1985)

\vspace{-0.3cm}
\item[9] S. N. Ahmed et al. : the SNO collaboration,  Phys. Rev. Lett. 92, 181301 (2004)
2004

\vspace{-0.3cm}
\item[10] B. Aharmim et al. : the SNO collaboration, Phys. Rev. C 81, 055504 (2010)

\vspace{-0.3cm}
\item[11] S. Turck-Chi\`eze  et al.,  Phys. Rev. Lett. 93, 211102 (2004)

\vspace{-0.3cm}
\item[12] S. Mathur, et al., \apj, 668, 594  (2007)

\vspace{-0.3cm}
\item[13] R. A. Garcia  et al.,, Science 316, 1591 (2007)

\vspace{-0.3cm}
\item[14] A. Gould \& G. Raffelt, ApJ. 352, 654 (1990)
\end{description}

\newpage

\section{Posters Highlights}

\subsection{Ayuki Kamada}

\vskip -0.3cm

\begin{center}

%
Institute for the Physics and Mathematics of the Universe, TODIAS,\\
The University of Tokyo, 5-1-5 Kashiwanoha, Kashiwa, Chiba 277-8583, Japan

\bigskip


{\bf Light sterile neutrino as warm dark matter and the structure of galactic dark halos} 

\end{center}

\medskip

%
%
%
%
%
%
{\bf Abstract}

We study the formation of nonlinear structure in $\Lambda$ Warm Dark Matter (WDM) 
cosmology using large cosmological N-body simulations.
We assume that dark matter consists of sterile neutrinos that are generated 
through nonthermal decay of singlet Higgs bosons 
near the Electro-Weak energy scale.
Unlike conventional thermal relics, the nonthermal WDM has a peculiar velocity 
distribution, which results in a characteristic shape of the matter
power spectrum.
We perform large cosmological N-body simulations for the nonthermal 
WDM model. We compare the radial distribution of subhalos 
in a Milky Way size halo 
with those in a conventional thermal WDM model.
The nonthermal WDM with mass of 1 keV predicts
the radial distribution of the subhalos that is
remarkably similar to the observed distribution
of Milky Way satellites.

\medskip

{\bf Introduction}


Alternative models to the standard $\Lambda$ Cold Dark Matter model have been suggested as a solution 
of the so-called 'Small Scale Crisis'.
One of them is $\Lambda$ Warm Dark Matter cosmology, in which Dark Matter particles had non-zero velocity dispersions. 
The non-zero velocities smooth out primordial density perturbations below its free-streaming length of sub-galactic sizes. 
The formation of subgalactic structure is then suppressed. Moreover, The large phase space density may prevent dark matter 
from concentrating into galactic center. Particle physics models provide promising WDM candidates such as gravitinos, 
sterile neutrinos and so on. 
Gravitinos with mass of $\sim$keV in generic Supergravity theory are produced in thermal bath immediately 
after reheating and are then decoupled kinematically
from thermal bath similarly to the Standard Model (SM) neutrinos 
because gravitinos interact with SM particles only through gravity. 
The thermal relics have a Fermi-Dirac (FD) momentum distribution. 
Sterile neutrinos are initially proposed in the see-saw mechanism to explain the masses of the SM neutrinos. 
If the mass of sterile neutrinos is in a range of $\sim$keV 
and their Yukawa coupling is of order $\sim 10^{-10}$, then 
it cannot be in equilibrium with SM particles 
throughout the thermal history of the universe. 
There are several peculiar production mechanisms for sterile neutrinos,
such as Dodelson-Widrow (DW) mechanism~[1] and EW scale Boson Decay (BD)~[2]. 
In DW mechanism, sterile neutrinos 
are produced through oscillations of active neutrinos, and its velocity distribution has a Fermi-Dirac form 
just like gravitinos.
In the BD case, sterile neutrinos are produced via decay 
of singlet Higgs bosons and they have generally a nonthermal velocity distribution. 
The nonthermal velocity distribution imprints particular features in the transfer function of the density fluctuation 
power spectrum~[3]. The transfer function has a cut-off at the corresponding free-streaming length,
but it decreases somewhat slowly than thermal WDM models.
In this article, we study the formation of nonlinear structure for a cosmological model with nonthermal sterile neutrino WDM.
We perform large cosmological N-body simulations. 
There are several models of nonthermal WDM. For example, gravitinos can be produced via decay of inflaton~[4] 
or long-lived Next Lightest Supersymmetric Particle (NLSP)~[5]. 
Our result can be generally applied to the formation of nonlinear structure in these models as well. 



\medskip

{\bf Nonthermal sterile neutrino}

The clustering properties of the above BD sterile neutrino model is investigated by Boyanovsky~[3],
who solved the linearized Boltzmann-Vlasov equation in the matter dominant era when WDM has already 
become non-relativistic. 
Firstly, solving the Boltzmann equation for sterile neutrinos produced by the singlet boson decay process, 
we find the most of contribution to the present number of sterile neutrinos comes when the temperature 
decreases to the EW scale. We then get the velocity distribution of sterile neutrinos. 
The BD distribution has a distinguishable feature from usual Fermi-Dirac distributions at small velocities, $y=P/T(t)$:
\begin{align}
f_{\rm BD}(y)\propto \frac{1}{y^{\frac{1}{2}}}, \ \ \ \ \ f_{\rm FD}(y)\propto 1. \notag
\end{align}
The velocity distribution is imprinted in the power spectrum, which decreases slowly across the free-streaming length scale. 
Following Boyanovsky~[3], we solve the linearlized Boltzmann-Vlasov equation. We define the comoving free-streaming wavenumber as
\begin{align} 
 k_{\rm fs}={\Big[}\frac{3H_{0}^2\Omega_{M}}{2\langle\vec{V}^2\rangle(t_{\rm eq})}{\Big]}^{\frac{1}{2}} \notag
\end{align}
akin to the Jeans scale at the matter-radiation equality.
In the BD case, the present energy density of dark matter is determined
by the Yukawa coupling.
Assuming the relativistic degree of freedom $\bar{g}$ is the usual SM value 
$\bar{g}\simeq 100$ for the $m=1$ keV sterile neutrino dark matter, we find
\begin{align}
k_{\rm fs}^{\rm BD}=18 \ h \ \mathrm{Mpc^{-1}}, \notag
\end{align}
while for the $m=1$ keV gravitino dark matter, which has the FD distribution, 
we should take $\bar{g}\simeq1000$ to get the present energy density of dark matter.
Then, for $m=1$ keV gravitinos, 
\begin{align}
k_{\rm fs}^{\rm FD}=32 \ h \ \mathrm{Mpc^{-1}}. \notag
\end{align}
Although these two models have almost the same free-streaming length, 
their linear power spectra show appreciable differences, as seen in Fig. 1. 
There, we adopt the numerically fit transfer function in Bode et al.~[6]
for the linear power spectrum of the gravitino dark matter.
 
The enhancement of the velocity distribution in the low velocity region leads 
to the slower decrease of the linear power spectrum. This implies that we should care not only free-streaming 
scale or velocity dispersion, but also the shape of the velocity distribution in studying the formation 
of the nonlinear object below the cut-off (free-streaming) scale.

Using these linear power spectrum, we have performed direct numerical simulations to study observational 
signatures of the imprinted velocity distribution in the subgalactic structures. 
We start our simulations from a redshift of $z=9$.
We use $N = 256^{3}$ particles in a comoving volume of $10 \ h^{-1}$ Mpc on a side. The mass of a dark 
matter particle is $4.53\times10^{6} \ h^{-1} \ \mathrm{M_{\odot}}$ and the gravitational softening
length is $2 \ h^{-1}$ kpc. 

\medskip

{\bf Simulation Results}

Fig. 2 shows the projected distribution of dark matter in and round a Milky Way size halo at $z=0$. 
The plotted region has a side of $2 \ h^{-1}$ Mpc. Dense regions appear bright. 
Clearly, there are less subgalactic structures for the WDM models. 
The 'colder' property of the nonthermal WDM shown in the linear 
power spectrum (see Fig. 1) can also be seen in the abundance of subhalos.


In Fig. 3, we compare the cumulative radial distribution of the subhalos in our 'Milky Way' halo at $z=0$ 
with the distribution of the observed Milky Way satellites~[7]. 
Interestingly, the nonthermal WDM model reproduces the radial distribution of the observed Milky Way satellites 
in the range above $\sim 40$ kpc. 
Contrastingly, the CDM model overpredicts the number of subhalos by a factor of $5$ than the 
observed Milky Way satellites. This is another manifestation of the so-called 
'Missing satellites problem'. The thermal WDM model 
surpresses subgalactic structures perhaps too much, by a factor of $2-4$ than the observation.

\medskip

{\bf Summary}

We have studied the formation of the nonlinear structures in a $\Lambda$WDM cosmology using 
large cosmological N-body simulations. We adopt the sterile neutrino dark matter produced via the decay of 
singlet Higgs bosons with a mass of EW scale. The sterile neutrinos have a nonthermal velocity distribution, 
unlike the usual Fermi-Dirac distribution. The distribution is a little skewed to low velocities. 
The corresponding linear matter power spectrum decreases
slowly across the cut-off scale compared to the thermal WDM, such as gravitino dark matter.
Both of the two models have the same mass of $1$ keV and an approximately the same cut-off (free-streaming) scale 
of a few 10 $h \ \mathrm{Mpc^{-1}}$.  We have shown that this 'colder' property of the nonthermal WDM can be seen 
in richer subgalactic structures. The nonthermal WDM model with mass of $1$ keV appears to reproduce the radial 
distribution of the observed Milky Way satellites. 
\\
\\

\bigskip

{\bf References}

\begin{description}

\item[1]
  S.~Dodelson and L.~M.~Widrow,
  Phys.\ Rev.\ Lett.\  {\bf 72}, 17 (1994)
  [arXiv:hep-ph/9303287].

\vspace{-0.3cm}

\item[2]
  K.~Petraki and A.~Kusenko,
  Phys.\ Rev.\  D {\bf 77}, 065014 (2008)
  [arXiv:0711.4646 [hep-ph]].

\vspace{-0.3cm}

\item[3]
  D.~Boyanovsky,
  Phys.\ Rev.\  D {\bf 78}, 103505 (2008)
  [arXiv:0807.0646 [astro-ph]].

\vspace{-0.3cm}

\item[4]
  F.~Takahashi,
  Phys.\ Lett.\  B {\bf 660}, 100 (2008)
  [arXiv:0705.0579 [hep-ph]].

\vspace{-0.3cm}

\item[5]
  K.~Sigurdson and M.~Kamionkowski,
  Phys.\ Rev.\ Lett.\  {\bf 92}, 171302 (2004)
  [arXiv:astro-ph/0311486].
  
\vspace{-0.3cm}





\item[6]
  P.~Bode, J.~P.~Ostriker and N.~Turok,
  Astrophys.\ J.\  {\bf 556}, 93 (2001)
  [arXiv:astro-ph/0010389].

\vspace{-0.3cm}



  

  
 
 \item[7]
  E.~Polisensky and M.~Ricotti,
  Phys.\ Rev.\  D {\bf 83}, 043506 (2011)
  [arXiv:1004.1459 [astro-ph.CO]].

\clearpage


\begin{figure}[htbp]
\begin{center}
\includegraphics[height=45mm]{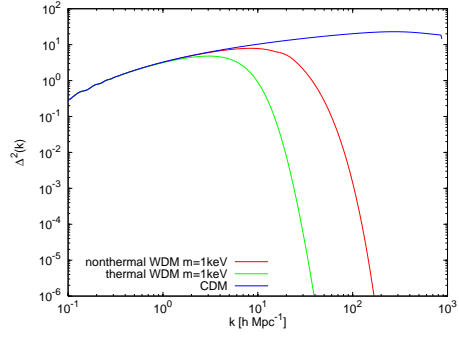}
\end{center}
\caption{The linear power spectra for the nonthermal WDM (red line), thermal WDM (green line) and CDM (blue line). 
The dark matter mass of the both WDM model is $m=1$ keV.}
\end{figure}

\begin{figure}[htbp]
\begin{minipage}{0.32\hsize}
\begin{center}
\includegraphics[width=35mm,height=35mm]{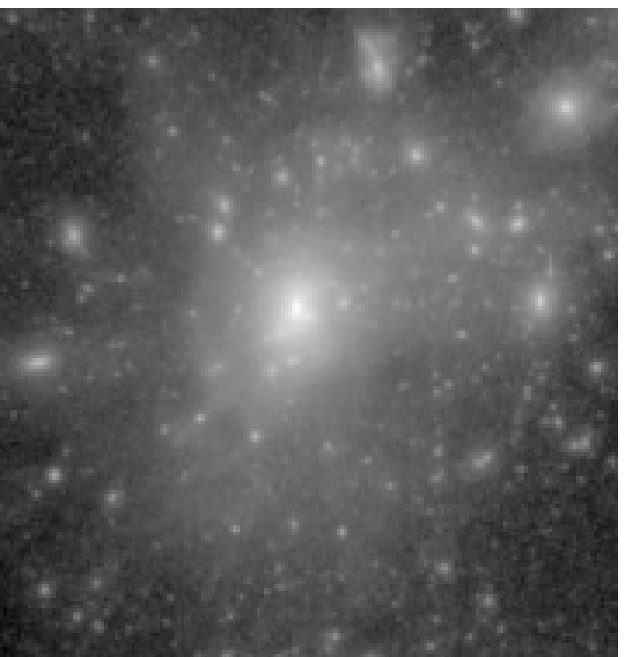}
\end{center}
\end{minipage}
\begin{minipage}{0.32\hsize}
\begin{center}
\includegraphics[width=35mm,height=35mm]{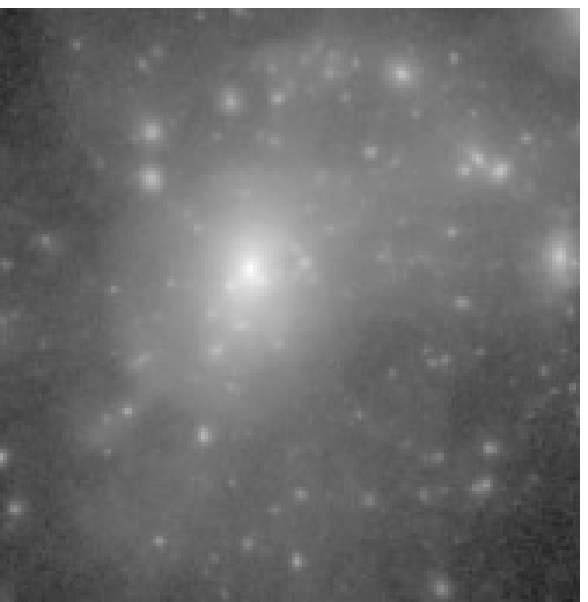}
\end{center}
\end{minipage}
\begin{minipage}{0.32\hsize}
\begin{center}
\includegraphics[width=35mm,height=35mm]{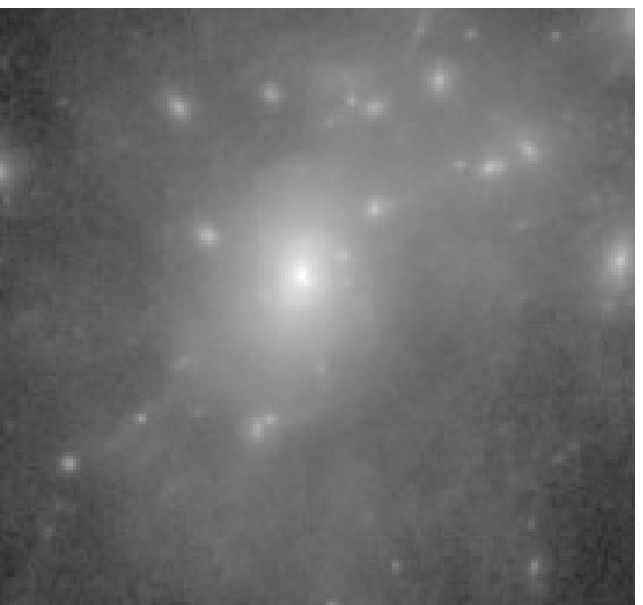}
\end{center}
\end{minipage}
\caption{The projected distributions of the substructures in our 'Milky Way' halo at $z=0$. 
The sidelength of the shown region is $2 \ h^{-1}$ Mpc. 
The results are for three dark matter models, CDM,  nonthermal WDM with 1 keV mass and thermal WDM with 1keV mass, respectively, 
from left to right.}
\end{figure}



\begin{figure}[htbp]
\begin{center}
\includegraphics[height=50mm]{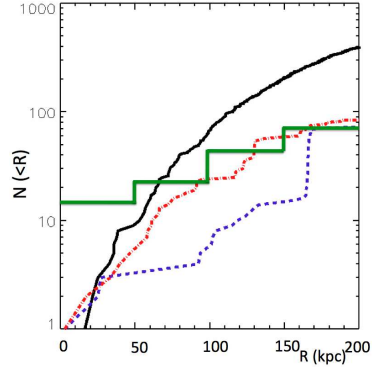}
\end{center}
\caption{The radial distribution of the subhalos in our 'Milky Way' halo at $z=0$. 
For simulation results for CDM (black), 1keV nonthermal WDM (red), and 
1keV thermal WDM (blue). Green bars show the distribution of the observed Milky Way satellites.}
\end{figure}

\end{description}

\newpage

\subsection{Sinziana Paduroiu$^{1}$, Andrea Macci\`o$^{2}$, Ben Moore$^{3}$,Joachim Stadel$^{3}$,Doug Potter$^{3}$,George Lake$^{3}$, Justin Read$^{4,5}$ \& Oscar Agertz$^{6}$}

\vskip -0.3cm

\begin{center}

$^1$ Observatory of Geneva, CH-1290, Geneva, Switzerland
$^2$ Max-Planck-Institute for Astronomy, K\"onigstuhl 17, D-69117
Heidelberg, Germany
$^3$ Institute for Theoretical Physics, University of Z\"urich, CH-8057 Z\"urich, Switzerland
$^4$ Department of Physics \& Astronomy, University of Leicester, Leicester LE17RH, United Kingdom
$^5$ Institute for Astronomy, Department of Physics, ETH Z\"urich, Wolfgang-Pauli-Strasse 16, CH-8093 Z\"urich, Switzerland.
$^6$ Kavli Institute for Cosmological Physics, University of
Chicago, 5640 South Ellis Avenue, Chicago, IL, USA 60637

\bigskip

{\bf The effects of free streaming on warm dark matter haloes: a test of the Gunn-Tremaine limit} 

\end{center}

\medskip

The free streaming of warm dark matter particles dampens the 
fluctuation spectrum, flattening the mass function
of haloes and imprinting a fine grained phase density (PSD) limit for dark matter structures.
The Gunn-Tremaine limit is expected to
imprint a constant density core at the halo center. 
In a purely cold dark matter model the fine grained phase space density is effectively infinite in the initial conditions and would therefore be infinite everywhere today. However when the phase space density profiles are computed using coarse grained averages that can be measured from N-body simulations, the value is finite everywhere and even falls with radius with a universal power law slope within virialised structures (Taylor \& Navarro 2001). The coarse grained phase space is an average over mixed regions of fine grained phase space, so this behaviour is as expected (Tremaine \& Gunn 1979).

Using high resolution simulations of structure
formation in a warm dark matter universe (movies available: http://obswww.unige.ch/$\sim$paduroiu ) we explore these effects on structure formation and the properties of warm dark matter halos.
\begin{itemize}
\item{The finite initial fine grained PSD is a also a maximum of the coarse grained PSD, resulting in PSD profiles of WDM haloes that are similar to CDM haloes in the outer regions, however they turn over to a constant value set by the initial conditions}
\item{The turn over in PSD results in a constant density core with characteristic size that is 
in agreement with the simplest expectations } 
\item{We demonstrated that if the primordial velocities are large enough to produce a significant core in dwarf 
galaxies i.e. $\sim$ kpc, then the free streaming erases all perturbations on that scale and the haloes cannot form}
\item{Halo formation occurs top down on all scales with the most massive haloes collapsing first}
\item{The concentration - mass relation for WDM haloes is reversed with respect to that found for CDM}
\item{Warm dark matter haloes contain visible caustics and shells}

\end{itemize}

We ran two suites of simulations, first with a $160^{3}$ particles in
40 Mpc box, and the second one with $300^3$ in a 42.51 Mpc box.
We adopt a flat $\Lambda$CDM  cosmology with parameters from the first
year  WMAP results  (Spergel et al. 2003). The transfer function in WDM model has been computed using the fitting formula suggested by Bode, Turok and Ostriker (2001) where $\alpha$, the scale of the break, is a function of the WDM parameters (Viel et al (2005)), while the index $\nu$ is fixed. From these simulations several galaxy mass haloes were re-simulated at higher resolution, with and without thermal velocities. 
The density profile for a  $7\times10^{11}M_{\odot}$ is shown in Figure \ref{fig:1} in four different
contexts, the CDM case, the WDM case with just the cutoff in the
power spectrum as expected for a 200eV particle (WDM1), with velocities corresponding to the 200eV particle (WDM2), and with velocities artificially increased such that they correspond to a 20eV particle but with the power spectrum
of the 200eV case (WDM3).

Figure 2 shows the corresponding coarse grained PSD 
profiles calculated by spherical averaging the quantity
$\rho/\sigma^{3}$.

\begin{figure}   
\psfig{file=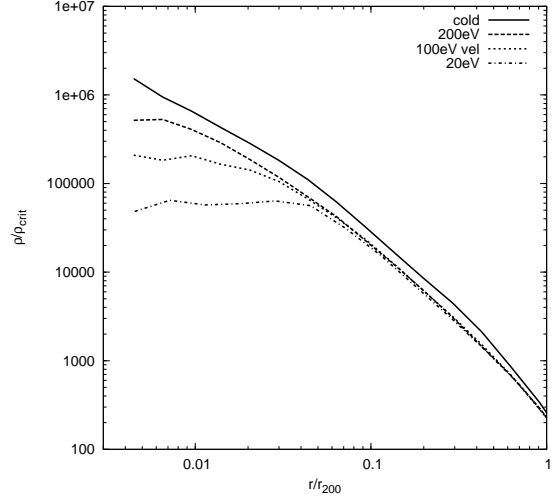,angle=-90,width=280pt}
\caption[]{The spherically averaged 
density profiles for CDM, WDM1, WDM2 \& WDM3 haloes.
The resolution limit is at approximately 0.5\% of the virial
radius (the softening radii are a 0.26\% of the virialized radius).
}
\label{fig:1}
\end{figure}        

\begin{figure}
\psfig{file=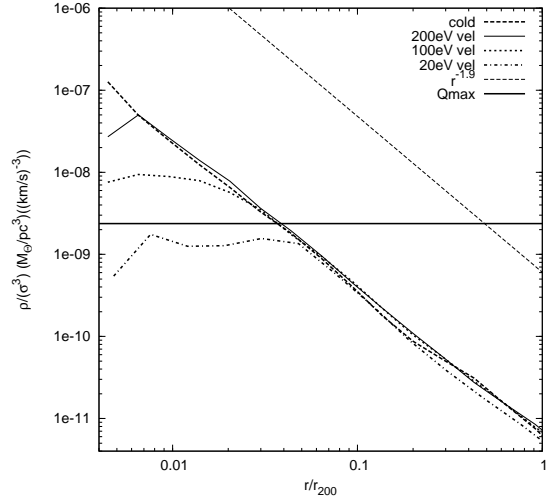,angle=-90,width=280pt}
\caption[]{``Phase-space density'' (PSD) profiles of the same haloes shown in Figure 1, calculated using $\rho/\sigma^{3}$.} 
\label{fig:2}
\end{figure}

In the case of our simulations, for a constant density in the initial conditions, the phase space density is:
\vspace{-0.3cm}
\begin{equation}
\label{eq=q0}
Q_0=\frac{\rho}{\sigma^{3}}=\rho_{crit}\Omega\left({\frac{m_{\nu}c^2}{KTc}}\right)^3
\end{equation}
\vspace{-0.3cm}
For a critical density $\rho_{crit}=1.4\times10^{-7}M_\odot/pc^3$ and $\Omega=0.268$ we find the phase space density:
\begin{equation}
\label{eq=q01}
Q_0=3\times10^{-4}M_\odot{\rm pc}^{-3}\left({\rm km/s}\right)^{-3}\left(\frac{m_{x}}{1keV}\right)^{3}
\end{equation} 
\vspace{-0.3cm}
The minimum core radius is described by
\vspace{-0.3cm}
\begin{equation}
\label{eq=core}
r_{c,min}^{2}=\frac{\sqrt{3}}{4\pi
  GQ_{0}}\frac{1}{ \langle {\sigma^{2} \rangle }^{1/2}},
\end{equation}

Thus for a particle with $m_{x}=20$ eV and $\sigma=150$ km/s the 
phase space density is $2.4\times 10^{-9}M_\odot{\rm pc}^{-3}\left({\rm km/s}\right)^{-3}$  
which gives a theoretical value for the core radius of 9.3 kpc. If we consider the core radius to be the radius where the density starts decreasing from the constant value, a ''by eye'' fit gives a slightly larger radius of ~6.5 kpc, while the radius where the density drops by a factor of 2 is ~9.4 kpc in agreement with the theoretical predictions.

\bigskip

{\bf References}

\begin{description}

\item[1] Bode, P., Ostriker, J. P., \& Turok, N. 2001, \apj, 556, 93

\vspace{-0.3cm}

\item[2] {Dalcanton}, J.~J. \& {Hogan}, C.~J. 2001, \apj, 561, 35

\vspace{-0.3cm}

\item[3] {Navarro}, J.~F., {Frenk}, C.~S., \& {White}, S.~D.~M. 1996, \apj, 462, 563

\vspace{-0.3cm}

\item[4]{Strigari}, L.~E., {Bullock}, J.~S., {Kaplinghat}, M.,  {Kravtsov}, A.~V.,
  {Gnedin}, O.~Y., {Abazajian}, K. \& {Klypin}, A.~A. 2006, \apj, 652, 306 

\vspace{-0.3cm}

\item[5]{Taylor}, J.~E. \& {Navarro}, J.~F. 2001, \apj, 563, 483

\vspace{-0.3cm}

\item[6]{Tremaine}, S. \& {Gunn}, J.~E. 1979, Phys.Rev.Let., 42, 407

\vspace{-0.3cm}

\item[7]Viel, M., Lesgourgues, J.,
Haehnelt, M.~G., Matarrese, S., \& Riotto, A.\ 2006, Physical Review
Letters, 97, 071301

\vspace{-0.3cm}

\item[8]{Wang}, J.,{White}, S.~D.~M. 2007 (astroph-0702575)

\end{description}

\newpage

\section{Summary and Conclusions of the Colloquium by
H. J. de Vega, M.C. Falvella and N. G. Sanchez}

\subsection{General view and clarifying remarks}

Participants came from Europe, North and South America, Russia, Ukraine, Japan, India, Korea.
Journalists, science editors and representatives of the directorates of several agencies were present in the Colloquium. 
Discussions and lectures were outstanding. The Standard Model of the Universe was at the center of this Colloquium with the last novelties in the CMB
data, high multipoles, CMB lensing detection, Sunayev-Zeldovich measurements and their
scientific implications, warm dark matter (WDM) advances with both theory and observations, neutrino masses and neutrino oscillations, keV sterile neutrinos as serious WDM candidates, with both theory and experimental search, heliosismology progresses, galaxy observations, clusters  and structure formation, the structure of the interstellar medium and star formation, and the links between these subjects, mainly through gravity and dark matter forming the structures of the cosmic web with the present vaccum energy being the cosmological constant (dark energy). Warm dark matter research evolve fastly in both astronomical, numerical, theoretical, particle and experimental research. 
The Colloquium allowed to make visible the work on WDM made by different groups over the world and cristalize 
WDM as the viable component of the standard cosmological model in agreement with CMB + Large Scale
Structure (LSS) + Small Scale Structure (SSS) observations, $\Lambda$WDM, in contrast to 
 $\Lambda$CDM which only agree with CMB+LSS observations and is plagued with SSS problems.  

The participants and the programme represented the different communities doing
research on dark matter:

\begin{itemize}
\item{Observational astronomers}
\item{Computer simulators}
\item{Theoretical astrophysicists not doing simulations}
\item{Physical theorists}
\item{Particle experimentalists}
\end{itemize}

\medskip

WDM refers to keV scale DM particles. This is not Hot DM (HDM). (HDM refers to eV scale DM particles, 
which are already ruled out). CDM refers to heavy DM particles (the so called wimps at the GeV scale or with any 
mass scale larger than the keV). 

\medskip 

It should be recalled that the connection between 
small scale structure features and the mass of the DM particle 
follows mainly from the value of the free-streaming
length $ l_{fs} $. Structures 
smaller than $ l_{fs} $ are erased by free-streaming.
WDM particles with mass in the keV scale 
produce $ l_{fs} \sim 100 $ kpc while 100 GeV CDM particles produce an
extremely small $ l_{fs} \sim  0.1 $ pc. While the keV WDM $ l_{fs} \sim 100 $ kpc
is in nice agreement with the astronomical observations, the GeV CDM $ l_{fs} $ 
is a million times smaller and produces the existence of too many
small scale structures till distances of the size of the Oort's cloud
in the solar system. No structures of such type have ever been observed.

\medskip
 
Also, the name CDM  precisely refers to simulations with heavy DM particles in the GeV scale.
Most of the literature on CDM simulations, do not make explicit
the relevant ingredient which is the mass of the DM particle (GeV scale wimps in the CDM case). 

The mass of the DM particle with the free-streaming length naturally enters in the initial power spectrum used in the N-body simulations and in the initial velocity. The power spectrum for large scales 
beyond 100 kpc is identical for WDM and CDM particles,
while the WDM spectrum is naturally cut off at scales below 100 kpc, 
corresponding to the keV particle mass free-streaming length. In contrast, the CDM spectrum 
smoothly continues for smaller and smaller scales till $\sim$ 0.1 pc, which gives rise to
the overabundance of CDM structures at such scales predicted by CDM.

CDM particles are always non-relativistic, the initial velocities
are taken zero in CDM simulations, (and phase space density is unrealistically infinity 
in CDM simulations), while all this is not so for WDM.

\medskip

Since keV scale DM particles are non relativistic for $ z < 10^6 $
they could also deserve the name of cold dark matter, although for historical reasons the name WDM
is used. Overall, seen in perspective today, the reasons why CDM does not work are simple: the heavy wimps 
are excessively non-relativistic (too heavy, too cold, too slow), and thus frozen, which preclude them to 
erase the structures below the kpc scale, while 
the eV particles (HDM) are excessively relativistic, too light and fast, (its free streaming length is too large), 
which erase all structures below the Mpc scale; in between, WDM keV particles produce the right answer.

\bigskip

Discussions and lectures were outstanding. Inflection points in several current 
research lines emerged.  
New important issues and conclusions arised and between them, it worths to highlight:

\bigskip

Results and the current state of missions and ongoing projects
were reported by their teams: Atacama Cosmology Telescope, Planck, Herschel, SPIRE,
ATLAS and HerMES surveys, the James Webb Telescope

\subsection{Conclusions}

Some conclusions are:

\begin{itemize}

\bigskip

\item{{\it James E. Webb} is honored with the most powerful space telescope ever designed, capable of observing the early universe within a few hundred million years of the Big Bang, revealing the formation of galaxies, stars, and planets, and showing the evolution of solar systems like ours. It is NASA project, with major contributions from the European and Canadian Space Agencies (ESA and CSA). It will have a 6.6 m diameter aperture (corner to corner), will be passively cooled to below 50 K, and will carry four scientific instruments: a Near-IR Camera (NIRCam), a Near-IR Spectrograph (NIRSpec), a near-IR Tunable Filter Imager (TFI), and a Mid-IR Instrument (MIRI). It is planned for launch in 2018 on an Ariane 5 rocket to a deep space orbit around the Sun - Earth Lagrange point L$_2$, about $1.5 \times 10^6$ km from Earth. The spacecraft will carry enough fuel for a 10 yr mission.  Four key topics were used to guide the design of the observatory:  
(i) The end of the dark ages: first light and reionization requires the largest feasible infrared telescope  (first objects of the universe are faint, rare, and highly redshifted), a wide wavelength range and a powerful multi-object infrared spectroscopy, to determine the physical conditions and redshifts of the earliest objects.(ii) The assembly of galaxies around dark matter concentrations. The interaction between dark matter, ordinary matter, black holes (when and how did they form), stars, winds, and magnetic fields is extraordinarily complex ``gastrophysics''. With its infrared imaging and spectroscopy, JWST will reveal details of galaxy formation and evolution. The JWST will also address some aspects of dark energy and dark matter. It can extend the Hubble measurements of distant supernovae; improve the calibration of the Hubble constant;  extend maps of the dark matter distribution to higher redshift.(iii) The birth of stars and protoplanetary systems uniquely observed in the infrared which complement radio observations very well.(iv) Planetary systems and the origins of life from the bright planets to the faint comets, asteroids, and dwarf planets of the outer solar system. JWST's image quality, field of view, and ability to track moving targets are essential to detect and analyze many such targets, among other feasible observations with JWST in this topic.}

\bigskip

\item{The $Herschel$ Gould Belt survey  
images the bulk of nearby ($d \sim 500$~pc) molecular clouds 
 in a variety of star-forming environments, allowing 
to clarify the physical mechanisms of prestellar cores's origin out of the diffuse interstellar medium (ISM), and the  stellar initial mass function (IMF)origin. A profusion of parsec-scale filaments in nearby  ISM molecular clouds is found and an intimate connection between the filamentary structure 
 and the dense cloud core formation process. Remarkably, filaments are omnipresent even in unbound, non-star-forming complexes and all appear to share a common width $\sim 0.1$pc:

In active star-forming regions prestellar cores are located within gravitationally unstable filaments which mass per unit length exceeds the critical value $M_{\rm line, crit} = 2\, c_s^2/G \sim 15\, M_\odot$/pc,  
where $c_{\rm s} \sim 0.2$~km/s is the isothermal sound speed for $T \sim 10$~K.  Core formation occurs in two main steps: First, an intricate ISM filaments network is generated by large-scale turbulence ; second, the densest filaments fragment into prestellar cores by gravitational instability. 

This  {\it explains} the star formation threshold at a gas {\it surface density}  $\Sigma_{\rm gas}^{\rm th} \sim $~130~$M_\odot \, {\rm pc}^{-2} $ found recently in both Galactic and extragalactic cloud complexes  : 
Given the typical  $\sim $~0.1~pc width of interstellar filaments, the threshold $\Sigma_{\rm gas}^{\rm th}$ 
corresponds to within a factor of $< 2$ to the critical mass per unit length  $M_{\rm line, crit}$ above which gas filaments at $T \sim 10$~K are gravitationally unstable.}

\bigskip

\item{Impressive CMB observations and their scientific implications have been reported on small angular scales, i.e.\ at the high-$\ell$ part of the CMB power spectrum. Also the Sunyaev-Zeldovich story continues to be interesting. The key modern frontiers being {\em polarization} and {\em the high resolution temperature power spectrum}.
The current direct limit on the primordial tensor to scalar ratio from QUIET is $r<0.9$ (95\%),thus promising well for the future measurements (with more data and a further frequency) from QUIET. Results on $r$ and  $n_s$, are crucial discriminators of the dynamics of inflationary models. Constraints from WMAP7 in the $(r,n_s)$ plane have already effectively ruled out a purely  (monomial) quartic inflation potential, $V(\phi)\propto \phi^4$, and the monomial $V(\phi)\propto \phi^2$ is under pressure. 

\medskip

High-$\ell$ power spectra were released from the Atacama Cosmology Telescope (ACT), and the South Pole Telescope (SPT): The CMB power spectrum impressively displays now 9 peaks clearly discerned. Various degeneracies in cosmic parameters can be now resolved with CMB-alone data. The high-$\ell$ part can also better constraint: (a) the tilt of the primordial spectrum, $n_s$; (b) a possible running  $n_{\rm run}$; (c) the primordial Helium abundance $Y_p$; and (d) the effective number of neutrino species at decoupling, $N_{\rm eff}$. The small-scale CMB can now be used to probe late-time physics  secondary effects.

ACT observed primordial helium at high significance, and detected the CMB lensing (photons gravitationally deflected-a few arcminutes-by large scale structures): lensing directly smooths the acoustic peaks and increases the small-scale power, and couples modes of different scales generating a non-zero lensing 4-point function  whose measure led to a 4$\sigma$ detection of the ACT lensing power: the first direct detection of CMB lensing. The CMB alone can now provide evidence for a dark energy component in the universe.

The amplitudes of the Sunayev-Zeldovich effect in clusters measured by WMAP7 were below a factor of 0.5-0.7 with respect to the standard X-ray models expectations for temperature and profile. However for the much larger sample of clusters surveyed by Planck (first released results) no such deficit has been found. It is unclear currently how to explain the apparent discrepancy in findings, and suggests that there is still quite a bit to learn in this subject.
ACT cluster number counts and scaling relations between cluster mass and SZ signal: further multi-wavelength observations are expected to better determine cluster masses and constraints. 

\medskip

The maps from the balloon-borne Spider experiment around Antarctica could be of great interest as providing $B$-mode polarisation observations which can be combined with those from the Planck Satellite. Planck is still on course to provide first cosmological information in January 2013, with the first release of primordial CMB polarization data likely in January 2014. Results from both the {\em KECK} array, the BICEP2 in the South Pole and ACTPol are eagerly awaited. ACTPol is due to begin observations in 2012 with improved sensitivity and polarization capabilities to measure the primordial power spectrum in polarization and the improved lensing signal.}

\bigskip

\item{The primordial CMB fluctuations are almost gaussian. The
effective theory of inflation \`a la Ginsburg-Landau predicts negligible primordial 
non-gaussianity, negligible running scalar index and the 
tensor to scalar ratio $ r $  in the range $0.021 < r < 0.053$, with the best value $\sim 0.04-0.05$ at reach of the next CMB observations.
Forecasted  $r$-detection probability for Planck with 4 sky coverages is border line. 
Improved measurements on $n_s$ as well as on TE and EE modes will improve 
these constraints on $r$ even if a detection will be lacking. 
Results from Planck are eagerly expected.
{\it Planck\/}  will improve the accuracy of current measures of a wide set of cosmological parameters by a factor from $\sim 3$ to $\sim 10$ and will characterize the universe geometry with unprecedented accuracy thanks to its excellent mapping and wide frequency coverage removal which allows all astrophysical emissions.  {\it Planck\/} will provide information and constraints on the early stages of the universe to the late phases of cosmological reionization, on the neutrino masses and effective species number, the primordial helium abundance, the stochastic field of gravitational waves through the B-mode angular power spectrum of the CMB anisotropies,   
{\it Planck\/} represents also an extremely powerful fundamental and particle physics laboratory.}

\bigskip

\item{Linear polarization of the cosmic microwave background (CMB) provides a direct test of inflation. Gravity waves generated during inflation impart a characteristic curl pattern (B-mode) in the linear CMB polarization.
Satisfying the simultaneous requirements of sensitivity, 
foreground discrimination, and immunity to systematic errors to detect such signal
is a technological challenge. The Primordial Inflation Explorer (PIXIE) is planned to detect and characterize 
such polarization signal. PIXIE will reach the confusion noise 
from the gravitational lensing E-mode signal and has the sensitivity and angular response
to measure even the minimum predicted B-mode power spectrum
at high statistical confidence. PIXIE will measure the CMB linear polarization
to sensitivity of 70 nK per $1\deg \times 1\deg$ pixel,
including the penalty for foreground subtraction. Averaged over the cleanest 75\% of the sky,
PIXIE can detect B-mode polarization to 3 nK sensitivity,
well below the 30 nK predicted from large-field inflation models.
The sensitivity is comparable to the ``noise floor'' of gravitational lensing,
and allows robust detection of primordial gravity waves to limit $r < 10^{-3}$ at more than 5 standard deviations. In addition, PIXIE provides a critical test for keV dark-matter candidates:
 primordial dark matter annihilation distorts the CMB spectrum away from the blackbody shape with 
The resulting chemical potential allows to determine the dark matter mass particle in the keV scale.
PIXIE measurements of the $y$ distortion determine the temperature of the intergalactic medium
at reionization.}

\bigskip

\item{Large-scale peculiar flows of clusters of galaxies provide an important test of the physics and initial conditions in the early Universe.  On sufficiently large scales, ${>\atop{\sim}} 100$ Mpc, this leads to a robust prediction of the amplitude and coherence length of these velocities independently of cosmological parameters or evolution of the Universe. One way to measure the peculiar velocities is from the kinematic component of the Sunyaev-Zeldovich (KSZ) effect: this had lead to measure a CMB KSZ dipole signal which is not from the primary CMB: this KSZ dipole implies a bulk flow of about 600-1,000 km/sec remaining coherent out to at least $\sim 750$ Mpc. (There may be a systematic overestimate smaller than $20 \%$ of this flow). The signal can be confirmed with the existing {\it public} cluster data in 5- and 7-yr WMAP data. These measurements do not agree with CDM structure formation. Perhaps, they are indicative of structures  well beyond the present-day horizon left over from pre-inflationary epochs (as fast-roll). The {\it SCOUT} experiment ({\bf S}unyaev-Zeldovich {\bf C}luster {\bf O}bservations as probes of the {\bf U}niverse's {\bf T}ilt) will catalog up to $\sim2,000$ clusters extending to $z\sim 0.7$, double the depth of the measured flow and its accuracy and improve the current calibration uncertainties, upon application to forthcoming CMB data releases.}

\bigskip

\item{Substantial observation efforts are being made to quantify both the statistical(number counts, correlations, spatial distributions...) and the internal
properties (star formation histories, dark matter density profiles,...) of dSph galaxies. 
Recent discovery surveys have accentuated the {\bf Satellite Problem} of CDM: Although some 25 dSph are now known assciated with each of the Milky way and M31, their numbers remain orders of magnitude too
low compared to simple LCDM predictions. Further, the spatial distributions of these satellites are concentrated in sheets/groups, more so than is predicted. Dominant Baryonic Feedback on galaxy formation is required in LCDM to solve the Satellite Problem, and to reproduce something like the observed
galaxy luminosity function. Feedback is not a free parameter, to be
sub-grid fixed, but it is a consequence of the star formation rate, which
is determinable from the chemical abundance distribution function
of the earliest stars. Substantial progress is being made in quantifying the early histories of the local dSph: In all cases, {\it very low star formation rates are required by observation}. This implies : (i)
{\bf Very low baryonic feedback.} (ii)  The field stars
in the Milky Way  were {\bf not} formed in now tidally destroyed dSph.
Appropriate high-resolution simulation efforts conclude
that {\bf baryon feedback does not affect DM structure} and that CDM
still cannot produce realistic galaxies with realistic feedback and
star formation recipes.

\medskip

Direct kinematic probing of dSph density profiles is making
significant progress: Very precise kinematics for faint
stars in dSph, and interestingly, the total mass enclosed within a
half-light radius is a robust parameter.  The existence of multiple populations inside a single dSph
galaxy is used to determine the mass enclosed with the half-light radius of
each population, thus providing several integrated mass determinations
inside a single profile. This new kinematic and chemical abundance work has considerable potential to
provide direct determinations of primordial DM-dominated density profiles.}

\bigskip

\item{Cosmology and oscillation data show that at least one neutrino mass 
should be in the interval 0.04 - 0.30 eV. The smallness of this neutrino mass,
which is realized in a seesaw mechanism,
can be due to the existence of a large mass scale (as GUT scale). 
Lepton mixing may be explained by the tri-bimaximal mixing scheme (TBM).
The issue of the existence of sterile neutrinos becomes crucial in the theory
of neutrino masses and mixing. Both laboratory experiments and cosmological observations
favour a 3+1 scheme with one sterile neutrino in the $ \sim $eV scale.}

\bigskip

\item{Sterile neutrinos with mass in the keV scale (1 to 10 keV) 
 emerge as leading candidates for the dark matter (DM)
particle from theory combined with astronomical observations.

DM particles in the keV scale (warm dark matter, WDM)
naturally reproduce (i) the observed
galaxy structures at small scales (less than 50 kpc), (ii) the observed
value of the galaxy surface density and phase space density
(iii) the cored profiles of galaxy density profiles seen in
astronomical observations.

Heavier DM particles (as wimps in the GeV mass scale) do not
reproduce the above important galaxy observations and run into
growing and growing serious problems (they produce satellites problem,
voids problem, galaxy size problem, unobserved density cusps and other
problems). 

Minimal extensions of the Standard Model of particle physics 
include keV sterile neutrinos which are very weakly coupled to the standard model particles
and are produced via the oscillation of the light (eV) active neutrinos, with their mixing
angle governing the amount of generated WDM. The mixing angle theta between active and sterile neutrinos
should be in the $ 10^{-4} $ scale to reproduce the average DM density in the
Universe.\\  
Sterile neutrinos are necessarily produced out of thermal equilibrium. 
The production can be non-resonant (in the absence of lepton asymmetries) 
or resonantly ennhanced (if lepton asymmetries are present). The usual X ray bound together
with the Lyman alpha bound forbids the non-resonant mechanism in the $\nu$MSM model.}

\bigskip
 
\item{Warm Dark Matter particles feature a non-vanishing velocity dispersion which leads to a 
cutoff in the matter power spectrum as a consequence of free streaming. 
The essential ingredient is the distribution function of the WDM candidate
solution of the collisionless Boltzmann equation from which the transfer function and power spectra 
are obtained. A WDM particle decoupled ultrarelativistic from the cosmological plasma has three evolution stages :the radiation dominated stages, when the particle is ultrarelativistic and then after when is non-relativistic, and the matter dominated stage. WDM sterile neutrinos produced non-resonantly via two different mechanisms yield very different distribution functions and power spectra which show two characteristic features: (i) a quasidegeneracy between the distribution function and the mass of the WDM particle (a less massive WDM candidate with a distribution function favoring low momenta has a power spectrum similar to that of a more massive particle but with a near thermal distribution function),(ii) the emergence of warm dark matter oscillations at scales of the order of the free streaming scale. Feature (i) suggests that constraints on the mass from the Lyman-$\alpha$ must be taken with a caveat since a reliable assessment requires knowledge of the distribution function, and feature (ii) suggests that typical power laws fits to the power spectra do not reliable describe small scales. Non-vanishing velocity dispersion and free streaming lead to a redshift dependence of the matter and velocity power spectra which affects peculiar velocities: Using the $z=0$ power spectra for N-body simulations with initial conditions at large $z$ incur in substantial errors underestimating the power spectrum of peculiar velocities.

\medskip

Sterile neutrinos with mass in the $\sim \,\mathrm{keV}$ range are suitable warm dark matter 
candidates. These neutrinos can decay into an active-like neutrino and an X-ray photon. 
Abundance and phase 
space density of dwarf spheroidal galaxies constrain the mass to be in the 
$ \sim $ keV range.  
Small scale aspects of sterile neutrinos and different mechanisms of their production 
were presented: 
The transfer function and power spectra are obtained by solving the 
collisionless Boltzmann equation 
during the radiation and matter dominated eras: as a consequence,  
the power spectra features new WDM acoustic
oscillations on mass scales $ \sim 10^8-10^{9} \, M_{\odot} $.} 

\bigskip

\item{A right-handed neutrino of a mass of a few keV appears as the most interesting candidate to 
constitute dark matter. A consequence should be Lyman alpha emission and absorption at around a few 
microns; corresponding emission and absorption lines might be visible from molecular Hydrogen H$_2$  
and H$_3$  and their ions, in the far infrared and sub-mm wavelength range.  The detection at very 
high redshift of massive star formation, stellar evolution and the formation 
of the first super-massive black holes would constitute the most striking and testable prediction of 
this dark matter particle. This particle allows star formation very early, near redshift 80, 
and so also allows the formation of supermassive black holes by agglomerating massive stars; here 
merging starts at masses of a few million solar masses, ten percent of the baryonic mass in the 
initial DM clumps.  This readily explains the supermassive black hole mass function.  
The formation of the first super-massive stars might be detectable among the point-source 
contributions to the fluctuations of the MWBG at very high wave-number (Atacama); their contribution 
is independent of $z$.  The corresponding gravitational waves are not constrained by any existing 
limit, and could have given a substantial energy contribution at high $z$.}

\bigskip

\item{Dark matter may be possibly visible (indirectly) via decay in odd properties of energetic particles
 and photons:  The following discoveries have been reported:  
(i) an upturn in the CR-positron fraction, (ii) 
an upturn in the CR-electron spectrum, iii) a flat radio emission component near the Galactic Center 
(WMAP haze), iv) a corresponding IC component in gamma rays (Fermi haze and Fermi bubble), v) the 
511 keV annihilation line also near the Galactic Center (Integral), and most recently, vi) an 
upturn in the CR-spectra of all elements from Helium (CREAM), with a hint of an upturn for Hydrogen, 
vii)  A flat $\gamma$-spectrum at the Galactic Center (Fermi), and viii) have the complete cosmic ray 
spectrum available through $10^{15}$ to $10^{18}$ eV (KASCADE-Grande). All these features can be 
quantitatively well  explained by the action of cosmic rays accelerated in the magnetic winds of very 
massive stars when they explode. This approach does not require any significant free parameters, it is 
older and simpler than adding Wolf Rayet-star supernova CR-contributions with pulsar wind nebula 
CR-contributions, and implies that Cen A is our highest energy physics laboratory accessible to direct 
observations of charged particles. All this allows with the galaxy data to derive the key properties 
of the dark matter particle: this clearly points to a keV mass particle (keV warm dark matter).

\medskip

This particle has the advantage to allow star formation very early, near redshift 80, and so also allows 
the formation of supermassive black holes: they possibly formed out of agglomerating massive stars, in 
the gravitational potential well of the first DM clumps, 
whose mass in turn is determined by the properties 
of the DM particle.  Black holes in turn also merge, but in this manner start their mergers at masses of a 
few million solar masses, about ten percent of the baryonic mass inside the initial dark matter clumps.  
This readily explains the supermassive black hole mass function .  The formation of the first 
super-massive stars might be detectable among the point-source contributions to the fluctuations of the 
MWBG at very high wave-number (Atacama); their contribution is independent of redshift.  
The corresponding gravitational waves are not constrained by any existing limit, and could have
 given a substantial energy contribution at high redshift. }

\bigskip

\item{All the observations of cosmic ray positrons and 
electrons and the like are due to normal astrophysical sources and processes, and do not require 
hypothetical decay or annihilation of heavy DM particles. 
The models of annihilation or decay of cold dark matter (wimps) become increasingly 
tailored and fine tuned to explain these normal astrophysical processes and their ability 
to survive observations is more and more reduced.
Pulsar winds are perfect positron sources to power the positron excess. The spectra inferred 
from observations work fine. The data are expected to improve soon with AMS-02 to confirm that 
the positron excess is as pronounced as shown by Pamela. As a conclusion on this issue it appeared: 
``Let us be careful to get too excited about spectral features (positrons, nuclei,...): Some of these 
features also appear due to fluctuations in the source activity or locations. There is a lot of work 
to be done before we actually figure out the details of CR propagation and acceleration. Excesses 
should be compared with how well we understand such details. This is especially to be kept in mind 
when invoking unconventional explanations to CR excess, 
such as those based on cold dark matter annihilation.
The CDM explanation to the positron excess was not the most natural: The signal from wimps is naturally 
too small but the theory was contrived (leptophilic DM, boost factors, Sommerfeld enhancement, ...) 
for the sole purpose of fitting one set of data (the positron fraction and the absence of 
antiproton anomalies)''.}

\bigskip

\item{Lyman-$\alpha$ constraints
have been often misinterpreted or superficially invoked in the past
to wrongly suggest on a tension with WDM, but those constraints have been by now clarified
and relaxed, and such a tension does not exist: keV sterile neutrino dark matter (WDM) 
is consistent with Lyman-alpha constraints within a 
{\it wide range} of the sterile neutrino model parameters. {\bf Only} for sterile
neutrinos {\bf assuming} a {\bf non-resonant} (Dodelson-Widrow model) 
production mechanism, Lyman-alpha constraints provide a lower bound for the 
mass of about 4 keV. For thermal WDM relics (WDM particles decoupling at
thermal equilibrium) the Lyman-alpha lower particle mass bounds are 
smaller than for non-thermal WDM relics (WDM particles decoupling out of
thermal equilibrium). The number of Milky-Way satellites indicates lower bounds 
between 1 and 13 keV for different models of sterile neutrinos.

\medskip

WDM keV sterile neutrinos can be copiously produced in the supernovae cores. Supernova stringently 
constraints the neutrino mixing angle squared to be $ \lesssim 10^{-9}$ for sterile neutrino masses
$ m > 100$ keV (in order to avoid excessive energy lost) but for smaller sterile neutrino masses the 
SN bound is not so direct. Within the models 
worked out till now, mixing angles are essentially unconstrained by 
SN in the favoured WDM mass range, namely 
$ 1 < m  < 10 $ keV. Mixing between electron and keV sterile neutrinos could help SN explosions, case 
which deserve investigation.}

\bigskip
        
\item {The possibility of laboratory detection of warm dark matter
is extremely interesting. Only a direct detection of the DM particle can give a clear-cut answer
to the nature of DM and at present. At present,  only the {\bf Katrin and Mare experiments}
have the possibility to do that for sterile neutrinos.
{\bf Mare} bounds on sterile neutrinos are placed from the beta decay of Re187 and EC decay of Ho163,
{\bf Mare} keeps collecting data in both. The possibility that {\bf Katrin} experiment can look to 
sterile neutrinos in the tritium decay was discussed.
Katrin experiment have the potentiality to detect warm dark matter if its set-up 
would be adapted to look to keV scale sterile neutrinos.
KATRIN experiment concentrates its attention right now
on the electron spectrum near its end-point
since its goal is to measure the active neutrino mass.
Sterile neutrinos in the tritium decay will affect the electron
kinematics at an energy about $m$ below the end-point of the
spectrum ($m$ = sterile neutrinos mass). KATRIN in the future
could perhaps adapt its set-up to look to keV scale sterile neutrinos.
It will be a a fantastic discovery to detect dark matter
in a beta decay.}

\bigskip

\item{Astronomical observations strongly indicate that
{\bf dark matter halos are cored till scales below 1 kpc}. 
More precisely, the measured cores {\bf are not} hidden cusps.
CDM Numerical simulations -with wimps (particles heavier than $ 1 $ GeV)-
without {\bf and} with baryons yield cusped dark matter halos.
Adding baryons do not alleviate the problems of wimps (CDM) simulations,
on the contrary adiabatic contraction increases the central density of cups
worsening the discrepancies with astronomical observations. 
In order to transform the CDM cusps into cores, the baryon+CDM simulations
need to introduce strong baryon and supernovae feedback which produces a
large star formation rate contradicting the observations.
None of the predictions of CDM simulations at small scales
(cusps, substructures, ...) have been observed. The discrepancies of CDM 
simulations with the astronomical observations at small scales $ \lesssim 100 $ kpc {\bf is staggering}: 
satellite problem (for example, only 1/3 of satellites predicted by CDM
simulations around our galaxy are observed), the surface density problem (the value 
obtained in CDM simulations is 1000 times larger than the
observed galaxy surface density value),  the voids problem, size problem 
(CDM simulations produce too small galaxies).}

\bigskip

\item{The use of keV scale WDM particles in the simulations instead of the GeV CDM wimps, alleviate all the
above problems. For the core-cusp problem, setting
the velocity dispersion of keV scale DM particles seems beyond
the present resolution of computer simulations. 
However, the velocity dispersion is negligible
for $ z < 20 $ where the non-linear regime and structure formation starts.
Analytic work in the
linear approximation produces cored profiles for keV scale DM particles
and cusped profiles for CDM. Model-independent analysis of DM from phase-space density
and surface density observational data plus theoretical analysis
points to a DM particle mass in the keV scale.
The dark matter particle candidates with high mass (100 GeV, `wimps') 
are strongly disfavored, while cored (non cusped) dark matter halos and warm (keV scale mass) 
dark matter are strongly favoured from theory and astrophysical observations.
As a conclusion, the dark matter particle candidates with large mass 
($ \sim 100$ GeV, the so called `wimps') are strongly disfavored,
while light (keV scale mass) 
dark matter are being increasingly favoured both from theory, numerical 
simulations and a wide set of astrophysical observations.}

\bigskip

\item{Recent $\Lambda$WDM N-body simulations have been performed by different groups. 
High resolution simulations for different types of DM (HDM, WDM or CDM), allow to visualize the effects of 
the mass of the corresponding DM particles: free-streaming lentgh scale, initial velocities and associated 
phase space density properties: for masses in the eV scale (HDM), halo formation occurs top down on all scales 
with the most massive haloes collapsing first; if primordial velocities are large enough, 
free streaming erases 
all perturbations and  haloes  cannot form (HDM). The concentration-mass halo relation for mass of
 hundreds eV  is reversed with respect to that found for
CDM wimps of GeV mass. For realistic keV WDM these simulations deserve investigation: 
it could be expected from 
these HDM and CDM effects that combined free-streaming and velocity effects in keV WDM 
simulations could produce 
a bottom-up hierarchical scenario with the right amount of sub-structures (and some scale at 
which transition 
from top-down to bottom up regime is visualized).

\medskip

Moreover, interestingly enough, recent large high resolution $\Lambda$WDM N-body simulations allow to 
discriminate among thermal and non-thermal WDM (sterile neutrinos): Unlike conventional thermal relics, 
non-thermal WDM has a peculiar velocity distribution (a little skewed to low velocities) which translates 
into a characteristic linear matter power spectrum decreasing slowler across the cut-off free-streaming scale 
than the thermal WDM spectrum. As a consequence, the  radial distribution of the subhalos predicted by WDM 
sterile neutrinos remarkably reproduces the observed distribution of Milky Way satellites in the range above 
$\sim 40$ kpc, while the thermal WDM supresses subgalactic structures perhaps too much, by a factor $2-4$ 
than the observation. Both simulations were performed for a mass equal to 1 keV. Simulations 
for a mass larger than 1 keV  (in the range between 2 and 10 keV, say) should still improve these results.}

\bigskip

\item{The effect of keV WDM can be also observable in the statistical 
properties of cosmological Large Scale Structure. Cosmic shear 
(weak gravitational lensing) does not strongly depend on baryonic 
physics and is a promising probe. First results in a simple thermal 
relic scenario indicate that future weak lensing surveys could see a WDM 
signal for $m_{WDM} \sim 2$ keV or smaller. The predicted limit beyond 
which these surveys will not see a WDM signal is $m_{WDM} \sim 2.5$ keV 
(thermal relic) for combined 
Euclid + Planck. More realistic models deserve investigation and are 
expected to relaxe such minimal bound. With the real data, the 
non-linear WDM model should be taken into account.

\medskip

The predicted galaxy distribution in the local universe for $\Lambda$WDM (ie,
CLUES numerical simulations with warm dark matter
of mass of $ m_{\rm WDM}=1 $ keV) well reproduces 
the observed one in the ALFALFA survey. On the
contrary, the $\Lambda$CDM model predicts a steep rise in the velocity
function towards low velocities and thus forecasts much  more
sources both in Virgo-direction as well as in anti-Virgo-direction
than the ones observed by the ALFALFA survey. These results add 
problems for CDM, also the spectrum of mini-voids points to  
a problem of the $\Lambda$CDM model. The $\Lambda$WDM model
provides a natural solution to this problem.}

\bigskip

\item{An `Universal Rotation Curve' (URC) of spiral galaxies emerged from  
3200 individual observed Rotation Curves (RCs) and 
reproduces remarkably well out to the virial radius the Rotation Curve of 
any spiral galaxy. The URC is the observational counterpart of the  circular  
velocity profile from cosmological  simulations.
CDM numerical simulations give the NFW cuspy halo profile. A careful analysis 
from about 100 observed high quality rotation curves has now {\bf ruled out} the disk + NFW halo 
mass model, in favor of {\bf cored profiles}. 
The observed galaxy surface density (surface gravity acceleration) appears to be universal within 
$ \sim 10 \% $ with values around $ 100 \; M_{\odot}/{\rm pc}^2 $
, irrespective of galaxy morphology, luminosity and Hubble types, spanning over 14 magnitudes in 
luminosity and mass profiles determined by several independent methods.}

\bigskip

\item{Interestingly enough, a constant surface density  (in this case column density) with value around 
$ 120 \; M_{\odot}/{\rm pc}^2 $ similar to that found for galaxy systems  is found too for the interstellar 
molecular clouds, irrespective of size and compositions over six order of magnitude; this universal surface 
density in molecular clouds
is a consequence of the Larson scaling laws. This suggests the role of gravity on matter (whatever DM or 
baryonic) as a dominant underlying mechanism to produce such universal surface density 
in galaxies and molecular 
clouds. Recent re-examination of different and independent (mostly millimeter) molecular cloud data sets show 
that interestellar clouds do follow Larson law $Mass \sim (Size)^{2}$ exquisitely well, and therefore very 
similar projected mass densities at each extinction threshold. Such scaling and universality should play a key 
role in cloud structure formation.}

\bigskip

\item{A key part of any galaxy formation process and evolution involves dark matter. Cold gas accretion and 
mergers became important ingredients of the CDM models but they have little observational evidence. 
DM properties 
and its correlation with stellar masses are measured today up to $z =2$; at $z > 2$ observations are much less 
certain. Using kinematics and star formation rates, all types of masses -gaseous, stellar and dark- 
are measured 
now up to $z =1.4$. The DM density within galaxies declines at higher redshifts.  Star formation is observed 
to be more common in the past than today. More passive galaxies are in more 
massive DM halos, namely most massive DM halos have lowest fraction of stellar mass. CDM predicts high
overabundance of structure today and under-abundance of structure in the past with respect to observations. 
The size-luminosity scaling relation is the tightest of all purely photometric correlations used to 
characterize galaxies; its environmental dependence have been highly debated but recent findings show that the 
size-luminosity relation of nearby elliptical galaxies is well defined by a fundamental line and is 
environmental 
independent. Observed structural properties of elliptical galaxies appear simple and with no environmental 
dependence, showing that their growth via important mergers -as required by CDM galaxy formation- is not 
plausible. Moreover, observations in brightest cluster galaxies (BCGs) show little changes in the sizes of 
most massive galaxies since $z =1$ and this scale-size evolution appears closer to that of radio galaxies over 
a similar epoch. This lack of size growth evolution, a lack of BCG stellar mass evolution is observed too, 
demonstrates that major merging is not an important process. Again, these observations put in serious trubble 
CDM `semianalytical models' of BCG evolution which require about $70\%$ of the final BCG stellar mass to be 
accreted in the evolution and important growth factors in size of massive elliptical massive galaxies.}

\bigskip

\item{The sun seismic model allows a prediction of the gravity mode frequencies that are in very good agreement with the first observation of dipole gravity modes. Their period differences are smaller than 0.3 minute. It also produces results in agreement with the detection of all the different neutrino installations . So the central temperature, central density and core density profile of the Sun must be very near from those of the seismic model predictions, even  the detailed physics of the radiative zone is not totally under control: possible existence of a fossil field, possible bad determination of the transfer of energy, possible effect of dark matter. The new solar detections cannot  constrain  cold dark matter annihilation models as the estimated cross section is too small to produce any effect on the density profile. The presence of  non-annihilating WIMPs in the Sun for masses $\le$ 10 GeV  and spin dependent cross sections $> 5 \,10^{-36}$ cm$^2$ is excluded. No signature is visible in the obtained profile in the solar core. The seismic model has a central temperature higher than the standard solar model (SSM),   (T$_C$= 15.51, SSM neutrino prediction is not compatible with boron neutrino detection),  but a density slightly smaller by 2 or 3\%. This remark could favor other kinds of processes like a different gravitational effect due to WDM (sterile neutrinos ?) acting on the whole radiative zone.  The radiative zone is certainly more complex than thought previously: a good energetic balance and all the potential dynamical effects need to be included before any conclusion on this subject.
In the coming years helio- and asteroseismology will provide complementary  results  to extend these studies and to check these probes of dark matter.}

\bigskip

\item{ {\it As an overall conclusion}, CDM represents the past and WDM represents the future in the DM research. 20 year old CDM research, namely: CDM simulations and their proposed baryonic solutions, 
and the CDM wimps candidates 
($ \sim 100$ GeV, the so called `wimps') are strongly pointed out by the galaxy observations as the {\it wrong} solution to DM. Theoretically, and placed in perspective after more than 20 years, the reason why CDM does not work appears simple and clear to understand and directly linked to the excesively heavy and slow CDM wimp, which determines an excessively small (for astrophysical structures) free streaming length, and unrealistic overstructure at these scales.  On the contrary, new keV WDM research, keV WDM simulations, and keV scale mass  
WDM particles are strongly favoured by galaxy observations and theoretical analysis, they naturally {\it work} and agree with the astrophysical observations at {\it all} scales, (galactic as well as cosmological scales). Theoretically, the reason why WDM works  so well
is clear and simple, directly linked to the keV scale mass and velocities of the WDM particles,
and free-streaming length. The experimental search for serious WDM particle candidates (sterile neutrinos) appears urgent and important: it will be a fantastic discovery to detect dark matter in a beta decay. There is a formidable WDM work to perform ahead of us, these highlights point some of the directions where it is worthwhile to put the effort.}

\end{itemize}

\subsection{The present context and future in Dark Matter and Galaxy Formation research.}

\begin{itemize}
\item{Facts and status of DM research: Astrophysical observations
point to the existence of DM. Despite of that, proposals to
replace DM  by modifing  the laws of physics did appeared, however
notice that modifying gravity spoils the standard model of cosmology
and particle physics not providing an alternative.
After more than twenty active years the subject of DM is mature, (many people is involved in this problem, 
different groups perform N-body cosmological simulations and on the other hand direct experimental particle 
searches are performed by different groups, an important
number of conferences on DM and related subjects is held regularly). DM research  
appears mainly in three sets:
(a) Particle physics DM model building beyond the standard
model of particle physics, dedicated laboratory experiments,
annhilating DM, all concentrated on CDM and CDM wimps.
(b) Astrophysical DM: astronomical observations, astrophysical models.
(c) Numerical CDM simulations. The results of (a) and (b)
do not agree and (b) and (c) do not agree neither at small scales.
None of the small scale predictions of CDM simulations 
have been observed: cusps and over abundance of substructures
differ by a huge factor with respect to those observed.
In addition, all direct {\it dedicated} searchs of CDM wimps from more than twenty years 
gave {\it null results}. {\it Something is going wrong in the CDM research and the right answer is: 
the nature of DM is not cold (GeV scale) but warm (keV scale)}.}

\bigskip

\item{Many researchers continue to work with heavy CDM candidates
(mass $ \gtrsim 1 $ GeV) despite the {\bf staggering} evidence that these
CDM particles do not reproduce the small scale astronomical observations
($ \lesssim 100 $ kpc). Why? [It is known now that the keV scale DM particles naturally produce the
observed small scale structure]. Such strategic
question is present in many discussions, everyday and off of the record (and on the record) talks in the field. 
The answer deals in large part with the inertia (material, intellectual, social, other, ...) that structured research 
and big-sized science in general do have, which involve huge number of people, huge budgets, longtime planned 
experiments, and the `power' (and the conservation of power) such situation could allow to some of 
the research lines following the trend; as long as budgets will allow to run wimp experimental searches and CDM simulations 
such research lines could not deeply change, although they would progressively decline.

\medskip

Notice that in most of the DM litterature or conferences, wimps are still `granted' as ``the'' DM particle, 
and CDM as ``the'' DM; is only recently that the differences and clarifications are being clearly recognized 
and acknowledged. While wimps were a testable hypothesis at the beginning of the CDM research, 
one could ask oneself why they continue to be worked out and `searched' experimentally in spite 
of the strong astronomical and astrophysical  evidence against them. \\
Similar situations (although do not such extremal as the CDM situation) happened in other branches of 
physics and cosmology:
Before the CMB anisotropy observations, the issue of structure formation was plugged with several alternative 
proposals which were afterwards ruled out. Also, string theory passed from being considered "the theory of 
everything" to "the theory of nothing" (as a physical theory), as no physical experimental evidence have been 
obtained and its cosmological implementation and predictions desagree with observations. (In despite of all 
that, papers on such proposals continue -and probably will continue- to appear. 
But is clear that  big dedicated 
experiments are not planned or built to test such papers).  In science, what is today `popular' 
can be discarded 
afterwards; what is today `new' and minoritary can becomes `standard' and majoritarily accepted if verified 
experimentally. }

\end{itemize}

\bigskip

\begin{center}

 {\bf \em ``Things are beginning to hang together, and we can now make quite specific 
predictions as a consequence of the keV DM model.  
If the right-handed neutrino were this particle, star formation and the first super-massive 
black holes could be formed quite early, possibly earlier than redshift 50.  
A confirmation would be spectacular.''}

\medskip

[Peter Biermann, his conclusion of the 14th Paris Cosmology Colloquium Chalonge 2010 in
 `Live minutes of the Colloquium',  arXiv:1009.3494].  

\bigskip

 {\bf \em  ``Examine the objects as they are and you will see their true nature;
look at them from your own ego and you will see only your feelings;
because nature is neutral, while your feelings are only prejudice and obscurity.''}

\medskip

[Gerry Gilmore quoting Shao Yong, 1011-1077 in the 14th Paris Cosmology Colloquium Chalonge 2010
http://chalonge.obspm.fr/Programme\_Paris2010.html, arXiv:1009.3494].

\medskip

 {\bf \em ``Let us be careful to get too excited about spectral features (positrons, nuclei,...): 
This is especially to be kept in mind when invoking unconventional explanations to CR excess, 
such as those based on cold dark matter annihilation.
The CDM explanation to the positron excess was not the most natural: The signal from wimps is naturally 
too small but the theory was contrived (leptophilic DM, boost factors, Sommerfeld enhancement, ...) 
for the sole purpose of fitting one set of data (the positron fraction and the absence of 
antiproton anomalies).''}

\medskip

[Pasquale Blasi, his conclusion in his Lecture at the  15th Paris Cosmology Colloquium Chalonge 2011.]

\end{center}

\bigskip

\begin{center}

The Lectures of the Colloquium can be found at:

\bigskip

{\bf http://chalonge.obspm.fr/Programme$_{}$Paris2011.html}

\bigskip

The photos of the Colloquium can be found at:

\bigskip

{\bf http://chalonge.obspm.fr/album2011/album/index.html}

\bigskip

The photos of the Open Session of the Colloquium can be found at:

\bigskip

{\bf http://chalonge.obspm.fr/albumopensession2011/index.html}

\end{center}

\bigskip

Best congratulations and acknowledgements to all lectures and participants which 
made the 15th Paris Cosmology Colloquium 2011 so fruitful and interesting, the 
Ecole d'Astrophysique Daniel Chalonge looks forward to you for the next Colloquium of this series:

\begin{center}

The 16th Paris Cosmology Colloquium 2012 devoted to 

\bigskip

THE NEW STANDARD MODEL OF THE UNIVERSE: LAMBDA WARM DARK MATTER ($\Lambda$WDM) THEORY AND OBSERVATIONS

\bigskip

Observatoire de Paris, historic Perrault building,  25, 26, 27 JULY 2012.

\bigskip

http://www.chalonge.obspm.fr/colloque2012.html

\end{center}

\newpage

\section{Photos of the Colloquium}

\bigskip

Photos of the Colloquium are available at:

\bigskip

http://www.chalonge.obspm.fr/colloque2011.html

\bigskip

\begin{figure}[ht]
\includegraphics[height=14cm,width=18cm]{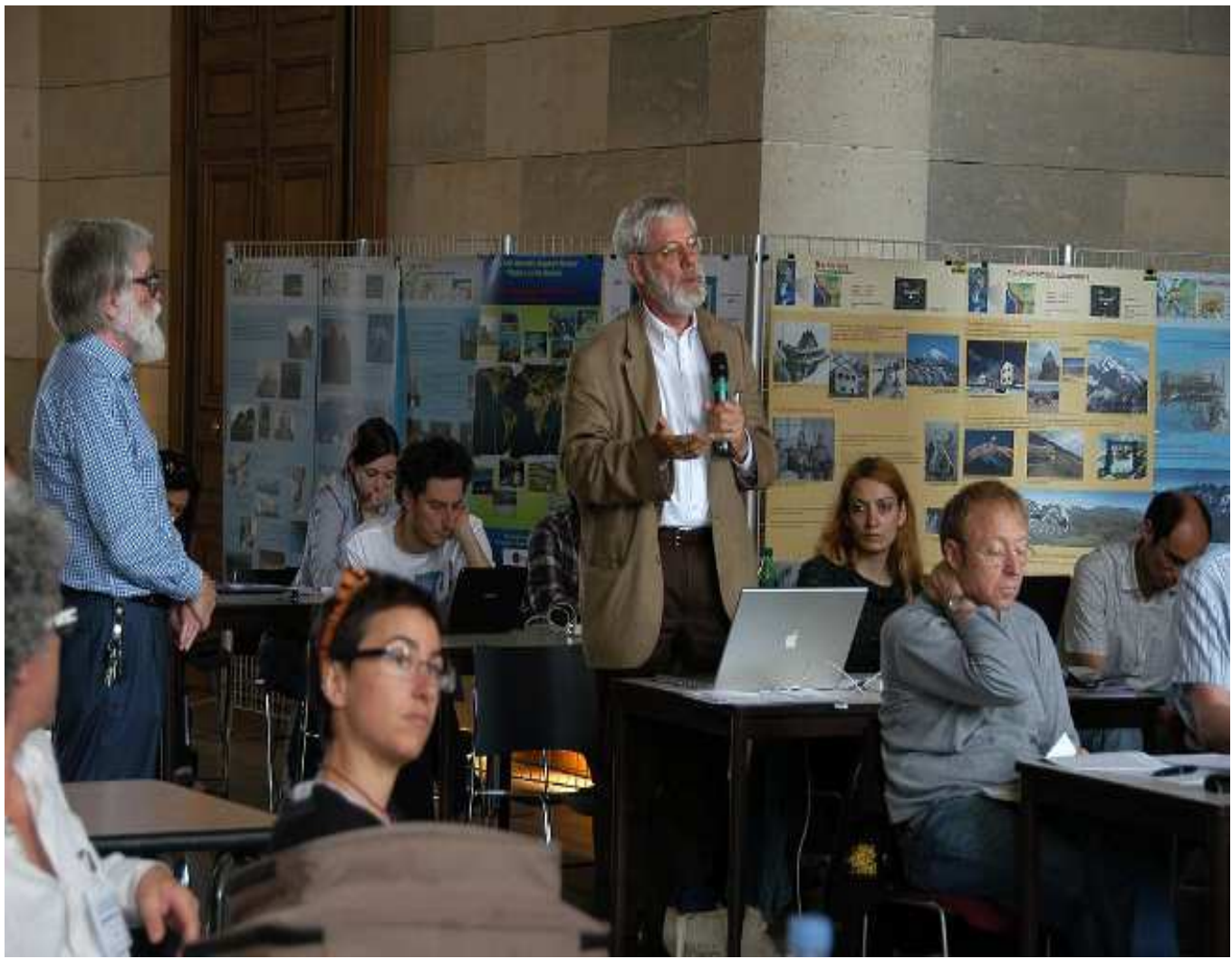}
\caption{Discussion time during a session of the  Chalonge School atthe  Salle Cassini}
\end{figure}

\begin{figure}[ht]
\includegraphics[scale=.9]{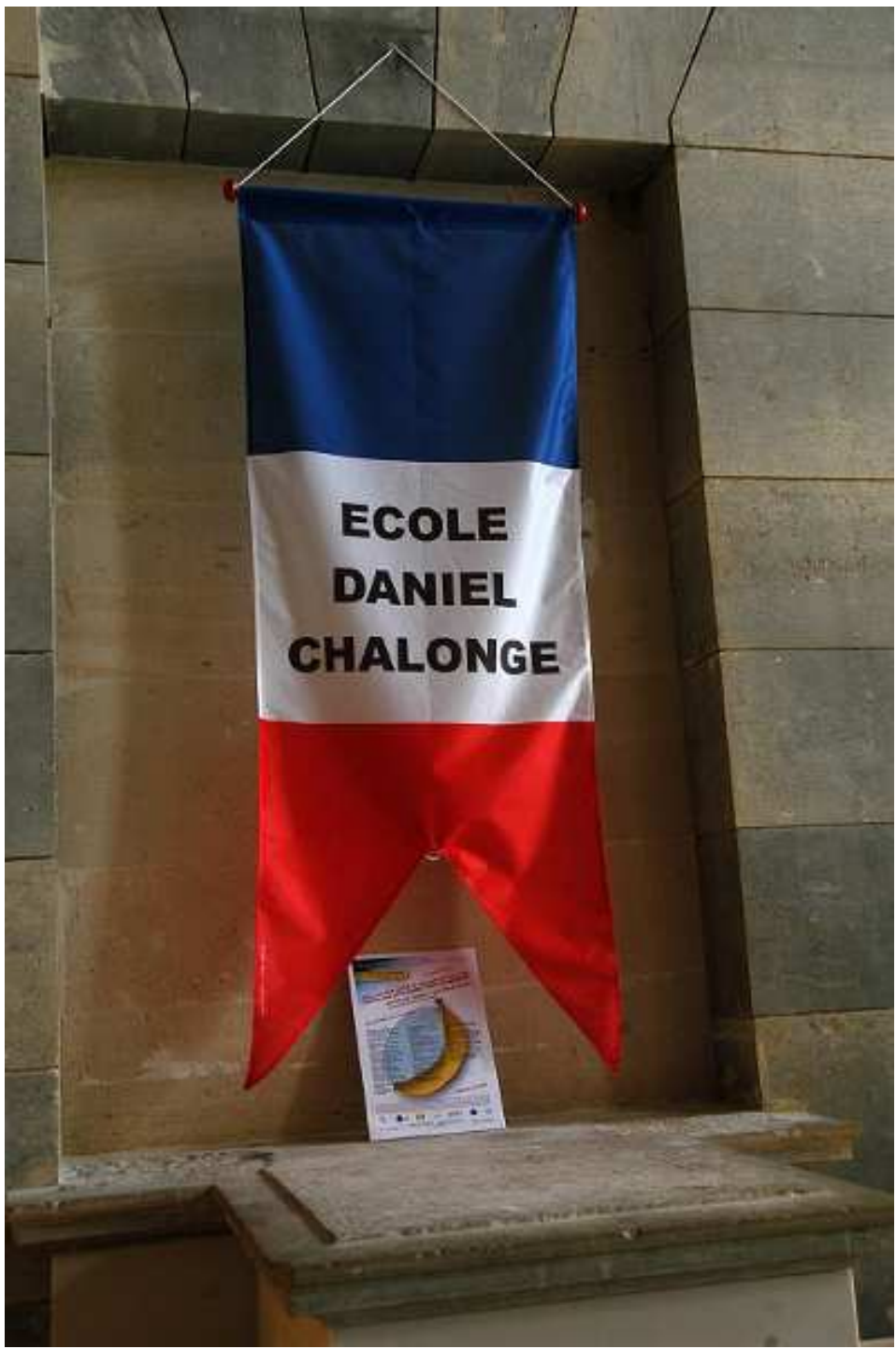}
\caption{At the Salle Cassini}
\end{figure}

\begin{figure}[ht]
\includegraphics[height=14cm,width=18cm]{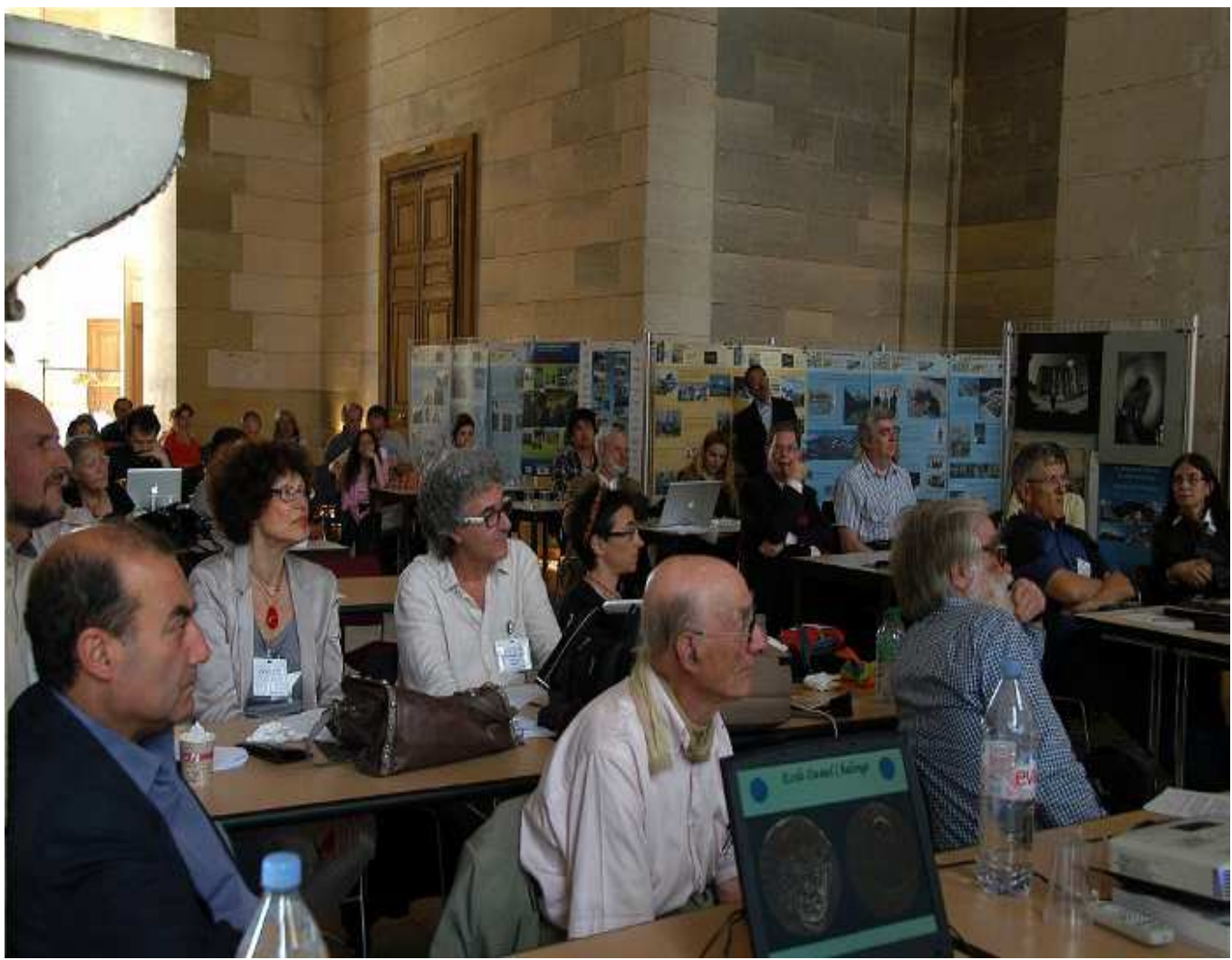}
\caption{Open Session of the Chalonge School at the Salle Cassini}
\end{figure}

\begin{figure}[ht]
\includegraphics[height=12cm,width=17cm]{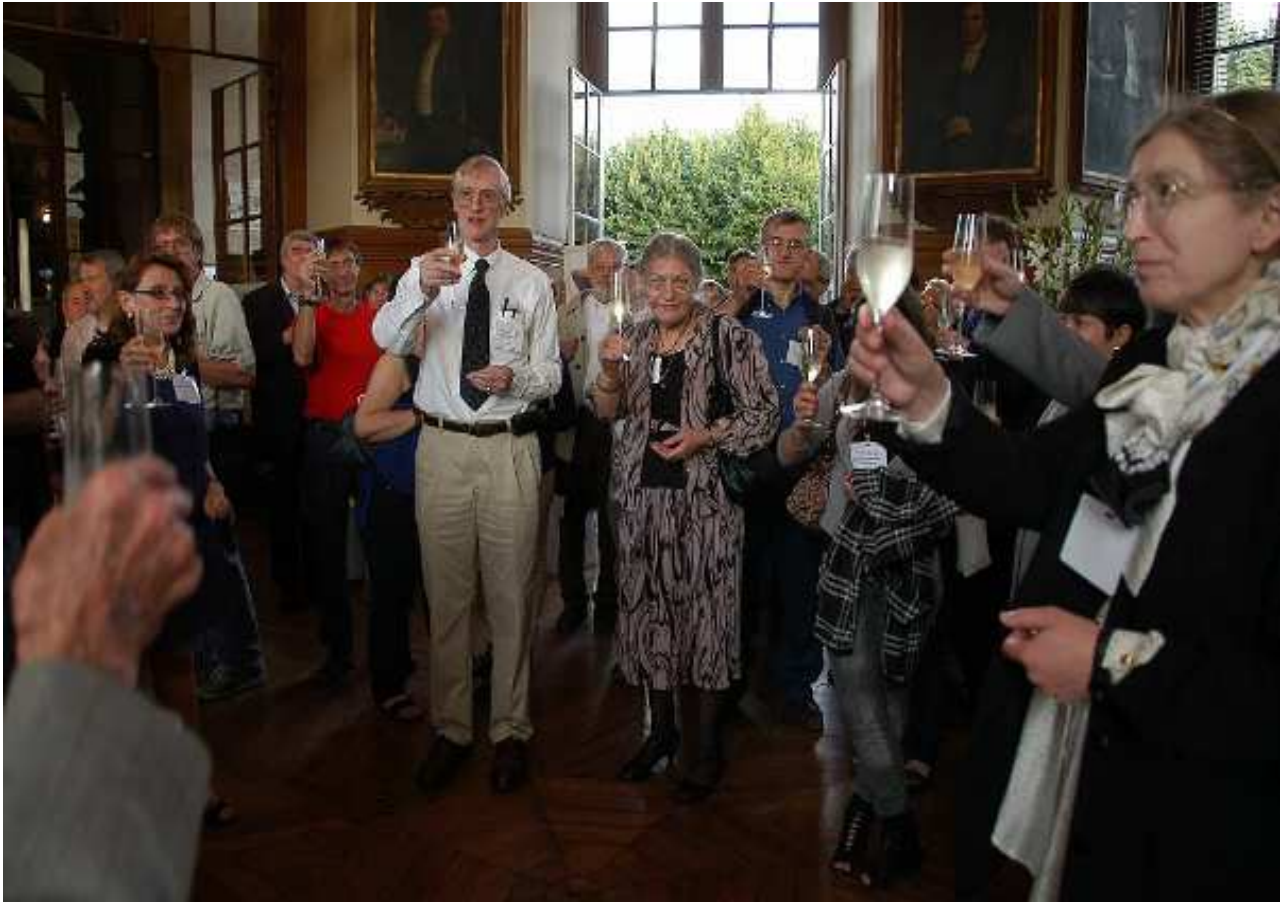}
\caption{Champagne at the Salle du Conseil}
\end{figure}

\begin{figure}[ht]
\includegraphics[height=12cm,width=17cm]{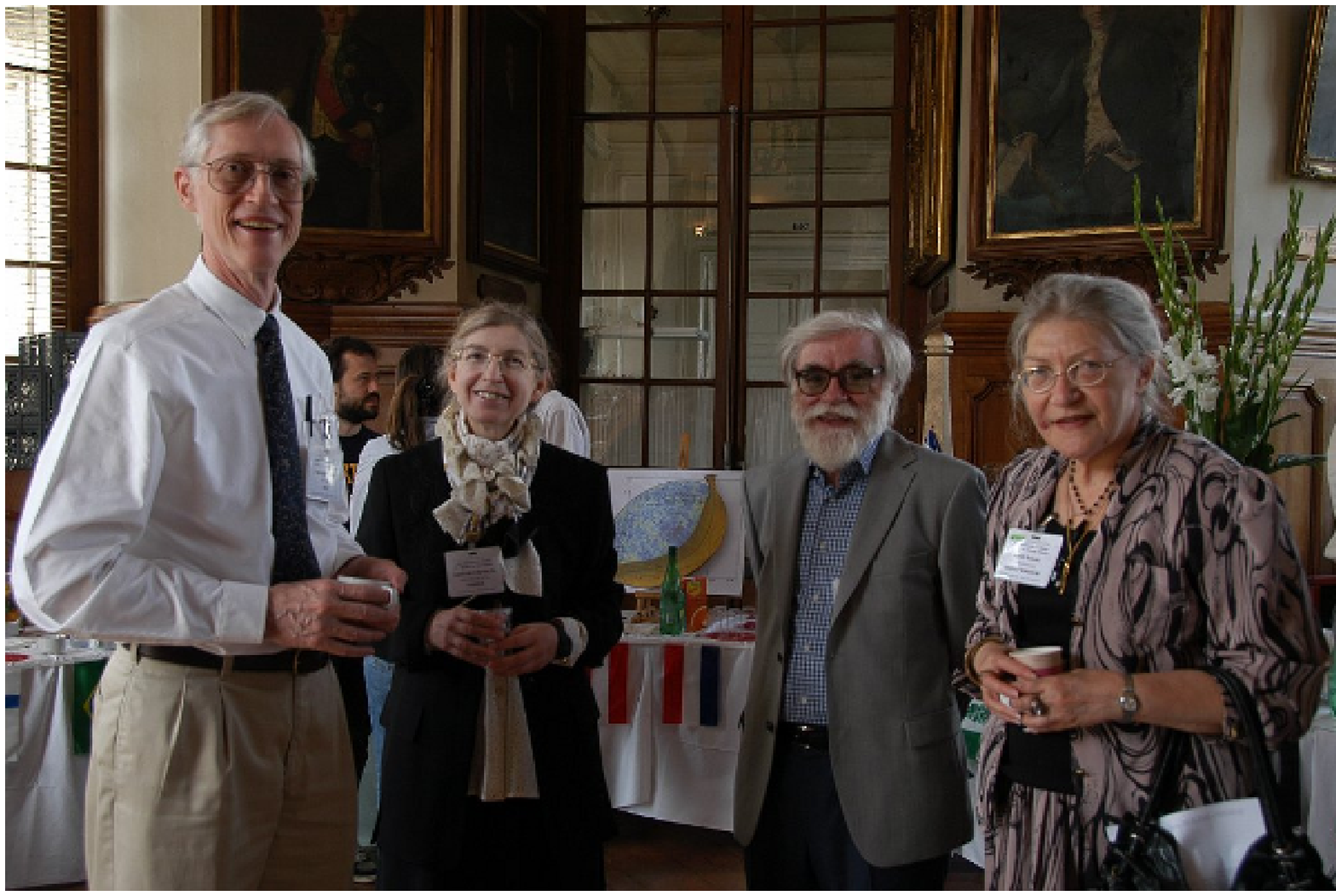}
\caption{Coffee break at the Salle du Conseil}
\end{figure}

\newpage

\section{List of Participants}

\bigskip

\noindent
\medskip
ALEXANDROVA  Alexandra,	Karl-Francens University Graz,	Graz,	Austria \\
\medskip
AMES  Susan,  Univ of Oxford, Astrophysics Group,	 Oxford, 	England\\
\medskip
ANDRE  Philippe,	CEA / DSM / IRFU, Saclay, 	Gif-sur-Yvette, 	France\\
\medskip
ANTOLINI  Claudia,	SISSA/ISAS,	Trieste, 	Italy\\
\medskip				
ARKHIPOVA Natalia, Astro Space Center of the Lebedev Physical Institute, Moscow, Russia\\
\medskip
BIERMANN Peter,	MPIfR Bonn, Germany \& Univ of Alabama, Tuscaloosa, AL,  USA\\
\medskip
BLASI	 Pasquale,	INAF/Arcetri Astrophysical Observatory,	Firenze,	Italy\\
\medskip
BOYANOVSKY  Daniel,	University of Pittsburgh, Dept of Physics \& Astron	
Pittsburgh, PA,	USA\\
\medskip
Dr Marcela Madrid BOYANOVSKY, Pittsburgh, PA,	USA\\
\medskip
Miss Natalie BOYANOVSKY, Pittsburgh, PA,	USA \\
\medskip
BOZEK Brandon,	Johns Hopkins University,	Baltimore,	USA\\
\medskip
BURGOA  ROSSO, Karen Luz	Universidade Federal de Lavras  UFLA,	Lavras-MG Brazil\\
\medskip
BURIGANA Carlo,	INAF-IASF Bologna, 	Bologna,	Italy\\
\medskip
CAPISTRANO Abraao,	Federal University of Tocantins	Palmas,	Brazil\\
\medskip
CHAUDHURY Soumini,	Saha Institute Of Nuclear Physics	Kolkata,	India\\
\medskip
CHINAGLIA Mariana,	Universidade Federal de São Carlos (UFSCar),	Sào Carlos,	Brazil\\
\medskip
CNUDDE Sylvain,	LESIA Observatoire de Paris,	Meudon,	France\\
\medskip
COORAY Asantha,	University of California,Irvine	Irvine, CA	USA\\
\medskip
CRUCIANI Angelo,	Università La Sapienza - Institut Neel	Roma - Grenoble	Italia - France\\
\medskip
DAS Sudeep,	University of California, Berkeley,	United States\\
\medskip
DESTRI Claudio,	INFN Dipt di Fisica G. Occhialini, Univ Milano Bicocca,	Milano,	Italy\\
\medskip
DE VEGA Héctor J.,	CNRS LPTHE UPMC Paris,	Paris	France\\
\medskip
DUNKLEY	Joanna,	Oxford University, Astrophysics Group, Oxford,	United Kingdom\\
\medskip	 	 	 	 
DVOEGLAZOV Valeriy,	Universidad de Zacatecas,	Zacatecas,	Mexico\\
\medskip
ERRARD Josquin,	Lab.Astroparticule and Cosmologie APC,	Paris,	France\\
\medskip
FABBIAN Giulio,	Lab.Astroparticule and Cosmologie APC,	Paris,	France\\
\medskip			
FALVELLA Maria Cristina, Italian Space Agency,	Rome,	Italy\\
\medskip
FEDOROV Nikolay,	Lebedev Physical Institute,	Moscow,	Russia\\
\medskip
GILMORE Gerard F., Institute of Astronomy, Cambridge Univ.,	Cambridge,	United Kingdom\\
\medskip
IVANOV Mikhail,	Central Institute of Aviation Motors,	 Moscow,	Russia\\
\medskip
IVASHCHENKO	Ganna,	Astronomical Observatory of the Taras Shevchenko N	Kyiv,	Ukraine\\
\medskip
JOURNEAU Philippe,	Discinnet Labs,	Boulogne 92100,	France\\
\medskip
KAMADA	Ayuki,	Institute for the Physics and Mathematics of the Universe	Kashiwa,	Japan\\
\medskip
KARCZEWSKA Danuta,	University of Silesia,	Katowice,	Poland\\
\medskip
KASHLINSKY Alexander,	NASA Goddard Space Flight Center,	Greenbelt, MD	USA\\
\medskip
KIRILOVA	Daniela,	Institute of Astronomy, Bulgarian Academy of Sciences,	Sofia,	Bulgaria\\
\medskip	 	 	 	 
KOGUT Alan,	NASA Goddard Space Flight Center,	Greenbelt, MD	USA\\
\medskip
KOPTYEVA Olena,	Dniepropetrovsk National University,	Dniepropetrovsk, Ukraine\\
\medskip
KUCUK Hilal,	UCL- University College London, 	London,	UK\\
\medskip
LASENBY Anthony N.,		Cavendish Laboratory, Univ. of Cambridge,	Cambridge	United Kingdom\\
\medskip
Dr Joan LASENBY, Cambridge,	United Kingdom\\
\medskip
LETOURNEUR Nicole,	CIAS Observatoire de Paris, Meudon, France\\
\medskip
LINDNER Manfred,	Max-Planck-Institut f\"ur Kernphysik, HD	Heidelberg,	Germany\\
\medskip
MACHADO Ulisses,	Universidade de S\~ao Paulo- USP,	S\~ao Paulo,	Brazil\\
\medskip
MATHER John,	NASA GSFC	Greenbelt, MD	USA\\
\medskip
MAZEVET Stephane,	LUTH Observatoire de Paris,	Paris,	France\\
\medskip
MIRABEL F\'elix,	CEA-Saclay \& IAFE-Buenos Aires,	Gif-sur-Yvette,	France\\
\medskip				
MOHANTY Subhendra,	Physical Research Laboratory,	Ahmedabad,	India\\
\medskip
NOGALES VERA Jose Alberto C.,	Universidade Federal de Lavras - UFLA,	Lavras-MG,	Brazil\\
\medskip
NOH	Hyerim, Korea Astronomy and Space Science, Institute	Taejon,	South Korea\\
\medskip	 	 	 	 
NOVIELLO	Fabio,	Institut d'Astrophysique Spatiale,	Orsay,	France\\
\medskip
PADUROIU Sinziana,	Geneva Observatory, University of Geneva,	Geneva,	Switzerland\\
\medskip
PFENNIGER Daniel,	University of Geneva, Geneva Observatory,	Sauverny,	Switzerland\\
\medskip
RAMON MEDRANO Marina,	Univ. Complutense,	Madrid,	Spain\\
\medskip
RECOULES Vanina,	CEA-DAM-DIF,	Bruyeres Le Chatel,	France\\
\medskip
ROCHUS Pierre,	Centre Spatial CSL et Univ  de Li\`ege,	Li\`ege,	BELGIUM\\
\medskip
SANCHEZ Norma G.,	CNRS LERMA Observatoire de Paris,	Paris,	France\\
\medskip
SANTOS Jonas,	Federal University of S\~ao Carlos,	S\~ao Carlos,	Brazil\\
\medskip
SHARINA	Margarita,	Special Astrophysical Observatory, Russian Academy,	Nizhnij Arkhyz,	Russia\\
\medskip
SIELLEZ	Karelle,	Student IN2P3 and Observatoire de Paris- Meudon,	France\\
\medskip	 	 	 	 
SMIRNOV Alexei,	Abdus Salam ICTP,	Trieste,	Italy\\
\medskip
Mrs SMIRNOV, Trieste,	Italy \\
\medskip
SMOOT George F., BCCP/LBL Berkeley, IEU Seoul, U Paris Diderot,	Berkeley/Seoul, USA\\
\medskip				
STEFANESCU Petruta,	Institute for Space Science,	Bucharest,	Romania\\
\medskip
TEDESCO Luigi,	Dipartimento di Fisica di Bari \& INFN di Bari,	Bari,	Italy\\
\medskip
TIGRAK	Esra,	The Kapteyn Astronomical Institute, Rijksuniversit	Groningen,	 Netherlands\\
\medskip
TROMBETTI	Tiziana,	Univ. Roma La Sapienza / INAF-IASFBO,	Bologna, Italy\\
\medskip
TURCK-CHIEZE Sylvaine,	CEA	Gif sur Yvette,	France\\
\medskip
VALDES Marcos,	Scuola Normale Superiore di Pisa,	Pisa,	Italy\\
\medskip
VAN ELEWYCK Veronique,	APC and Universite Paris 7,	Paris,	France\\
\medskip				
VERMA Murli Manohar,	Lucknow University,	Lucknow,	India\\
\medskip
WEINHEIMER Christian,	Institut f\"ur Kernphysik Universit\"at M\"unster, M\"unster, Germany\\
\medskip
YANG Jongmann,	Ewha Womans University,	Seoul,	Korea\\
\medskip
ZANINI Alba,	INFN-Sezione di Torino,	Turin,	Italy\\
\medskip	 	 	 	 
ZHOGIN Ivan,	ISSCM SB RAS,	Novosibirsk,	Russia\\
\medskip
ZIAEEPOUR Houri, Max-Planck Institut f\"ur Extraterrestrische Physik, Garching bei M\"unchen, Germany\\
\medskip				
ZIDANI Djilali,	Observatoire de Paris - CNRS	Paris,	France\\
\medskip

\end{document}